\documentclass[fleqn]{2020SCGE}
\usepackage{titlesec}

\titleformat{\paragraph}
{\normalfont\normalsize}{\theparagraph}{1em}{}
\titlespacing*{\paragraph}
{0pt}{3.25ex plus 1ex minus .2ex}{1.5ex plus .2ex}

\begin{document}

\ensubject{subject}

\ArticleType{Article}
\SpecialTopic{SPECIAL TOPIC: }
\Year{2025}
\Month{xx}
\Vol{xx}
\No{x}
\DOI{??}
\ArtNo{000000}
\ReceiveDate{January xx, 2025}
\AcceptDate{xx xx, 2025}

\title{Introduction to the Chinese Space Station Survey Telescope (CSST)}

\author[1,2,3]{CSST Collaboration: Yan Gong}{gongyan@bao.ac.cn}
\author[1]{Haitao Miao}{}
\author[1,4]{Hu Zhan}{}
\author[5,6]{Zhao-Yu Li}{}
\author[4,7]{Jinyi Shangguan}{}
\author[1]{\\Haining Li}{}
\author[1,2]{Chao Liu}{}
\author[8,9,10]{Xuefei Chen}{}
\author[11,12]{Haibo Yuan}{}
\author[13,14]{Jilin Zhou}{}
% ...
\AuthorMark{Gong Y.}

\AuthorCitation{Gong Y,, Miao H., Zhan H., et al.}


\address[1]{National Astronomical Observatories, Chinese Academy of Sciences, 20A Datun Road, Beijing 100101, China}
\address[2]{School of Astronomy and Space Sciences, University of Chinese Academy of Sciences (UCAS), 19A Yuquan Road, Beijing 100049, China}
\address[3]{Science Center for CSST, NAOC, 20A Datun Road, Beijing 100101, China}
\address[4]{Kavli Institute for Astronomy and Astrophysics, Peking University, Beijing 100871, China}
\address[5]{Department of Astronomy, School of Physics and Astronomy, Shanghai Jiao Tong University, Shanghai 200240, China}
\address[6]{Shanghai Key Laboratory for Particle Physics and Cosmology, Shanghai 200240, China}
\address[7]{Department of Astronomy, School of Physics, Peking University, Beijing 100871, China}
\address[8]{Yunnan Observatories, China Academy of Sciences, Kunming 650216, China}
\address[9]{ Key Laboratory for the Structure and Evolution of Celestial Objects, Chinese Academy of Sciences, Kunming 650011, China}
\address[10]{ Center for Astronomical Mega-Science, Chinese Academy of Sciences, 20A Datun Road, Chaoyang District, Beijing 100012, China}
\address[11]{ School of Physics and Astronomy, Beijing Normal University, Beijing 100875, China}
\address[12]{ Institute for Frontiers in Astronomy and Astrophysics, Beijing Normal University, Beijing 102206, China}
\address[13]{ School of Astronomy and Space Science, Nanjing University, Nanjing, 210023, China}
\address[14]{ Key Laboratory of Modern Astronomy and Astrophysics, Ministry of Education, Nanjing, 210023, China}
% ...

\abstract{The Chinese Space Station Survey Telescope (CSST) is an upcoming Stage-IV sky survey telescope, distinguished by its large field of view (FoV), high image quality, and multi-band observation capabilities. It can simultaneously conduct precise measurements of the Universe by performing multi-color photometric imaging and slitless spectroscopic surveys. The CSST is equipped with five scientific instruments, i.e. Multi-band Imaging and Slitless Spectroscopy Survey Camera (SC), Multi-Channel Imager (MCI), Integral Field Spectrograph (IFS), Cool Planet Imaging Coronagraph (CPI-C), and THz Spectrometer (TS). Using these instruments, CSST is expected to make significant contributions and discoveries across various astronomical fields, including cosmology, galaxies and active galactic nuclei (AGN), the Milky Way and nearby galaxies, stars, exoplanets, Solar System objects, astrometry, and transients and variable sources. This review aims to provide a comprehensive overview of the CSST instruments, observational capabilities, data products, and scientific potential.
}


\maketitle

\begin{multicols}{2}
\section{Introduction}\label{sec:1}

\begin{table*}
\begin{center}
  \caption{Summary of CSST instruments and surveys.}
  \label{tab:CSST_sum}
  \includegraphics[scale=1.1]{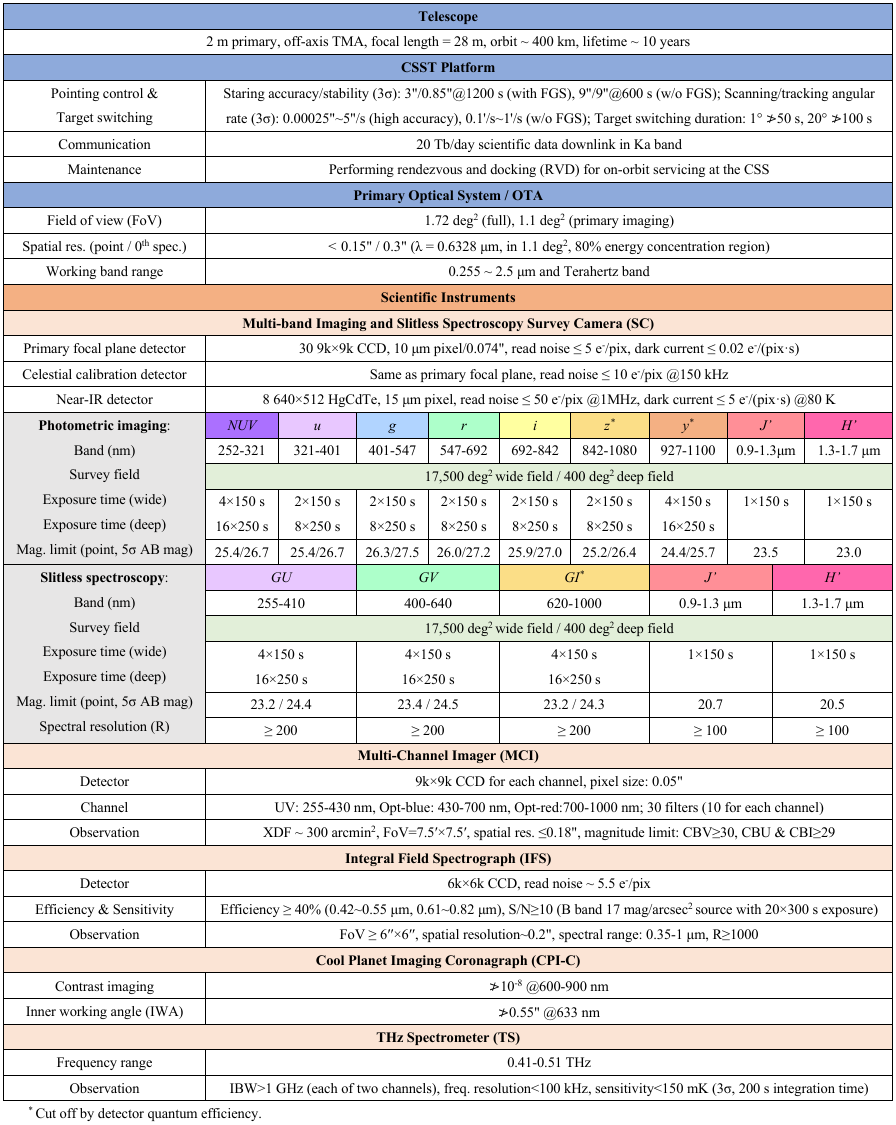}
  \end{center}
\end{table*}

Sky survey is an important means of exploring the Universe. In the past two decades, the development of several sky survey projects has greatly advanced our understanding of the Universe, including the Two-degree-Field Galaxy Redshift Survey (2dF or 2dFGRS) \cite{Cole05}, Six-degree Field Galaxy Survey (6dF or 6dFGS) \cite{Jones09}, Sloan Digital Sky Survey (SDSS) \cite{Eisenstein05}, WiggleZ Dark Energy Survey (WiggleZ) \cite{Parkinson12}, SDSS-III Baryon Oscillation Spectroscopic Survey (BOSS, e.g. \cite{Anderson12}), SDSS-IV extend BOSS (eBOSS, e.g \cite{Gil-Marin20}), etc. Especially in recent years and the near future, with the operation of next-generation photometric and spectroscopic survey projects, e.g. the Dark Energy Spectroscopic Instrument (DESI, e.g. \cite{DESI Collaboration16,DESI Collaboration24}), {\it Euclid} \cite{Euclid24}, Nancy Grace Roman Space Telescope (RST or WFIRST) \cite{Green12}, and Vera Rubin Observatory's Legacy Survey of Space and Time (LSST) \cite{LSST19}, the understanding and solutions to important issues in astronomy are expected to be further promoted.

As a Stage-IV survey telescope, the Chinese Space Station Survey Telescope (CSST) is an off-axis three-mirror anastigmat (TMA) telescope with a primary mirror of 2 meters in diameter, designed to probe the Universe with extremely high levels of precision \cite{Zhan11, Zhan21, Cao18, Gong19, Gong25}. The CSST is scheduled to be launched in 2027, which will be co-orbiting with the China Space Station (CSS or {\it Tiangong}) at an altitude of about 400 kilometers, and it can rendezvous and dock with the CSS for on-orbit servicing (OOS). Aiming at different scientific goals and astronomical targets, it is equipped with five scientific instruments, i.e. Multi-band Imaging and Slitless Spectroscopy Survey Camera (SC), Multi-Channel Imager (MCI), Integral Field Spectrograph (IFS), Cool Planet Imaging Coronagraph (CPI-C), and THz Spectrometer (TS). The design parameters of the telescope and scientific instruments are summarized in Table~\ref{tab:CSST_sum}.

By using these instruments, the CSST can perform photometric imaging and spectroscopic surveys in different fields, including 17,500 deg$^2$ wide field, 400 deg$^2$ deep field, 9 deg$^2$ ultra-deep field (UDF) by CSST-SC, and 300 arcmin$^2$ extreme-deep field (XDF) by CSST-MCI. It can obtain more than one billion galaxy images and one hundred million galaxy spectra, and discover millions of active galactic nuclei (AGN) and other astronomical objects in a large redshift range, covering near-ultraviolet (near-UV), optical, near-infrared (near-IR), and THz bands. The nature of dark energy and dark matter, theory of gravity, and cosmic large-scale structure (LSS) will be fully investigated by precise measurements using several cosmological probes, e.g. weak gravitational lensing, galaxy clustering, galaxy clusters, cosmic voids, Type Ia supernovae (SNe Ia), baryon acoustic oscillation (BAO), etc. The morphology, formation and evolution of galaxies will be carefully studied, as well as AGNs and supermassive black holes (SMBHs). Besides, the CSST will explore the Milky Way (MW) and nearby galaxies, stars, exoplanets, Solar System objects, and transients and variable sources.

In this review, we briefly introduce the CSST instruments, observation strategy, data processing, and discuss its scientific potential. The paper is organized as follows. In Section~\ref{sec:2}, we list the scientific objectives of the CSST. In Section~\ref{sec:3}, the design of the survey platform, primary optical system, and the five scientific instruments are introduced. In Section~\ref{sec:4}, we discuss the CSST observations for different instruments. In Section~\ref{sec:5}, we introduce the CSST data processing and data release. In Section~\ref{sec:6}, the details of the CSST science are discussed. We summarize in Section~\ref{sec:7}.

\section{Scientific Objectives}\label{sec:2}

The CSST scientific objectives focus on important scientific issues in various fields of astronomy, which are divided into primary  and other scientific objectives. In this section, we discuss them in details.

\subsection{Primary scientific objectives}

        	CSST will perform 17,500 deg$^2$ wide-field and 400 deg$^2$ deep-field multi-color photometric imaging and slitless spectroscopic surveys in ten years, and plans to carry out 9~deg$^2$ UDF observations in the first two years. It is expected to obtain huge amounts of photometric and spectroscopic data of a large number of galaxies, AGNs and supernovae. The CSST primary scientific objectives are as follows:
\begin{enumerate}
        	\item {\bf Cosmology}
		\begin{enumerate}
		\item Measure the evolution history of the Universe, reveal the observational characteristics of dark energy, and test general relativity on cosmic scales.
		\item Measure the cosmic structure of dark matter at different scales, test the dark matter theory, and reveal the physical properties of dark matter.
		\end{enumerate}
		CSST will accurately measure the BAO signal in the wide-field survey and the light curves of about 2,000 SNe Ia at $z\lesssim1.3$ in UDF, so as to accurately illustrate the expansion history of the Universe, obtain the evolution information of the equation of state of dark energy, and test the theories of gravity. CSST also will perform high-precision measurements of weak and strong gravitational lensing, galaxy angular and redshift-space clustering, galaxy clusters, cosmic voids, etc. It can provide convincing answers to important cosmological problems, such as spatial distribution and properties of dark matter, neutrino mass, primordial non-Gaussianity, and Hubble constant.
	\item {\bf Galaxy and AGN}
	\begin{enumerate}
		\item Measure the galaxy structure parameters at different redshifts, understand the causes of their morphological diversity, reveal the assembly and mass growth process of galaxies, and establish a panoramic view of the cosmological evolution of galaxies.
		\item Construct a large AGN sample at different redshifts and in different environments, reveal the relationship between black hole activity and star formation and galaxy evolution, and test theoretical models of the co-evolution of massive black holes and host galaxies.
	\end{enumerate}
	CSST will obtain multi-band photometric and slitless spectroscopic data of a large number of galaxies and AGNs in a redshift range from 0 to 7. By measuring the morphology and structure parameters of large samples of galaxies in different environments at different redshifts, it will reveal the evolution law of the galactic bulge, disk and other structures, probe the intrinsic physical process of galaxy formation and the influence of the external environment, and provide an important observational basis for the theoretical model of galaxy formation and evolution. By constructing a large sample of AGN in different environments at different redshifts, the physical parameters of massive black holes and their host galaxies can be measured, and an accurate physical model of the co-evolution of massive black holes and host galaxies will be established.
\end{enumerate}
	
\subsection{Other scientific objectives}

\begin{enumerate}
        	\item {\bf Milky Way and nearby galaxies}
		\begin{enumerate}
		\item Measure the structure, formation, and evolution of the Milky Way and nearby galaxies.
		\item Investigate the interstellar medium (ISM) and galactic ecosystems of the Milky Way and nearby galaxies.
		\end{enumerate}
		CSST will perform resolved star observations for galaxies within 4 Mpc, and deeply explore the structure, formation and evolution of the Milky Way and nearby galaxies, as well as the ISM and galactic ecosystem. It will precisely describe the structure of the Milky Way's bulge, halo and nearby galaxies, and is expected to accurately measure the mass distribution of dark matter; by studying special stars in chemistry and dynamics, it will be able to trace the early history of the Milky Way and nearby galaxies. CSST will measure the interstellar extinction, dust and neutral carbon distribution of the Milky Way and nearby galaxies, and study the relationship between galactic dynamic characteristics and the ISM environment; it will deeply explore the disk-halo interaction, black hole-star feedback and large-scale structural accretion, characterize the baryon circulation process in the galactic ecosystem, and reveal the key factors of galaxy evolution.
	\item {\bf Stars}
		\begin{enumerate}
		\item Study the formation, structure and evolution of stars and explore stellar population.
		\item Detect compact objects and stellar activity.
		\end{enumerate}
		The CSST main survey will obtain photometric data of billions of stars and hundreds of millions of stellar spectra in the Milky Way and nearby galaxies, covering a wide range of star types and stellar populations. At the same time, CSST-IFS can observe protoplanetary disks with high spatial resolution. Based on the CSST main survey and IFS observations, it is expected to comprehensively explore the formation, structure and evolution of stars, accurately characterize the life course of massive stars, and deeply understand the physical mechanisms in stellar evolution; build a large sample of binary stars with high-precision stellar parameters to accurately reveal the key physical processes of binary star evolution; analyze the key physical mechanisms in the star formation process, reveal the characteristics of different stellar populations, discover rare stellar objects and new stellar populations. CSST will detect stellar magnetic fields and stellar coronal mass ejection events, carry out detection of faint stars and white dwarf candidates, greatly expand the sample of compact stars, and establish an optical and UV sample of accreting neutron stars and a sample of black hole binary candidates.\\ 

\begin{figure*}
    \centering
    \includegraphics[width=0.98 \linewidth]{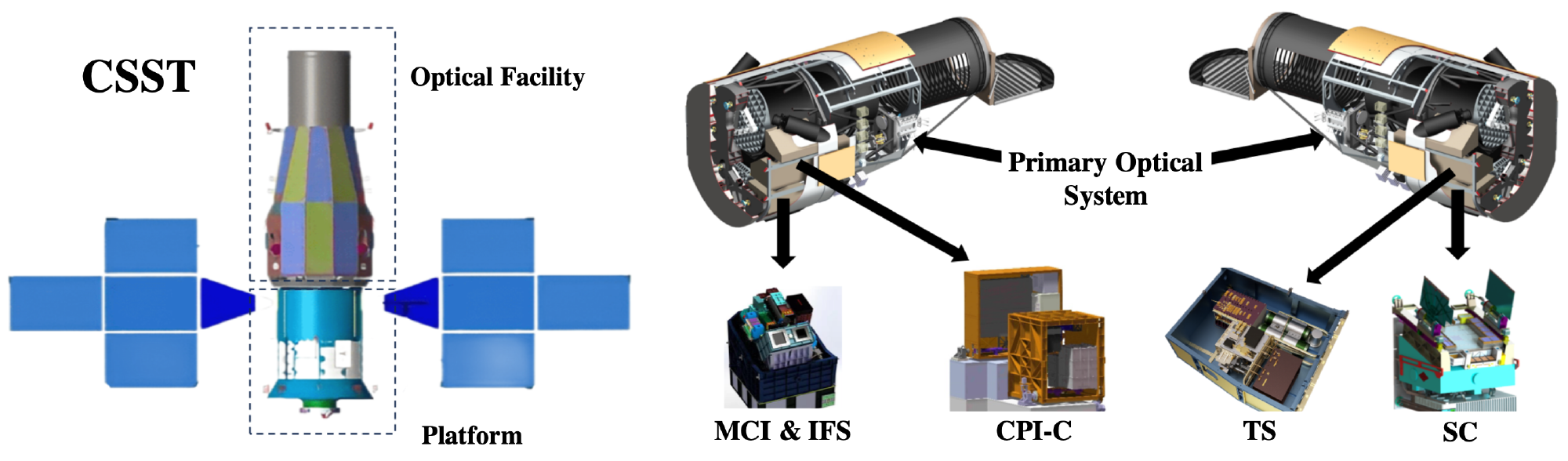}
    \caption{The CSST is composed of the platform and optical facility (left panel). The CSST platform can rendezvous and dock with the CSS for on-orbit servicing. The optical facility includes the primary optical system (or OTA) and five scientific instruments, i.e. SC, MCI, IFS, CPI-C, and TS (right panel).}
    \label{fig:Instruments}
\end{figure*}

	In addition, CSST will also explore exoplanets, Solar System objects, astrometry, transients and variable sources, as briefly described below:
	\item {\bf Exoplanets}
		\begin{enumerate}
		\item Directly image neighboring Jupiter-like planets and protoplanetary disks.
		\item Detect exoplanets in the direction of Galactic center and characterize the atmosphere of exoplanets, to study planetary formation and evolution.
		\end{enumerate}
		
		\item {\bf Solar System objects}
		\begin{enumerate}
		\item Explore the spatial distribution, physical and chemical properties and evolutionary history of small objects in the Solar System, and discover new asteroids, comets, irregular satellites, etc.
		\item Study the faint, irregular satellites of giant planets.
		\end{enumerate}
		
		\item {\bf Astrometry}
		\begin{enumerate}
		\item Achieve milliarcsecond-level positioning and orbit determination, and measure and classify the orbits and types of moving Solar System objects.
		\item Achieve milliarcsecond-level accuracy in the calculation of stellar proper motions and parallaxes, and provide a celestial reference frame for deeper and fainter of the sky.
		\end{enumerate}
		
		\item {\bf Transients and variable sources}
		\begin{enumerate}
		\item Study cosmological parameter measurement and star formation and evolution in the early Universe based on supernova observations.
		\item Investigate gravitational wave events, gamma-ray bursts (GRBs), fast radio bursts (FRBs), high-energy neutrino events, black hole tidal disruption events (TDEs), etc.
		\end{enumerate}
\end{enumerate}

\section{Telescope and instruments}\label{sec:3}

CSST consists of the survey platform and optical facility (see the left panel of Figure~\ref{fig:Instruments}), with a total weight of 15.5 t and a total length of about 16 m in observation state. It will operate independently in the same orbit as the CSS, and can rendezvous and dock with the CSS for on-orbit servicing during its 10-year lifetime. The optical facility is composed of a primary optical system or optical telescope assembly (OTA) and five scientific instruments, i.e. SC, MCI, IFS, CPI-C, and TS. These instruments share the primary optical system, as shown in the right panel of Figure~\ref{fig:Instruments}. The details of the CSST instrumental design can be found in Table~\ref{tab:CSST_sum}.

\subsection{CSST platform}

CSST platform and survey optical facility adopt integrated design and development‌ methodology for the CSST scientific objectives. It can provide high-precision and high-stability pointing for both staring and tracking/scanning modes, rapid attitude maneuvering and switching of astronomical targets, electrical power management, and space-ground communication support. The survey platform is responsible for the attitude control, autonomous orbit maintenance, active rendezvous and docking during the 10-year mission. It is the first long-life low Earth orbit (LEO) platform that takes the CSS as the spaceport and can accept on-orbit services.

\subsection{Primary optical system}

The CSST primary optical system adopts the Cook type off-axis TMA design. Its primary mirror has an aperture of 2 m and a focal length of 28 m with a full field of view (FoV) of 1.72 deg$^2$ and a FoV of $1.1^{\circ}\times1^{\circ}$ in the primary imaging area. Within this area, the dynamic image quality can reach $R_{\rm EE80}<0.15''$, where $R_{\rm EE80}$ is the radius of the point spread function (PSF) of 80\% energy concentration region. Around it, there are short-wave infrared areas, wavefront sensing areas, fine guidance sensor (FGS) areas, etc. As shown in the right panel of Figure~\ref{fig:Instruments}, the primary optical system is equipped with six image plane interfaces, which is symmetrically distributed on both sides of the primary mirror and tertiary mirror, corresponding to the five scientific modules including SC, MCI, IFS, CPI-C and TS, as well as one reserved interface. The switching between the focal planes is achieved by the module switching and the steering of the focusing mechanism. The CSST ensures the stability of the optical axis orientation during the observation through precise image stabilization systems (e.g. FGS and fast steering mirrors), passive vibration isolation and other means.

\subsection{Survey camera}

The SC is used to perform the main tasks of CSST multi-band photometric imaging and slitless spectroscopic surveys, which is composed of the main focal plane and the celestial calibration component, the short-wave infrared component, the filter component, the slitless spectroscopic component, calibration light source, etc. 

\begin{figure}[H]
\centering
\includegraphics[scale=0.45]{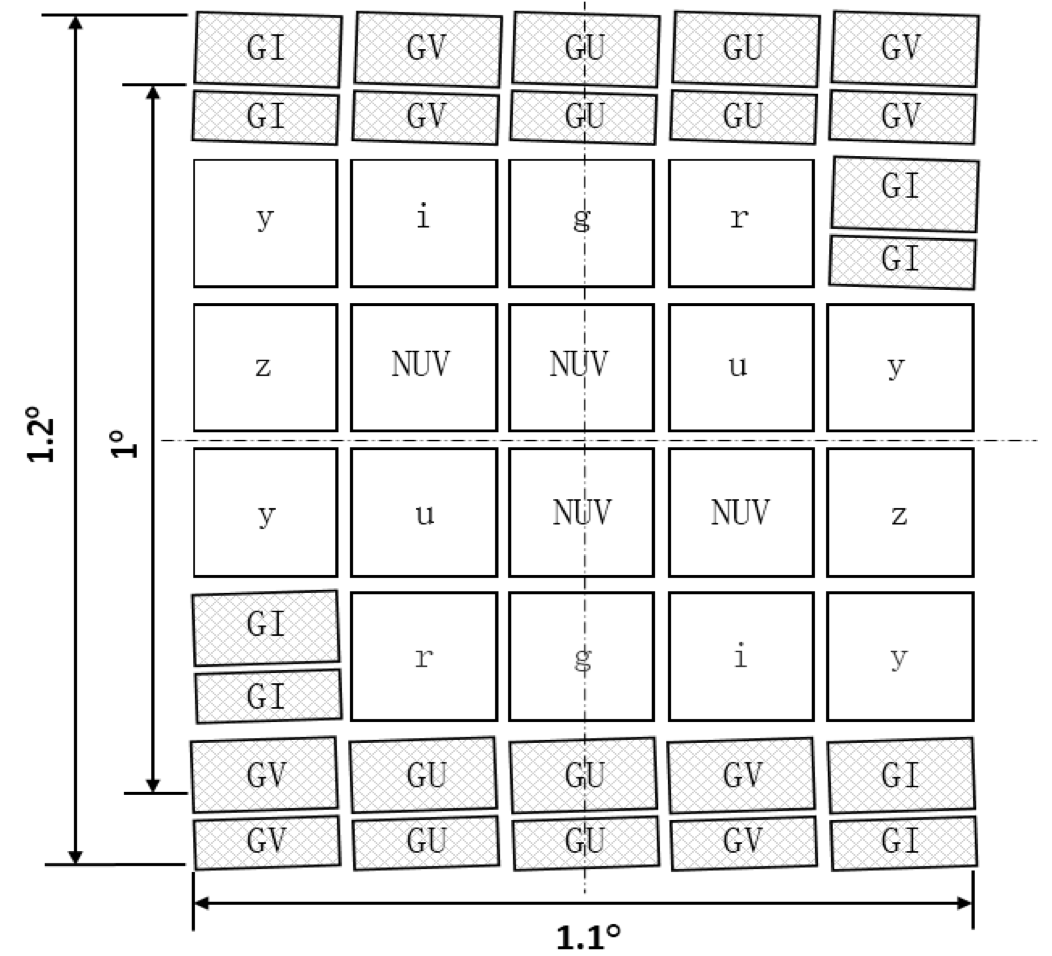}
\caption{The Layout of the main focal plane of the CSST-SC.} 
\label{fig:focus}
\end{figure}

\begin{figure}[H]
\centering
\includegraphics[scale=0.32]{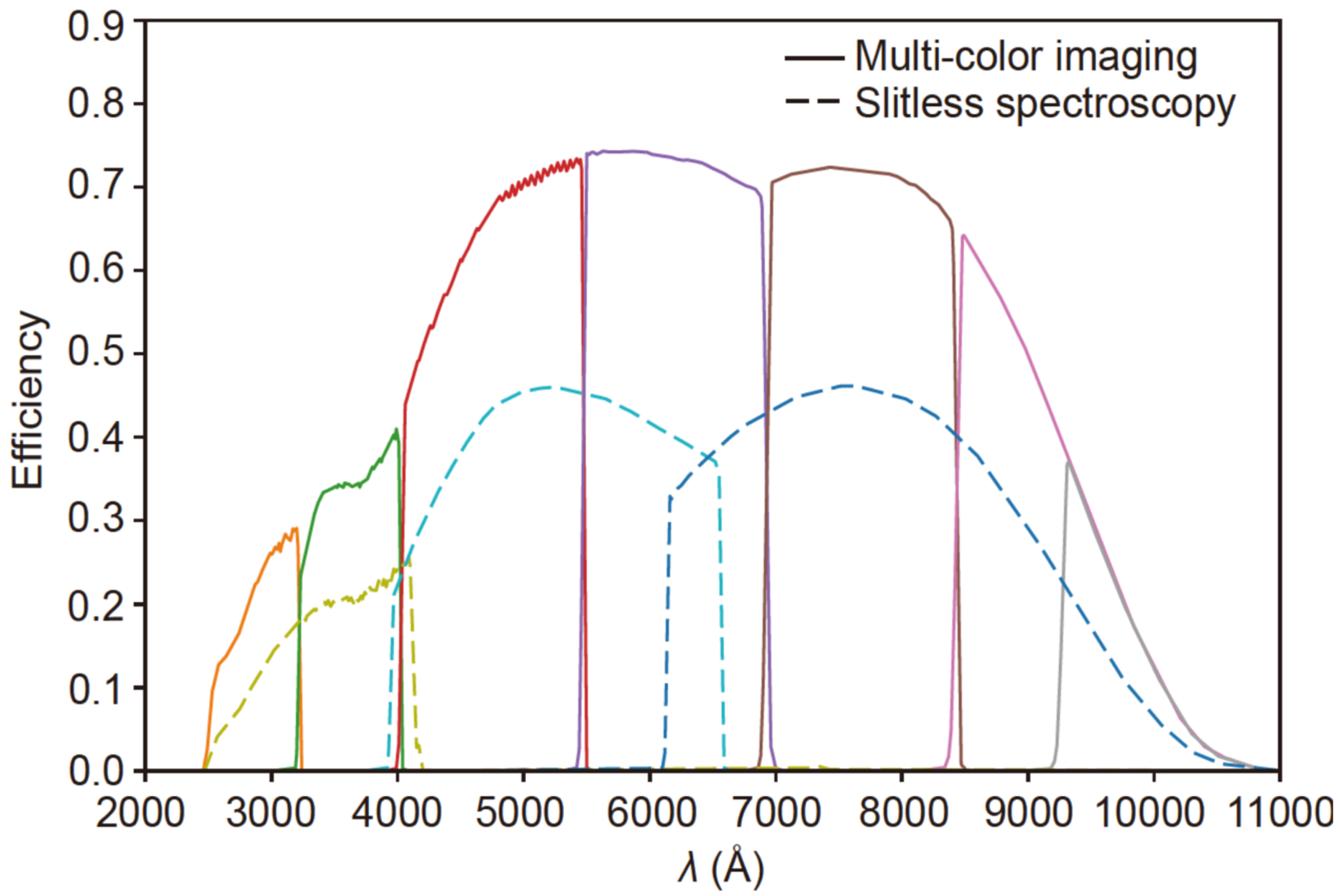}
\caption{The transmission curves of the seven photometric imaging bands, i.e. $NUV$, $u$, $g$, $r$, $i$, $z$, and $y$ (solid curves), and three slitless spectroscopic bands, i.e. $GU$, $GV$, and $GI$ (dashed curves) for the CSST-SC \cite{Zhan21}.} 
\label{fig:filter}
\end{figure}

The focal plane contains 2.6 billion pixels, corresponding to an angular size of $0.074''$ per pixel, a FoV of $\sim$1.1 deg$^2$ and dynamic image quality $R_{\rm EE80}<0.15''$. The layout of the main focal plane is shown in Figure~\ref{fig:focus}. The main focal plane is divided into 7 photometric imaging bands (i.e. $NUV$, $u$, $g$, $r$, $i$, $z$, and $y$) and 3 slitless spectroscopic bands ($GU$, $GV$, and $GI$, with an average spectral resolution of $R\ge200$), covering the wavelength range of 255-1000 nm. The transmission curves of the seven photometric bands and three slitless spectroscopic bands are shown in Figure~\ref{fig:filter}. The photometric imaging bands are located in the central area, and the slitless spectroscopic bands are arranged in the upper and lower ``L''-shaped areas. The charge-coupled devices (CCDs) of the main focal plane are fixedly installed with the filters or gratings.

Slitless spectroscopic observation is achieved by direct dispersion in the imaging optical path using the transmissive blazed gratings. Each CCD corresponds to two gratings, and the working order of each grating is biased towards the direction of the opposite grating, thereby ensuring that the spectra of all astronomical objects within the corresponding FoV of the CCD fall on the photosensitive surface of this CCD. To alleviate the problem of spectral overlap of astronomical objects, the dispersion directions of the upper and lower gratings are rotated relatively, and the rotation angle is only 2$^\circ$ due to structural limitations. In the case that the effective radius of most galaxies detectable by the slitless spectroscopy is less than $0.3''$ and the surface density is only a few per square arcminute, a 2$^\circ$ rotation can work effectively.

The CSST-SC plans to conduct cumulative observations for approximately 7 years during the 10-year orbital period, obtaining wide-field survey data of 7-band photometric imaging and 3-band low-dispersion slitless spectroscopy in a 17,500 deg$^2$ sky area, as well as deep-field survey data of 400 deg$^2$ in these bands. At the same time, the near-IR two-band (i.e. $J'$ and $H'$) imaging and slitless spectroscopic observations covering 0.9-1.7 $\rm \mu m$ will be carried out in small sky areas. 

The CSST wide-field survey mainly covers the mid-high galactic latitude and mid-high ecliptic latitude regions. The $g$-band magnitude limit can reach 26 (for point source, 5$\sigma$, AB magnitude). It is expected to obtain the photometric data of more than 1 billion galaxies and 1 billion stars, as well as hundreds of millions of spectra. It will be the most important dataset for CSST frontier research, and will also provide high-value observational targets for the other four CSST scientific instruments and other telescopes.

\subsection{Other instruments}

\subsubsection{MCI}

The MCI divides the optical path into three channels: near-UV (255-430 nm), optical-blue (430-700 nm) and optical-red (700-1000 nm), achieving the function of simultaneous imaging of the three channels. It also adopts a relay optical system for focus of expansion to obtain an angular size of $0.05''$ per pixel and a dynamic image quality $R_{\rm EE80}\le0.18''$. Each channel of MCI uses a CCD to cover the $7.5'\times7.5'$ FoV, and the filter is switched through a wheel to meet the observational requirements of different scientific goals \cite{Zheng25}.

The MCI is equipped with a total of 30 filters, with 10 filters for each channel. Among them, 4 filters cover the entire FoV, and 6 filters cover 1/4 of the FoV. The near-UV channel $CBU$ filter, optical-blue $CBV$ filter, and optical-red $CBI$ filter basically cover the spectral range of the channel, and will be used in the observation of the MCI XDF of $\sim$300 arcmin$^2$, with magnitude limits  $CBV\ge30$, and $CBU$ and $CBI\ge29$. The filters $NUV$, $u$, $r$, $z$ and $y$ can be used to assist the flux calibration of the SC. The remaining ones include the narrow-band, mid-band and broad-band filters, mainly used in scientific research such as high-redshift (high-$z$) and neighboring galaxies, and ionized gas \cite{Zheng25,Cao22, Li22}.

The key tasks of MCI include establishing high-precision flux calibration catalog of standard stars and conducting the UV and optical XDF observations. The former provides a guarantee for the CSST survey to achieve high-precision photometry, while the latter will achieve an unprecedented exposure depth in the UV and optical bands, and play a key role in the research of the joint evolution of galaxies and black holes, strong gravitational lensing of galaxy clusters, SN Ia cosmology, asteroids and drastically changing objects.

\subsubsection{IFS}

The IFS uses an image splitter with a cutting unit of $0.2''$ to divide the FoV into 32 units, and disperses these units within $0.35-1\ \rm \mu m$ in two bands of red and blue to obtain the spectrum of each unit. It has a FoV greater than $6''\times6''$, a spatial resolution of $\sim0.2''$, and a spectral resolution $R\ge1000$, covering the entire optical band of $0.35-1\ \rm \mu m$ in the three-dimensional spectroscopic observation.

The IFS transmission efficiency combined with the detector efficiency is more than 40\% on average in the two main red and blue bands. The detection capability of the surface source requires that the signal-to-noise ratio (S/N) of a single spectral resolution unit for the target source with a surface brightness of 17 mag/arcsec$^2$ in the B band should be $\ge$10 after $20\times300\ \rm{s}$  exposure stacking. The detector readout noise is approximately $5.5\, e^-\rm/pixel$, which is usually the main source of noise during IFS observations. In practice, the detectors can be combined for read out, which can provide the possibility of further improving the detection efficiency for sciences that do not require high spectral resolution or spatial resolution.

The observation of IFS can simultaneously obtain the two-dimensional structure and spectral information of the target, which is suitable for the research that requires spatially analyzing the chemical composition or physical properties of the observational target, such as the physical properties of the core region near the SMBH at the center of the galaxy, the co-evolution of galaxies and black holes, star formation in specific environments within galaxies, the dynamic properties of the source galaxy of the strong gravitational lensing and its dark matter distribution, Galactic nebulae, TDEs,  Solar System objects, etc.

\subsubsection{CPI-C}

The CPI-C adopts pupil modulation and high-precision wave aberration correction technology to suppress the diffraction photons and speckle noise generated by the primary optical system, achieving ultra-high contrast imaging of exoplanets ($\le10^{-8}$ in 600-900 nm, and the high-contrast inner working angle (IWA) is not greater than $0.55''$ at 633 nm). 

The CPI-C aims to search for mature Jupiter-like planets and super-Earths located in nearby solar-like stars from the habitable zone to the snow line (0.8-5 AU). It can perform ultra-high contrast direct imaging detection and multi-band photometry research on these targets, and carry out high-contrast imaging detection and studies on circumstellar disks and zodiacal dust, providing important observational evidence for the theory of planet formation and evolution.

\subsubsection{TS}

The TS mixes the $0.41-0.51$ THz signal received by the primary optical system with the signal generated by the local oscillator signal  on a superconducting mixer (the refrigerator provides an operating temperature of 8 K, and the mixer works in the superconducting state), and then performs Fourier transform on the intermediate frequency signal obtained by mixing to obtain the spectral information of the observed signal. The instantaneous bandwidth (IBW) of each of the two TS channels is over 1 GHz, the frequency resolution is better than 100 kHz, and the instrument sensitivity with 200 s integration time is better than 150 mK (3$\sigma$).

TS will be used to conduct spectral line surveys and CI mapping observations, detect the chemical composition of astronomical objects and ISM, search for new molecular species, reveal the atomic and molecular phase transition processes in galaxy evolution, study the formation and evolution mechanisms of molecular clouds, and combine with other multi-band observational data to deepen the understanding of the structure formation, dynamical evolution, chemical evolution and star formation process of nearby galaxies.

\section{Observation}\label{sec:4}

After CSST enters orbit, it will perform on-orbit testing to fully verify the system functions and performance, complete the initial calibration, and obtain the operation control parameters. After that, it will enter the scientific operation phase, including early scientific observations, multi-color imaging and slitless spectroscopic surveys, establishment of a high-precision flux standard star catalog, XDF observations, free application observations, and target of opportunity (ToO) observations.

\subsection{Observation strategy}

The survey strategy of the CSST main survey is determined by the survey scheduling algorithm. The algorithm assigns weights to various constraints and requirements (e.g. orbital environment, hardware, and operation control), selects the current optimal direction, forms a direction sequence of about 6 to 7 years, and adjusts the weights to meet the requirements of survey depth and area, as well as a small number of other observation requirements.

The MCI, IFS, CPI-C and TS each observe for about 4.5 months during the scientific operation period (excluding parallel observations), which mainly depends on the observation window of the targets or areas within one year and one orbit. Usually, several pointing directions are arranged in one orbit for observation. 

For the ToO observations, the CSST can point to the right position in a few minutes. Considering it may take at least several hours for a scientific module to start up and cool down, observations requiring fast response should use the currently working module.

\subsection{SC observation}

\begin{figure*}
\centering
\includegraphics[scale=0.52]{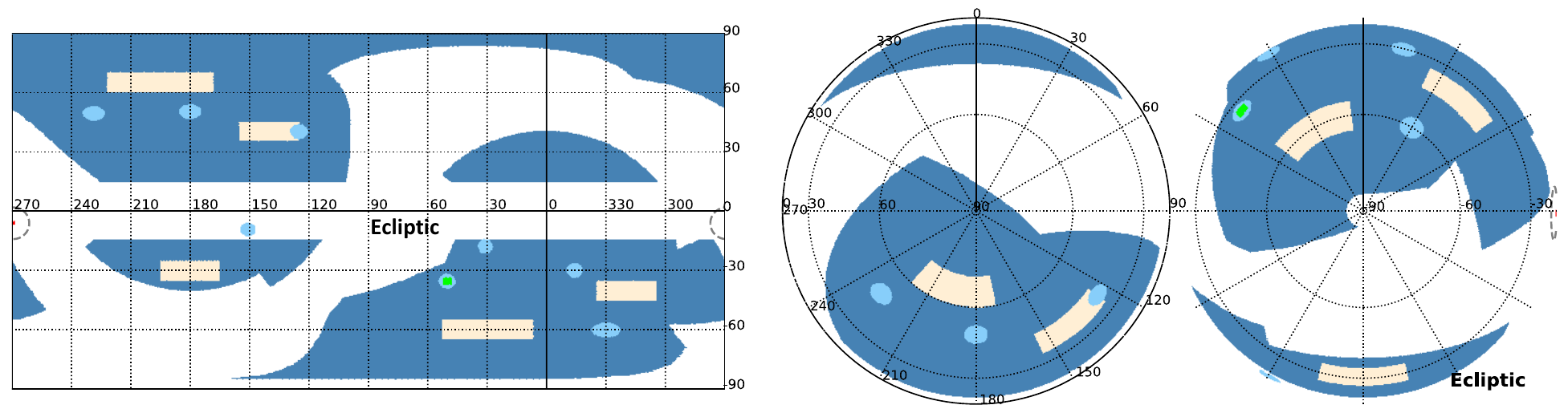}
\caption{The preliminary planned survey fields of the CSST-SC (ecliptic coordinates). The deep and light blue regions are the survey areas of the CSST 17,500 deg$^2$ wide-field and 400 deg$^2$ deep-field surveys, respectively. The yellow regions show the selected sky areas that {\it Euclid} has completed as CSST priority observations. The 9 deg$^2$ UDF and microlensing observations of Galactic bulge area are shown in green and red regions, respectively.} 
\label{fig:survey}
\end{figure*}

The CSST main survey performed by the SC aims to achieve the following goals:

\begin{enumerate}
\item {\bf 17,500 deg$^2$ multi-color imaging wide-field survey}: Focus on observing the middle and high Galactic and ecliptic latitudes (dark blue regions in Figure~\ref{fig:survey}). Each sky area is covered by each photometric band at least twice, with a single exposure time of approximately 150 s. The magnitude limit of the $g$ band is not less than 26 mag, and the average magnitude limit of the $NUV$, $u$, $r$, $i$, and $z$ bands is not less than 25.5 mag.

\item {\bf 17,500 deg$^2$ siltless spectroscopic wide-field survey}: The survey area is the same as that of the multi-color imaging wide-field observation, and the two observations are performed simultaneously. The magnitude limit of the $GU$ band is not less than 22 mag, and that of the $GV$ and $GI$ bands is not less than 23 mag.

\item {\bf 400 deg$^2$ multi-color imaging deep-field survey}: Multiple sky areas will be selected throughout the entire sky, with a total area of not less than 400 deg$^2$ (e.g. the light blue regions shown in Figure~\ref{fig:survey}). The magnitude limit is at least one magnitude deeper than the multi-color imaging wide-field survey. Each sky area is covered by each photometric band at least 8 times, and the single exposure time is approximately 250 s.

\item {\bf 400 deg$^2$ slitless spectroscopic deep-field survey}: The sky area is the same as that of the multi-color imaging deep-field survey, and both observations are performed simultaneously. The magnitude limit of the $GU$ band is not less than 23 mag, and that of the $GV$ and $GI$ bands is not less than 24 mag.
\end{enumerate}

The preliminary results of the survey simulation show that the SC can complete 19,500 deg$^2$ of wide-field and 450 deg$^2$ of deep-field in 7-year observation, with nearly 700,000 exposures, and an average of nearly 300 shots per day. In actual operation, it is expected that 10\% of the exposures will be used for calibration observations, and the area that can be completed will be reduced accordingly.

In addition, the SC may also perform other observations, such as selecting priority observations of some sky areas that {\it Euclid} has completed (e.g. the yellow regions in Figure~\ref{fig:survey}), 9 deg$^2$ UDF observations ($60\times250$ s, $g\ge28$, e.g. the green region in Figure~\ref{fig:survey}), microlensing observations of Galactic bulge area (red region in Figure~\ref{fig:survey}), and observations of the Milky Way and nearby galaxies.

\subsection{MCI observation}

The key tasks of the MCI are: 
\begin{enumerate}
\item Establish the CSST high-precision flux standard star catalog, assisting the SC and other scientific instruments to achieve high-precision flux measurement of astronomical objects.

\item Carry out XDF observations, with a depth of 0.2 magnitude deeper than the Hubble Space Telescope (HST) XDF, achieving the deepest observations in the near-UV and optical bands.
\end{enumerate}
In addition, MCI will also be widely used in the observation and research of static universe and time domain astronomy.

The flux standard stars include primary standard stars (e.g. GD71, GD153 and G191B2B) and MCI-selected secondary standard stars (18-19 mag, evenly distributed throughout the sky). The MCI will observe using the same $u$, $r$ and $z$ bands as the SC, and the cumulative S/N of each band after multiple observations is greater than 1000. The flux standard stars can be arranged for observation in the sunlit area and revisited once a year.

The MCI XDF observations will be conducted in the $CBU$, $CBV$ and $CBI$ bands in the Earth's shadow region, with a cumulative exposure of 510,000 s in each band and magnitude limits of no less than 29, 30, and 29, respectively. Six directions will be selected for the XDFs, totaling 300 arcmin$^2$, of which four directions will be used to observe galaxy clusters. The HST XDF is tentatively selected as one of the CSST XDFs, and is planned to be completed in the first two years of scientific operation. The remaining XDFs will be completed roughly simultaneously in the following seven years.
MCI can conduct observations in parallel with IFS or CPI-C, obtaining additional valuable observational data.

\subsection{IFS observation}

The IFS mainly observes brighter sources, such as SMBHs and their surrounding environments, star forming regions, gravitational lensing objects, and planetary nebulae, bulge regions and star clusters in the nearby Universe. The IFS FoV is only $6''\times6''$, and the direction needs to be accurately adjusted through the survey platform and fast steering mirror to ensure effective coverage of the targets. Most IFS targets require multiple exposures to accumulate the spectral S/N. If each target needs to be exposed for 2 hours, based on the IFS observation window in the current CSST observation arrangement, it is expected that about 600 targets can be observed.

IFS can carry out observations in parallel with MCI or CPI-C. When other modules are used as the main observation, IFS can capture the sky background or perform internal calibration.

\subsection{CPI-C observation}

The CPI-C uses thousands of FGK-type stars within 40 pc of the Sun as its main targets, searching for possible planets near these stars through high-contrast imaging observations. The distances of these stars from the Earth are much smaller than the thickness of the Milky Way disk, so they are basically evenly distributed throughout the sky.

The CPI-C adopts the staring mode for observation. In order to achieve better calibration of system aberrations and avoid FoV occlusion, the telescope can be rotated around the visual axis by a certain angle after completing a staring observation. Through the quasi-static aberrations at different rotation angles and the position changes of the exoplanet target, the quasi-static aberrations can be partially removed to improve the target S/N. Such continuous shooting conditions need to be considered when arranging observations.

\subsection{TS observation}

The TS performs observations mainly around the following three scientific goals:
\begin{enumerate}
\item Galactic gas cycle:  On-The-Fly (OTF) mapping observations of nearby molecular clouds, CO dark clouds, massive non-nebular cores, massive protostellar objects, and ultra-compact HII regions, with a total effective observation time of 450 hours.
\item THz spectral line survey: Single-point deep observation of high and low-mass star-forming regions, carbon stars, and low-mass protostars. The cumulative effective observation time for a single source is expected to be $50-100$ hours.
\item M31 neutral carbon survey: CI mapping observations of M31 are conducted with an area of $4^{\circ}\times 1^{\circ}$ in the OTF mode for approximately 450 hours.
\end{enumerate}

Candidate targets have been collected for these scientific objectives, and after simulation verification, observations of these sources can be completed within the allocated observation windows.

\subsection{ToO observation}

ToO are observations that are arranged in a short period of time for targets that cannot be pre-specified under certain triggering conditions, such as important astronomical events. CSST will arrange rapid response observations with appropriate time when necessary to seize the extremely rare scientific opportunities provided by time-domain astronomy.

\section{Data}\label{sec:5}

\subsection{Data processing}

The raw data generated by CSST in its ten-year operation is close to 5 PB. After processing, at least 30 PB of data products will be generated. The CSST scientific data processing system runs in a cloud computing environment and utilizes high-throughput computing capabilities to complete the processing of massive data.

The CSST Scientific Data Processing System generates level-1 and level-2 data products based on the original observational data (level-0 data). Level-1 data refers to scientific data that has been corrected for instrument effects and calibrated. Level-2 data refers to data products that have been further analyzed on the basis of level-1 data to form catalogs and complete the extraction of preliminary scientifically usable parameters, including model photometry of galaxies, photometric redshift (photo-$z$), stellar atmospheric parameters, etc. The CSST data products include:

\begin{enumerate}
\item {\bf SC multi-color imaging data}: To form level-1 scientific image data, the output images of each photoelectric detector are processed by instrumental effect correction and cosmic ray effect removal, and position and flux calibration are also performed. On this basis, level-2 merged image data (multi-color images of multiple observations of the same sky area are merged to increase the magnitude limit) and level-2 catalog data are formed. The catalog data will include the position, flux, morphology and other parameters of the detected astronomical objects and their errors.

\item {\bf SC astrometric data}: Carry out target extraction, centering and astrometry of single-frame images to generate level-1 astrometric scientific data. Carry out astrometric parameter estimation for the sky area that has accumulated multiple observation data. Identify the fast-moving Solar System objects in the image, and determine the orbits and classification for the targets of multiple observations. Finally, generate level-2 astrometric data products including astrometric catalogs and Solar System fast-moving object ephemerides with parameters such as the three-dimensional position and motion of astronomical objects.

\item {\bf SC slitless spectroscopic data}: The level-0 two-dimensional spectral image of each photodetector is processed to generate the level-1 two-dimensional spectral scientific image data. After flux and wavelength calibration, the level-1 one-dimensional spectral data are generated. On this basis, the level-2 catalog data are generated, which will include parameters such as redshift, information of emission line and their errors.

\item {\bf MCI data}: Instrumental effect correction, background light and cosmic ray removal, and calibration are performed to form level-1 scientific image data. Based on the above data, the merged image data, image subtraction data, and level-2 catalog data are generated according to scientific needs.

\item {\bf IFS data}: Perform instrumental effect correction, background light and cosmic ray removal, spectrum tracking and other processing to generate the level-1 single-exposure three-dimensional spectrum data cube, and perform multiple exposure stacking, wavelength and flux calibration to generate the level-1 three-dimensional spectrum data cube. Further perform spectral line and continuous measurement processing on the level-1 data, and use the measurement results to generate the level-2 scientific data cube.

\item {\bf CPI-C data}: Perform instrumental effect correction, background light and cosmic ray removal and other processing to generate level-1 image data.

\item {\bf TS data}: Pointing correction, frequency and speed correction, position correction and efficiency correction are performed to form the standard spectral line level-1 data with fixed integration time. The spectral line data of the same sky area and the same frequency are merged to generate the level-2 data product.

\item {\bf On-orbit calibration data}: 
	\begin{enumerate}
	\item On-orbit calibration of focal plane instruments: Each focal plane instrument compares the processing results of the fluxes and positions of the standard stars in each band with the known results, and obtains the corresponding calibration data as the on-orbit calibration data product for other observational data processing. After the reference image data collected by each focal plane instrument is processed by image merging, etc., the level-1 reference image data is formed and used in the observational data processing.
	\item Cross-calibration between instruments: The MCI can conduct dedicated on-orbit calibration observations of the SC standard stars for the calibration processing of the SC. The IFS needs to be accompanied by the MCI during observation to obtain calibration data. The data is processed according to the normal data processing pipeline of the MCI and forms the calibration data products required by the IFS.
	\end{enumerate}

\item {\bf Simulation data}: Before entering orbit, simulated CSST observations are performed based on the design and operation characteristics of the primary optical system and scientific modules. The output photometric images and spectral data are used to verify the effectiveness of the scientific data processing pipeline. After entering orbit, as the algorithms and parameters of the data processing pipeline are updated, the same simulation research of observational data still will be carried out.
\end{enumerate}

\subsection{Data management}

China Manned Space Program uses a space-based tracking and control system supported by the ``Tianlian'' relay satellites. In the CSST mission, the tracking and control communication system is responsible for both data downlink and tracking and control management. The scientific and engineering data packets of the CSST are downlinked to the ground stations through relay satellites, and then transmitted to the Space Application System. The tracking and control coverage is arranged according to demand, and the data downlink rate is 1.2 Gbps.

The Space Application System is responsible for establishing the ground facilities of CSST, including the Payload Operation Management Center, the Science and Application Data Center, the Scientific Data Processing System, and the Astronomical Data Science Center. The Payload Operation Management Center is the hub of CSST operation, and is responsible for CSST operation management and status monitoring. The Science and Application Data Center is responsible for the construction and operation of the public cloud, and the quality analysis, archiving, storage and external release of the level-0$\sim$level-2 scientific data products. The Scientific Data Processing System  generates the level-1 and level-2 data based on the public cloud platform.

In addition, the Digital Telescope System is also under development, providing digital support for understanding the CSST operating status and abnormal monitoring. To support CSST scientific data research, the CSST Data Science Center is established, which connects to the public cloud system. On the one hand, it can provide the storage and computing resources for CSST long-term tasks, and on the other hand, it can make full use of the artificial intelligence (AI) technology.

\subsection{Data release}

The data policy of CSST is similar to that commonly used in the international astronomical community. The specific regulations of CSST on the protection objects, protection period setting, and regular release of scientific data at all levels will be announced later. The data planned for regular release includes:
\begin{itemize}

\item Simulation data: Numerical simulation data of the CSST generated by the scientific data system, including survey strategy simulation, mock catalog, simulated observational data including instrument effects and various physical effects, etc.

\item Level-0 scientific data: Photometric images, spectra and other data generated by telescope observations after protocol unpacking and regularization, as well as auxiliary data from additional platforms and instruments, such as pointing, exposure time, shutter characteristics, detector parameters, FGS pointing solution, etc. Other downlink engineering data are generally not released separately, and scientific users can obtain them after submitting an application according to the process.

\item Level-1 scientific data: Single-exposure imaging data and two-dimensional slitless spectroscopic images after deducting various instrument effects and completing calibration, spectral data cube, etc. Level-1 data also includes relevant additional information, such as observation information (e.g. exposure time, observation quality, calibration file history of the observed object, etc.), PSF model, optical distortion image, shielding area (mask), etc.

\item Level-2 scientific data: Merged imaging data, one-dimensional spectral data; list of basic physical property parameters of astronomical objects, including photometric and spectral data, and basic physical parameters and their errors obtained after data processing.
\end{itemize}

After the latest raw observational data is downloaded and the above data products are generated, the intermediate products will be provided to the CSST scientific team during the internal testing period for data quality verification. The publication of papers during the internal testing period must go through a certain review process. The official data product is regularly released to domestic and foreign scientific teams every two years, and the final data includes all the observational data over a 10-year (or full life cycle) period.

The data collected by CSST for free application observations are subject to a certain applicant-exclusive research protection period, after which the data products are released publicly.

\section{Science}\label{sec:6}

The CSST Science Committee is organized to guide CSST scientific research and review top-level scientific policies. Five science centers have been established, i.e the Peking University Science Center, the NAOC (National Astronomical Observatories, Chinese Academy of Sciences) Science Center, the Yangtze River Delta Region Science Center, the Guangdong-Hong Kong-Macau Greater Bay Area Science Center, and CSST Data Science Center, to provide support platforms for the CSST scientific teams. The NAOC Science Center also serves as the CSST Joint Science Center, collaborating with all the science centers to jointly plan and manage CSST scientific projects.

Based on the CSST scientific objectives, CSST scientific teams will conduct extensive and in-depth scientific research in relevant fields, such as cosmology, galaxy and AGN, the Milky Way and nearby galaxies, stars, exoplanets, the Solar System objects, astrometry, and transients and variable sources. In this section, we discuss the scientific potential of the CSST in these fields.

\subsection{Cosmology}

Cosmology is the science that explores the origin and evolution of the Universe, the composition and properties of matter, and the laws of fundamental interactions. The development of theories and observations over the past century has greatly deepened the understanding of the Universe, and established a picture of the evolution of the Universe from the beginning to $\sim$14 billion years. In this framework, the Universe experienced an extremely early inflation phase dominated by the quantum vacuum field, and then entered the phase dominated by radiation and matter. At the beginning of the Big Bang, the Universe was in a high temperature and density state, and matter in the Universe existed in the state of relativistic elementary particles. As the Universe expanded, the temperature gradually dropped, and baryons, atomic nuclei, and atoms were formed. Subsequently, stars, galaxies, galaxy clusters and so on gradually emerged, eventually forming the current Universe with rich structures.

However, there are still many fundamental questions that remain unanswered at present, including the physical mechanism of the early inflation process, the nature of dark energy and dark matter, the mass and properties of neutrinos, and the fundamental properties of gravity, etc. These are major issues in cosmological research in the 21st century and the main scientific goals of various cosmological survey projects.

As one of the Stage-IV survey projects, CSST will utilize its high-resolution, multi-band photometry and slitless spectroscopic surveys to provide high-quality and massive data for cosmological research. It will conduct research on the above major scientific issues through cosmological probes, such as gravitational lensing, galaxy clustering, galaxy clusters, supernovae, etc., and collaborate with other observational projects.

\subsubsection{Dark energy}

Explaining the accelerating expansion of the Universe is one of the greatest challenges in modern astronomy. According to Einstein's general theory of relativity, a universe that only contains matter (e.g. baryonic matter, dark matter, photons, neutrinos, etc.) can only expand at a deceleration rate but not at an acceleration rate. Theoretically, accelerating expansion implies that either about 70\% of the energy in the Universe is provided by the so-called ``dark energy'' with negative pressure, or that general relativity needs to be modified on cosmological scales.

The physical property of dark energy is mainly determined by the ratio of its pressure $p$ to energy density $\rho$, i.e. the equation of state $w\equiv p/\rho$. For example, in the $\Lambda$CDM model, dark energy is the cosmological constant and can be seen as vacuum energy, that we have $w=-1$. On the other hand, in dynamical dark energy models, $w$ is a function of the redshift $z$, which can be seen as a scalar field, e.g. quintessence with $w>-1$ \cite{Ratra88}, phantom with $w<-1$ \cite{Caldwell02}, and quintom with $w$ crossing $-1$ \cite{Feng05}. Therefore, it is crucial to reconstruct the evolution history of $w$ from astronomical observations for studying the nature of dark energy.

One of the commonly used forms of $w$ is the Chevallier-Polarski-Linder (CPL) parameterization \cite{Chevallier01,Linder03}, which takes the form as
\begin{equation}
w = w_0 + w_a(1-a) = w_0 + w_a \frac{z}{1+z},
\end{equation}
where $a=1/(1+z)$ is the scale factor, and $w_0=w\, (a=1)$ and $w_a=-{\rm d}w/{\rm d}a$. Therefore, the two parameters have clear physical meanings, and different dark energy models can be naturally classified in the $w_0-w_a$ parameter space. Besides the parameterization method, we can also use the non-parameterization method to study the property of dark energy without assuming the form of $w(z)$ to obtain more accurate result in principle.

At present, various cosmological probes have been used to precisely measure the properties of dark energy, including SN Ia \cite{Brout22, Rubin25, Abbott25}, BAO \cite{Alam21, Abbott25b}, weak gravitational lensing \cite{Burger24, Gomes25}, cosmic microwave background (CMB) \cite{Aghanim18,Calabrese25,Camphuis25}, etc. In particular, the current constraint results from DESI BAO measurements tend to favor the dynamical dark energy model over the $\Lambda$CDM model \cite{Karim25,Lodha25,Gu25}, which is in tension with the results from previous measurements. This indicates that we need more accurate data from different cosmological probes to further investigate the evolution of dark energy with redshift.

CSST can explore the nature of dark energy with high precision by various cosmological probes in a large redshift range, e.g. weak and strong gravitational lensing, galaxy angular and redshift-space clustering power spectra, galaxy cluster, cosmic void, SN Ia, BAO, redshift-space distortion (RSD), etc.

\paragraph{Weak gravitational lensing}

The weak gravitational lensing is sensitive to both of the matter distribution and expansion history of the Universe, which are tightly related to the property of dark energy. By studying weak lensing, especially in different tomographic redshift bins, we can effectively constrain the equation of state of dark energy. The CSST photometric imaging survey can perform high-precision weak lensing measurements as its main cosmological probe. The high-resolution and multi-band observations of the CSST will provide high-quality data for the shear and photo-$z$ measurements.

The signal of weak gravitational lensing, as its name implies, is weak and particularly sensitive to systematics. Therefore, it is necessary to perform detailed studies on various possible observational and physical systematic errors, which come from the calibrations of shear and photo-$z$ measurements, galaxy intrinsic ellipticity and alignment, baryonic feedback and non-linear effects, etc. \cite{Yao24a}.

For the shear calibration, to achieve the capability of the CSST cosmological study, an accuracy at the level of one-thousandth is required, which is extremely challenging. We need to carefully study the systematic errors caused by undersampling, detector effects, etc. At the same time, detailed numerical simulations are required to provide accurate PSF models for helping construct the PSF using stars. Besides, CSST-MCI also can help to calibrate the shape of galaxies, since the magnitude limit of its XDF observation is about three magnitudes deeper than the SC wide-field survey and its spatial resolution also can match that of the SC.

\begin{figure*}
\centering
\includegraphics[scale=0.5]{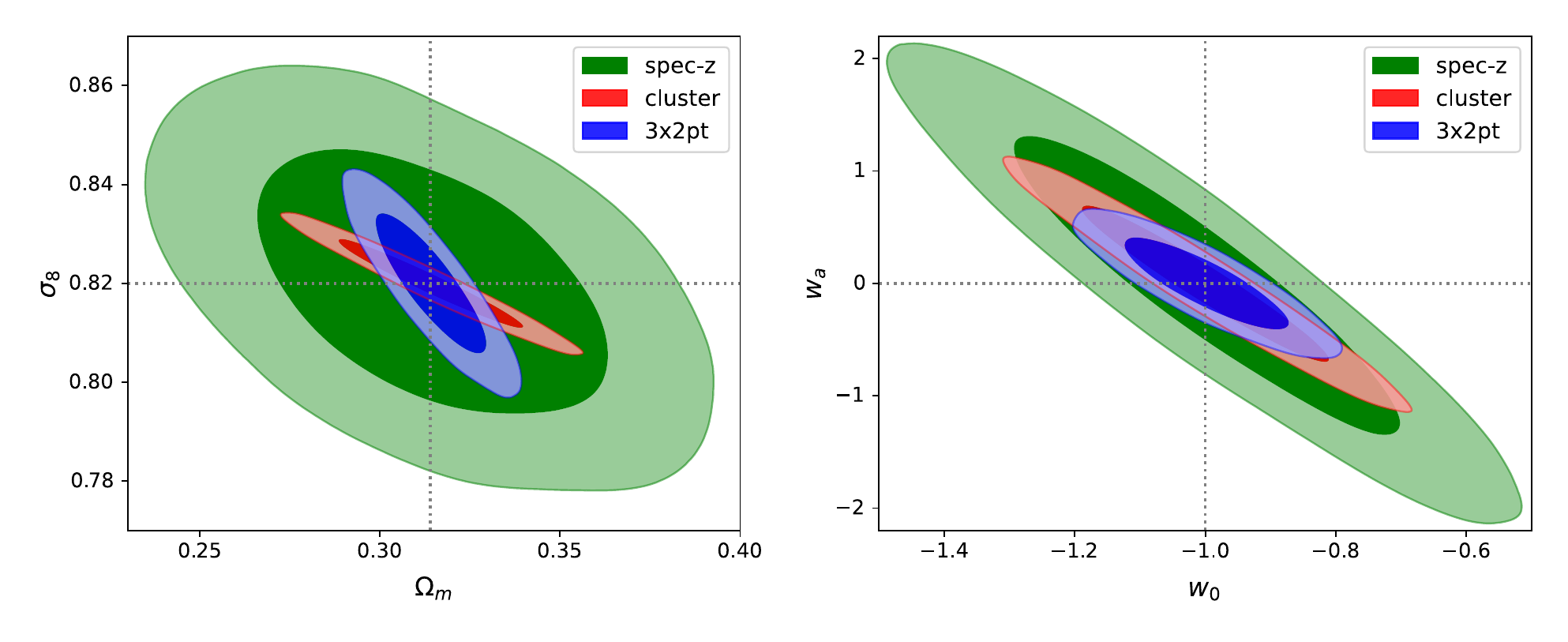}
\caption{The predicted contour maps of $\Omega_{\rm m}$ vs. $\sigma_8$ and $w_0$ vs. $w_a$ derived from the CSST 3$\times$2pt, galaxy spec-$z$, and galaxy cluster surveys. The 68.3\% and 95.5\% confidence levels (CLs) are shown. The gray dotted lines show the parameter fiducial values. The systematics have been considered in the constraint process, such as the shear calibration, photo-$z$ uncertainties, intrinsic alignment, galaxy bias, non-linear and instrumental effects.} 
\label{fig:constraint}
\end{figure*}

On the other hand, for the upcoming weak gravitational lensing surveys like CSST, the required photo-$z$ accuracy is extremely high, with the accuracies of the redshift bias $\Delta z$ being at the level of one thousandth and the scatter $\sigma_z$ is at the level of one percent. CSST has the wavelength coverage from near-UV to near-IR, with a total of seven bands, which has advantages for accurately measuring photo-$z$. Using the traditional spectral energy distribution (SED) fitting method, it is found that $\sigma_z$ can achieve $\sim 0.03$, and the outlier or fraction of catastrophic redshift is less than $5\%$ \cite{Cao18,Cao22} for the CSST photometric survey. When adopting the machine learning methods with appropriate training data from spectroscopic surveys, $\sigma_z$ can be better than $0.02$ and the outlier fraction will be less than $3\%$ \cite{Zhou21,Zhou22a, Zhou22b,Lu24,Luo24a,Luo24b}. When considering the measurements in the other bands from other telescopes, e.g. $Euclid$ $Y$, $J$ and $H$ bands, the photo-$z$ accuracy can be further significantly improved \cite{Cao18, Liu23}.

It is expected that the CSST 17,500 deg$^2$ wide-field photometric survey can obtain more than one billion galaxy images with a spatial resolution higher than $0.15''$ and a number density of $\sim28$ arcmin$^{-2}$, and the galaxy redshift distribution has a peak abound $z=0.6-0.7$ and can extend to $z>4$ \cite{Gong19}. Based on the results derived from Jiutian simulations \cite{Han25} and theoretical predictions, the shear two-point correlation functions or power spectra in different tomographic redshift bins can be precisely measured at $0<z<3$, and the constraint accuracy of $w$ can reach $\sim10\%$ level when only using the data of the CSST shear measurements  \cite{Xiong24,Miao23,Lin22,Lin24}. 

In addition, since the weak lensing effect is closely related to the LSS, the main signal comes from the nonlinear structures, which is highly non-Gaussian. Hence, the shear two-point correlation cannot reflect all the cosmological information. On the other hand, the peak statistics of weak lensing are concentrated in nonlinear regions and highly complementary to the shear two-point statistics \cite{Fan10, Liu15, Shan18}. The constraint strength of $w$ can be further improved when cooperating the two weak lensing probes.

The weak lensing survey also can cross-correlate with galaxy angular clustering measurement to obtain three two-point correlation functions (3$\times$2pt) for cosmological constraints, i.e. shear, galaxy angular and galaxy-galaxy lensing correlation functions or power spectra. The 3$\times$2pt data can effectively break the degeneracies between model parameters, and improve the constraints on the cosmological parameters and significantly suppress the effects of systematics, e.g. photo-$z$ uncertainties \cite{Lin22}. In Figure~\ref{fig:constraint}, we show the constraint results of $\Omega_{\rm m}-\sigma_8$ and $w_0-w_a$ from the CSST 3$\times$2pt surveys (see blue contours), where $\Omega_{\rm m}$ is the total matter density parameter and $\sigma_8$ is the amplitude of matter fluctuation on the scale of $8\ h^{-1}{\rm Mpc}$. The systematics from the shear calibration (multiplicative and additive errors), photo-z uncertainties (redshift bias and scatter), intrinsic alignment, galaxy bias, non-linear effects and instruments, have been considered in the analysis. The details can be found in \cite{Miao23}. We can find that the CSST 3$\times$2pt surveys can constrain $\Omega_{\rm m}$, $\sigma_8$, and equation of state of dark energy with accuracies $\sim3\%$, $\sim1\%$, and $\sim8\%$, respectively.

\paragraph{Galaxy redshift-space clustering}

Besides multi-band photometric imaging survey, CSST also will simultaneously perform the slitless spectroscopic wide-field survey with three bands, i.e. $GU$, $GV$, and $GI$. The main targets of the CSST spectroscopic survey are the emission line galaxies (ELGs) with H$\alpha$, [O III], and [O II] emission lines. The spectral resolution $R$ is higher than 200, and the spatial resolution is about $0.3''$. It is expected that the CSST slitless spectroscopic wide-field survey can obtain more than one hundred million galaxy spectra with a surface number density $2-3$ arcmin$^{-2}$ in 17,500 deg$^2$ survey area. The galaxy redshift distribution has a peak around $z=0.3$ and can extend to $z>1.5$ derived from the zCOSMOS catalog, which means that the galaxy number density $n_{\rm g}$ is greater than $10^{-2}\, h^3{\rm Mpc^{-3}}$ at $0<z<0.6$ and $10^{-3}\, h^3{\rm Mpc^{-3}}$ at $0.6<z<1.2$ \cite{Gong19}.

In a spectroscopic survey, the galaxy clustering is measured in redshift space, i.e. the redshift-space galaxy power spectrum $P^i_{\rm g}(k,\mu)$ at the linear regime in a spectroscopic redshift (spec-$z$) bin $i$, where $k$ is the wavenumber and $\mu$ is the cosine of the angle between $k$ and the line of sight. Theoretically, $P_{\rm g}(k,\mu)$ is dependent on several effects, e.g. the Alcock-Paczynski (AP) effect \cite{Alcock79}, Kaiser effect \cite{Kaiser87}, damping effect, and shot noise. The uncertainty of $P_{\rm g}(k,\mu)$ is tightly related to the galaxy selection function, survey volume, galaxy number density and the spec-$z$ accuracy. The spec-$z$ accuracy is one of the most important sample-selection criteria in the CSST cosmological study.

For a slitless spectroscopic survey with relatively low spectral resolution, it is challenging to obtain accurate galaxy spec-$z$ in data processing, which can be affected by different effects, e.g. spectrum blending, wavelength calibration, etc. Among the galaxy spectra measured by CSST, a considerable portion may not meet the required CSST spec-$z$ accuracy, i.e. $\sigma_z\simeq0.002-0.003$. This will not only affect the number density of available galaxies, but also may significantly change the galaxy power spectrum at a redshift, due to the interloper galaxies with incorrect redshifts. The spec-$z$ accuracy of the CSST slitless spectroscopic survey could be effectively improved by including the photometric data in the redshift analysis or making use of machine learning to achieve $\sigma_z\lesssim0.002$ \cite{Zhou21, Zhou24}.

In Figure~\ref{fig:constraint}, the contour maps of $\Omega_{\rm m}-\sigma_8$ and $w_0-w_a$ from the CSST slitless spectroscopic galaxy survey have been shown. We can find that the constraint accuracies of $\Omega_{\rm m}$, $\sigma_8$, and equation of state of dark energy are about $10\%$, $2\%$, and $20\%$, respectively, which are weaker than the results of the CSST 3$\times$2pt surveys. 

\paragraph{Galaxy cluster}

Galaxy cluster is the largest virialized system in the Universe with a typical mass $10^{14}-10^{15}\ M_{\odot}$, which is formed as a gravitationally bound system that has reached a state of equilibrium. Since the formation of galaxy clusters is tightly related to the LSS, they can be adopted to constrain the property of dark energy, e.g. using the number counts of galaxy clusters. 

Accurately measuring the redshift and mass of galaxy clusters is crucial in the cosmological study of galaxy clusters. CSST can simultaneously conduct photometric imaging and slitless spectroscopic surveys, which is highly advantageous for the identification and verification of galaxy clusters, as well as for measuring their redshift and mass. We note that the redshift and members of a cluster can be determined by the CSST spectroscopic observation, and the mass can be derived from the gravitational lensing and richness measurements in CSST photometric survey. Therefore, in principle, we can precisely perform the number counts of galaxy clusters in a mass range ($\gtrsim10^{14}\ M_{\odot}$) at a given redshift for the CSST surveys. 

At $0<z<1.5$, which is the redshift range covered by both CSST photometric and spectroscopic surveys, about 200,000-400,000 galaxy clusters can be detected according to the theoretical estimation based on the halo mass function, and the peak of cluster redshift distribution is around $z=0.6$ \cite{Miao23,Zhangy23}. The constraint results of $\Omega_{\rm m}-\sigma_8$ and $w_0-w_a$ from the CSST number counts of galaxy clusters are shown in red contours in  Figure~\ref{fig:constraint}. We can find that the constraint accuracies of $\Omega_{\rm m}$, $\sigma_8$, and equation of state of dark energy are $\sim 5\%$, $\sim1\%$, and $\sim13\%$, respectively, which are better than that from the CSST galaxy redshift-space clustering measurements. However, we should notice that, for other parameters, e.g. $\Omega_{\rm b}$ and $n_{\rm s}$, the constraint power of the CSST number counts of galaxy clusters could be worse than the galaxy redshift-space clustering, due to low sensitivity to these parameters \cite{Miao23}. 

\paragraph{Cosmic void}

In contrast to galaxy clusters, cosmic voids are regions of low matter density in the Universe. It is characterized by low density, large volume and linear evolution, which is suitable to study the evolution of the LSS, property of dark energy, and modified gravity theories.
Since the typical size of cosmic voids can range from 10 to 100 Mpc, it needs to explore extremely large survey volume to have sufficient statistical significance for studying relevant issues accurately.

The upcoming Stage-IV survey telescopes such as CSST will probe more than ten thousand square degrees of sky area with great depth and sensitivity. Hundreds of thousands of cosmic voids are expected to be identified in the CSST spectroscopic survey at $0<z<1.5$ \cite{Song24a,Song24b}. Using this  spec-$z$ void sample, we can precisely measure the void size functions \cite{Song24a}, void number counts \cite{Song24b}, and void power spectra \cite{Song24c} at different redshifts for investigating the equation of state of dark energy and other cosmological parameters.

Additionally, the CSST photometric survey also can be used to measure the power spectra of 2-dimensional (2D) voids in different photo-$z$ tomographic bins in the range $1\lesssim z\lesssim 2$ \cite{Song25}. Since the cosmic voids identified in spectroscopic surveys are usually limited by low number density and low redshift range, the 2D voids identified in photometric surveys can be adopted to perform the cosmological constraints for eliminating these limitations. As the result derived from Jiutian simulations, the constraint power on $w$ from the 2D void power spectrum is comparable to that from the galaxy angular clustering in the CSST photometric survey, with an accuracy of $\sim15\%$. The constraint accuracy can be further improved in the void and galaxy joint constraints, using the 2D void auto, galaxy angular auto, and void-galaxy cross power spectra \cite{Song25}. 

\paragraph{SN~Ia}

The nature of dark energy determines the background evolution of the late Universe, and SNe~Ia at different redshifts serve as ``standard candles'' for measuring the background evolution of the Universe. By utilizing SNe~Ia, the distances at different redshifts can be measured, so that the evolution history of the Universe can be accurately derived, and the nature of dark energy could be investigated with high precision.

CSST-SC preliminarily plans to conduct a 9 deg$^2$ UDF survey during the first two years, utilizing a single exposure time of 250 s and a total of 60 exposures. This means CSST can probe a patch of sky in the UDF averagely every $\sim12$ days, which is close to the best cadence found in \cite{Li23}. The magnitude limit of a single exposure can reach $i=25.9$ AB mag (5$\sigma$ detection for point sources) in the photometric observation \cite{Gong25}. So the CSST UDF photometric survey can effectively discover SNe Ia and precisely measure their light curves. 

Using simulated CSST-SC SN~Ia observational data, it is found that after removing the contamination of core-collapse supernovae (CCSNe) through methods such as light curve fitting, the photometric observations of the CSST UDF can obtain high-quality data of light curves of more than 2,000 SNe~Ia within two years \cite{Wang24}. It is found that the fraction of high-$z$ SNe~Ia in the CSST UDF survey is $\sim80\%$ at $z>0.5$ and $\sim15\%$ at $z>1$, which is is very advantageous in constraining the dynamical evolution of dark energy. Using this SN~Ia sample, the precision of constraints on cosmological parameters such as the equation of state of dark energy can be improved by almost twice compared to the current SN~Ia observation data \cite{Wang24}.

In addition, the CSST XDF planned by CSST-MCI also will make significant progress in SN~Ia cosmology. The CSST-MCI can capture full-band light variation events with a time scale of more than 5 minutes (1 hour) for astronomical objects at a depth of at least 26 (27) magnitude, which is very helpful in capturing high-$z$ supernovae in the CSST XDF, and can quickly and simultaneously determine their coordinates and approximate redshift information. The CSST-MCI XDF observation is expected to discover 100-200 SNe~Ia at $1.0<z<1.5$ in one year (2 hours of deep exposure per week), which can help us gain a decisive understanding of the equation of state of dark energy.

Besides the CSST UDF and XDF surveys, the CSST wide-field survey also can detect a large number of SNe in ten years. It is estimated that approximately 5 million different types of SNe will be observed, among which there are about 7,000 SNe~Ia as the ``gold'' sample that can be followed-up by time-domain telescopes \cite{Liu24}. This is obviously can promote the SN~Ia studies, help to measure the expansion history of the Universe and the equation of state of dark energy.

\paragraph{BAO and RSD}

BAO and RSD can be used to measure the cosmic background expansion and structure growth history, respectively, and provide observational support for the study of dark energy. Therefore, the measurement of BAO and RSD is one of the key scientific goals of almost all large galaxy spectroscopic surveys.

BAO is caused by the fact that in the early Universe, baryons and photons were coupled together, and under gravity and pressure, oscillation was formed. This oscillation caused the local matter fluctuations formed and  propagated outward as sound waves, i.e. BAO. It can be seen as the ``standard ruler'', which is roughly 150 Mpc, for measuring the cosmological distance and the property of dark energy.

RSD is a measure of the anisotropy of three-dimensional clustering caused by the peculiar motion of galaxies. Affected by the local gravitational potential, galaxies have peculiar velocities, and this information is reflected in the redshift of the galaxies. In spec-$z$ surveys, the redshift of galaxies can be precisely measured. If the sample is large enough, we can use statistical methods to obtain information about the gravitational potential and structure formation at a given redshift. The RSD data can be directly used to constrain cosmological parameters, especially those related to structure formation, such as the equation of state of dark energy.

The CSST slitless spectroscopic survey can observe 17,500 deg$^2$ sky area, and obtain more than one hundred million galaxy spectra and about four million AGN spectra \cite{Miao24}. This indicates that CSST can not only measure BAO with galaxies but also with AGNs over a larger redshift range. By considering the reconstruction technique, it is found that the CSST spec-$z$ survey can measure the two BAO scaling parameters $\alpha_{\parallel}$ and $\alpha_{\perp}$ with precisions $\sim 1\%$ using galaxies at $0.3<z<1.2$. When using AGNs, the precisions can reach $\sim 3\%$ at $z<3$. By combining the BAO data from CSST spec-$z$ galaxy and AGN surveys, the constraint accuracy of $w$ is expected to achieve $\sim 9\%$ \cite{Miao24}.

We should note that, since the CSST-SC spectral resolution $R\gtrsim200$ is relatively low, the spec-$z$ accuracy may not be higher than $\sigma_z = 0.002$, which may affect the measurement of $\alpha_{\parallel}$ in the radial direction. Hence, it requires $\sigma_z<0.005$ for precisely measuring $\alpha_{\parallel}$ in the CSST slitless spectroscopic survey \cite{Miao24,Shi25}. Otherwise, we can only measure $\alpha_{\perp}$ accurately, or perform a 2D BAO measurement with the CSST photometric data \cite{SongR24,Ding24}.

For the RSD observation, based on theoretical estimation, CSST can measure the structure formation factor $f\sigma_8$ with a precision of $\sim2\%$ at $0.4<z<1$, where $f$ is the growth rate. Compared with DESI, CSST RSD observation can constrain $f\sigma_8$ more accurately by a factor of 2 at least at $z<0.6$ \cite{Aghamousa16}, since CSST spectroscopic survey has larger survey area and deeper magnitude limit. This means that the CSST spec-$z$ survey can be complementary to other spec-$z$ surveys such as DESI. By combining the CSST and other spec-$z$ surveys, we expect to measure the equation of state of dark energy at a high level of precision in a large redshift range.

\paragraph{Strong gravitational lensing}

The strong gravitational lensing systems detected by CSST also can be used to constrain the property of dark energy \cite{Cao24}. There are several commonly used strong lensing methods in cosmological study, such as the velocity dispersion of lens galaxies and time delay between multiple images.

In the study of the velocity dispersion of strong lensing on the galaxy scale, by combining the strong lensing of lens galaxies (mostly early-type galaxies) with the observational data of stellar dynamics, the velocity dispersion (VD) of lens galaxies can be used as a statistic to estimate the cosmological parameters and the mass density profile of lens galaxies \cite{Futamase01}. By generating and analyzing the mock strong lensing sample in the CSST photometric survey and considering DESI survey, it is found that the CSST lens galaxy VD data can provide the constraint on $w$ with an accuracy higher than $20\%$, which is several times higher than the result from the current measurements.

For the strong lensing time delay measurement, due to the luminosity variation of the variable source, e.g. AGNs, supernovae, kilonovae, etc., we can measure the time delay between different images reaching the Earth. At the same time, through high-resolution imaging of the host galaxy light arc, the gravitational potential of the lens also can be measured. With the information of time delay and gravitational potential, we can get a measurement of the so-called ``time delay distance'' \cite{Treu10,Treu16}. This quantity is a combination of three angular diameter distances, depends on the redshift of the source and the lens, and reflects the space-time geometry information. The time delay distance is highly dependent on the Hubble constant, and also dependent on other cosmological parameters such as the dark energy equation of state. Due to the high spatial resolution, large FoV and multi-bands, CSST is very beneficial for the search of these sources. When cooperating with follow-up observations of other telescopes, the CSST strong lensing time delay measurement will also provide strong constraints on cosmic expansion history and the nature of dark energy.

Besides, the Multiple Source Plane (MSP) system, which is a special strong gravitational lensing system and has two or more background sources at different redshifts in the line of sight, also can be used to constrain the cosmological parameters, such as $w$ and $\Omega_{\rm m}$. Since the background sources at different redshifts correspond to different Einstein radii in the lens plane, this type of system is also called a multiple Einstein ring system. With the launch and operation of the new generation of space survey telescopes, such as {\it Euclid} and CSST, it is expected that about 100,000 strong gravitational lensing systems will be observed, including about 2,000 Double Source Plane (DSP) systems \cite{CaoX24} and dozens of triple source plane systems \cite{Collett14}. The key information contained in the MSP system is the relative distance scaling factor $\beta$ in the multi-plane lens equation. It is independent to the Hubble constant and the distance between the observer and the lens, and can reflect the ratio of the critical surface density corresponding to different background sources and the same foreground lens.

It is expected that, about 60\% of the strong lensing systems observed by CSST are located at $z>0.5$, and about 15\% at $1<z<2$ \cite{CaoX24}. Therefore, The MSP system in the CSST strong lensing sample can provide a supplementary tool for studying the proportion and evolution of dark matter components in the Universe at $z<2$. Besides, we can not only use MSP systems independently, but also combine the MSP system with other cosmological probes (such as CMB, BAO, time delay, etc.) to more accurately constrain $\Omega_{\rm m}$ and $w$.\\

By combining all the CSST data discussed above and perform a joint fit, the constraint accuracy of $w$ is expected to be higher than $5\%$, and the redshift evolution of $w$ can be precisely measured by the CSST cosmological survey.

\subsubsection{Dark matter}

Dark matter is the most important component of the Universe besides dark energy, accounting for more than 80\% of the total amount of  matter, far more than ordinary baryonic matter that can produce electromagnetic radiation. Baryonic matter falls into the dark matter gravitational potential well under the action of gravity, and is considered to be the driving factor for the formation of cosmic structures from galaxies to galaxy clusters and even larger structures like superclusters and filaments. Evidence for the existence of dark matter comes from its gravitational effects on photon propagation and baryonic matter. It was first discovered in 1933 in the measurement of galaxy velocity dispersion within galaxy clusters. Since then, it has been confirmed that it is ubiquitous on various scales from galaxies to the entire Universe, but its physical nature has always been an unsolved mystery. With the help of gravitational lensing observations, we can observe how these non-luminous matter are distributed in the Universe and infer their nature through the structures they form. 

The CSST survey will accurately measure the dark matter structure at various scales since the redshift $z=1.5$, trying to answer several of the most important questions about dark matter: Is dark matter cold? Do dark matter particles have self-interactions? What is the structure evolution of dark matter on cosmic scales? How do galaxies form and evolve in dark matter halos?

At small scales, the main method for measuring dark matter is through the dynamics of stars, gas or globular clusters and strong gravitational lensing, while at large scales, weak gravitational lensing is used as the main means to obtain dark matter distribution.

\paragraph{Cold and warm dark matter}

Since the 1980s, the cold dark matter (CDM) model has achieved great success. This type of dark matter had only very small kinetic energy in the early Universe, so it is called ``cold'' dark matter. In such a universe, small structures are formed first and grow into large structures through mergers. At present, the cold dark matter model can well explain most cosmological and galaxy formation observations, but the corresponding CDM particles have not yet been discovered. Weakly interacting massive particle (WIMP) is currently the most promising CDM candidate, but no confirmation of the existence of WIMP has been found. 

In recent years, with the observations such as X-ray emission lines in the center of galaxy clusters and the problems such as ``missing satellite'' problem, the possibility of another type of dark matter model, i.e. ``warm'' dark matter (WDM), has been considered. This type of dark matter model can give observational predictions of galaxy formation and structure formation that are very similar to those of cold dark matter. The only difference is that dark matter halos with very small masses will not form in the universe filled with WDM.

CSST sky surveys can detect a large number of strong gravitational lensing systems that have never been discovered before (estimated to be about hundreds of times the current number), which can provide information on the distribution of dark matter at small scales. Using the abundance of these small-scale dark matter structures, it can be effectively distinguished whether dark matter is cold (such as WIMPs) or warm (such as sterile neutrinos).

\paragraph{Axion dark matter}

Leaving aside the debate over the ``temperature'' of dark matter, the CDM model itself also contains a variety of different candidates, such as the axion. The axion is a hypothetical scalar particle proposed in the 1970s to solve the strong CP problem in quantum chromodynamics \cite{Peccei77,Weinberg78,Wilczek78}. The mass of the axion is weakly constrained in theory, and has a very wide possible mass range. The axion is a good candidate particle for cold dark matter. 

In cosmology, the ultra-light axion (ULA) with a mass of about $10^{-22}$ eV has attracted much attention. For ULA dark matter, due to the extremely small particle mass, its de~Broglie wavelength can reach the order of magnitude of the dwarf galaxy or galaxy scale, so the ULA is consistent with the prediction of the standard CDM model at large scales ($>10$ kpc); but at smaller scales, the ULA dark matter model shows some unique characteristics. It can affect the transfer function of matter power spectrum and suppress the small-scale structure formation.

\begin{figure}[H]
\centering
\includegraphics[scale=0.6]{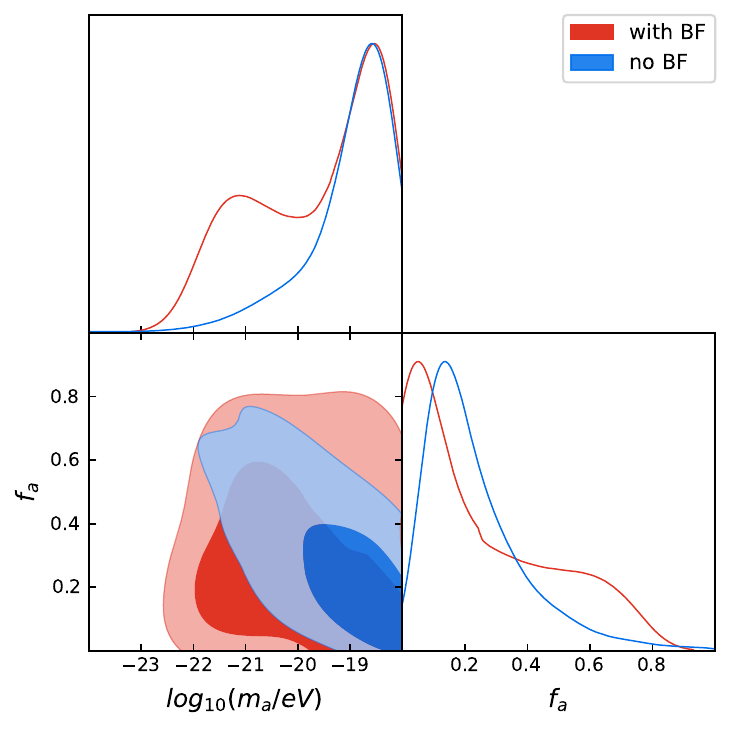}
\caption{The prediction of the constraint results of $m_a$ and $f_a$ with (red) and without (blue) baryonic effect in the CSST 3$\times$2pt surveys \cite{Lin24}. The contours show the 68.3\% (1$\sigma$) and 95.5\% (2$\sigma$) CLs.} 
\label{fig:ma_fa}
\end{figure}

Since ULAs have a suppression effect on the small-scale matter clustering, their particle mass $m_a$ and the fraction of total dark matter $f_a$ can be effectively constrained through observations such as weak gravitational lensing and galaxy clustering. By using the CSST 3$\times$2pt data, considering the effects of neutrino and baryonic feedback, and including the systematics from shear calibration, intrinsic alignment, galaxy bias and photo-$z$ calibration, it is found that $m_a$ and $f_a$ can be well constrained \cite{Lin24}. As shown in Figure~\ref{fig:ma_fa}, $m_a$ can be constrained with an upper limit $10^{-22}$ eV (with baryonic effect) and $10^{-20}$ eV (without baryonic effect) in 1$\sigma$ CL for the CSST 3$\times$2pt surveys. 

\paragraph{Self-interacting dark matter}

For a fundamental particle, the two most important properties are mass and interaction cross section. It is known that dark matter basically does not interact with visible matter, so do dark matter particles collide with themselves, and what is the cross section? These questions are crucial for understanding the nature of dark matter. Self-interacting dark matter (SIDM) has a self-scattering cross section that is comparable to the nucleon-nucleon scattering cross section \cite{Spergel00}. 

In a dark matter halo, when the surface density of dark matter particles multiplied by the self-scattering cross section exceeds a certain threshold, collisions between dark matter particles will lead to complex structure evolution. In the initial stage of this process, due to the scattering of dark matter particles, the central density of the dark matter halo will decrease, forming a structure similar to a galactic bulge. At the same time, scattering will also strip off the dark matter halo from the small dark matter clumps surrounding the large-scale structure, making these substructures more susceptible to tidal gravity and reducing their number. The structure formation order of SIDM is the same as that of CDM, but it will slowly change the distribution of dark matter in high-density areas.

In astronomy, merging galaxy clusters are natural large particle colliders, providing ideal laboratories for measuring dark matter self-interactions \cite{Harvey19, Adhikari22}. During the process of galaxy cluster merging, the non-gravitational interactions between dark matter and standard model particles will be more significant. In addition to galaxy cluster mergers, we can also study SIDM at the scale of galaxy clusters and their substructures. Considering that cluster subhalos and main cluster halos probe different regions of velocity space, these studies at different scales are helpful for exploring the potential velocity dependence of the self-interaction cross section of dark matter. These effects can be precisely measured in the CSST gravitational lensing and galaxy cluster observations.\\

In addition to explore the property of dark matter particle, CSST also can use various cosmological probes, such as weak and strong gravitational lensing, galaxy clustering, galaxy clusters, etc., to weight total dark matter in the Universe, i.e. constrain the matter density parameter $\Omega_{\rm m}$ (e.g. see the left panel of Figure~\ref{fig:constraint}), since the baryon density parameter $\Omega_{\rm b}$ can be precisely determined by Big Bang nucleosynthesis (BBN) and CMB. 

\subsubsection{Neutrino}

Neutrinos are one of the important links between nuclear physics, particle physics, astrophysics and cosmology \cite{Kamionkowski99,Balantekin13}, and the mass of neutrinos has been one of the research focuses in the past decades. Neutrinos affect almost every aspect of the early Universe, and various cosmological probes can constrain the sum of the three mass eigenstates of neutrinos. These observational constraints from cosmology complements the results of neutrino oscillation experiments.

There are three generations of neutrinos in the Standard Model of particle physics. Research on neutrino properties has always been an important area in particle physics. In 1998, experiments confirmed the existence of neutrino oscillation, which shows that the mass of neutrinos is not zero \cite{Fukuda98}. At present, experiments on atmospheric neutrino and solar neutrino have measured the difference between the squares of the masses of three generations of neutrinos \cite{Ashie05,Araki05,Abe08}, but the measurement of the absolute mass of neutrinos is very difficult and no conclusion has been reached yet. 
Based on the measurements, there are three possible orders of neutrino mass: 1. Normal hierarchy; 2. Inverted hierarchy; 3. Quasi degenerate, which is called ``hierarchy'' problem of neutrino mass. In addition, is there a fourth generation of neutrinos besides the three generations? These are all problems to be solved in the field of particle physics. 

By observing the influence of neutrinos left over from the Big Bang on the LSS and the CMB, it is possible to measure the sum of neutrino mass $\Sigma m_{\nu}$ and effective generations $N_{\nu}$, and solve the neutrino hierarchy problem, which are difficult to achieve in ground-based particle physics experiments. The change of the effective number and total mass of neutrinos mainly affects the position of the CMB sound wave peak and the suppression of the small-scale power spectrum, which degenerate with other cosmological parameters. The observational results of the LSS obtained from the CSST survey can effectively break these degeneracies, thereby obtaining better constraints on $\Sigma m_{\nu}$ and $N_{\nu}$.

Massive neutrinos mainly change the scale dependence of the growth of cosmic structure. After decoupling, as the expansion rate of the Universe continues to decay, neutrinos become non-relativistic particles. Freely diffused neutrinos will smooth out the fluctuations in the neutrino density field that are smaller than the free streaming scale. Dark matter interacts with neutrinos through gravity. Since neutrinos do not form clusters at small scales, they also affect the growth rate of dark matter fluctuations at small scales. Therefore, the net effect is that freely diffused neutrinos will limit the growth rate of small-scale structure, which will cause the matter power spectrum to have a suppression effect at small scales, and the magnitude of this suppression is proportional to $\Sigma m_{\nu}$.

\begin{figure}[H]
\centering
\includegraphics[scale=0.67]{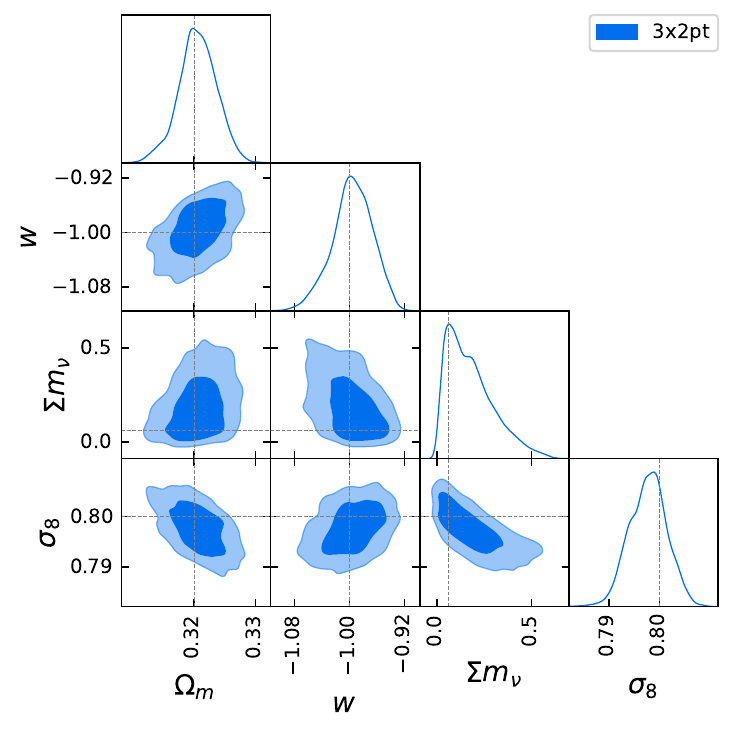}
\caption{The prediction of the constraint results of $\Sigma m_{\nu}$ in the CSST 3$\times$2pt surveys \cite{Lin22}. The contours show the 68.3\% (1$\sigma$) and 95.5\% (2$\sigma$) CLs.} 
\label{fig:mv}
\end{figure}

Using the CSST 3$\times$2pt data, we can effectively constrain $\Sigma m_{\nu}$, considering other cosmological and systematical parameters. In Figure~\ref{fig:mv}, the constraint result of $\Sigma m_{\nu}$ and some of other cosmological parameters have been shown. The systematics from shear and photo-$z$ calibration, intrinsic alignment, galaxy bias, and baryonic feedback effect have been considered in the analysis. It is found that the CSST 3$\times$2pt surveys can provide a stringent constraint on the total neutrino mass with $\Sigma m_{\nu} \lesssim 0.36\ (0.56)$ eV at 68\% (95 \%) CL, which is about one order of magnitude tighter than the current 3$\times$2pt results \cite{Porredon22,Lin22}. By including other cosmological data, such as CMB and BAO, the degeneracies between parameters can be effectively broken, and the constraint accuracy of $\Sigma m_{\nu}$ can be further improved \cite{Abbott25b,Quintero25,Chebat25}. Hence, CSST is expected to measure  $\Sigma m_{\nu}$ with extremely high precision, and can decisively distinguish the neutrino mass hierarchy.

\subsubsection{Gravity}

As discussed above, within the framework of general relativity, we can introduce dark energy to explain the accelerating expansion of the Universe. Another possibility is that general relativity is no longer applicable on cosmological scales, that is, a new modified gravity theory can be introduced. However, even if we can accurately measure the expansion history of the Universe, e.g.  using SNe Ia, BAO, gravitational waves, etc., it is difficult to distinguish between dark energy and modified gravity. Under low energy conditions, most modified gravity theories can be approximated by an equivalent ``scalar tensor theory'' and generate the ``fifth force'' through an additional scalar field, thereby changing the formation history of the LSS. Therefore, in order to strictly distinguish dark energy from modified gravity theory on cosmological scales, we need to use CSST weak gravitational lensing and galaxy clustering surveys, combined with the RSD information in the CSST slitless spectroscopic  survey, to reconstruct the structure growth history of the Universe with high precision.

A series of experiments in the inner Solar System and observations of binary pulsar timing have shown that Einstein's general relativity is strictly applicable in these two environments. Therefore, any modified gravity theory must rely on a special ``screening mechanism'' to fully restore general relativity in areas of high density or high gravitational potential. The currently popular screening mechanisms of modified gravity can be roughly divided into two categories: the first is the thin-shell screening mechanism based on the strength of the scalar field, such as the chameleon mechanism, the symmetron mechanism, and the dilaton mechanism; the second category is the kinetic screening mechanism based on the first or second derivatives of the scalar field, such as the Vainshtein mechanism and the K-mouflage mechanism. These two types of shielding mechanisms will have different predictions for the redshift and scale dependence of cosmic structure growth. By combining cosmological numerical simulations based on different modified gravity theories, the CSST survey can provide high-precision constraints on the modified gravity models.

The popular modified gravity theories in cosmological research include the models of $f(R)$, $f(T)$, DGP, Galileon, Horndeski, etc. For example, the $f(R)$ theory is a generalization of the Einstein-Hilbert action, which generalizes the action from the Ricci scalar to an arbitrary function of the Ricci scalar. For the translational gauge theory of gravity equivalent to general relativity, there is also a similar $f(T)$ theory that generalizes the torsion scalar to an arbitrary function of the torsion scalar \cite{Cai16}.

The CSST weak gravitational lensing or shear measurement can directly probe the dependence of dark matter clustering on scale and redshift in the Universe. The matter clustering predicted by most modified gravity theories is on a near-linear scale, and the S/N of the shear signal is strongest at this scale. Therefore, if the CSST shear measurement has any statistically significant deviation from the predictions of the general relativity+dark energy model, we believe that there may be evidence of modified gravity on the relevant scale. However, the baryonic process and neutrinos may also have a large impact on the matter power spectrum at this scale, thereby weakening the constraints of cosmic shear on modified gravity. In addition, we can also use the stronger signal in cosmic shear, i.e. the peak statistics of weak gravitational lensing, to constrain gravity models. 
For the $f(R)$ theory, it is found that the CSST weak gravitational lensing measurement can constrain $|f_{R0}|=|{\rm d}f/{\rm d}R|_{z=0}$ to an accuracy of $\sim 10^{-6}$. Besides, using the CSST SN~Ia and BAO data, we can also obtain stringent constraint on the $f(R)$ theory \cite{Yan24}.

Brans-Dicke theory is another typical scalar-tensor modified gravity theory that is widely discussed. The characteristic of Brans-Dicke theory is that the gravitational constant $G$ is no longer a constant, but is replaced by a scalar field that can change with time and space, i.e. $\phi\sim1/G$. This will theoretically affect the formation and evolution of the LSS, so galaxy surveys can be used to effectively constrain this gravity theory. It is found that the CSST spec-$z$ galaxy survey can constrain the absolute value of the characteristic parameter $\omega$ of the Brans-Dicke theory to $\sim1000$, and the relative change of the Newtonian gravitational constant over time, i.e., $\dot{G}/G$, to within $10^{-13}$ per year \cite{Chen22}.

Besides, using the data of the CSST photometric and spectroscopic surveys, the quantity $E_G$, which can be adopted to test the gravity theory, could be precisely measured \cite{Zhang07}. This quantity measures the ratio of the galaxy-galaxy lensing cross power spectrum to the galaxy-velocity cross power spectrum, and is therefore not affected by the uncertainty of the galaxy bias and the matter clustering amplitude $\sigma_8$. It is expected that the statistical error of $E_G$ measured by CSST can be reduced to less than 5\%.

Additionally, galaxy clusters also can be used to test the gravity theory and study the modified gravity. CSST wide-field  multi-band photometric survey can detect approximately 300,000 galaxy clusters with a mass greater than $10^{14}\ M_{\odot}$ at $z<1.5$, providing an excellent sample of massive dark matter halos for measuring the structure growth history of the Universe and testing modified gravity theories. Using the shear signal of galaxy clusters, CSST will accurately measure the density profiles of galaxy clusters with different satellite galaxy abundances between 0.1 $h^{-1}$Mpc and 30 $h^{-1}$Mpc, thereby detecting the density enhancement effect caused by some modified gravity models near the Virial radius of galaxy clusters \cite{Lombriser12}. The CSST weak gravitational lensing survey can also measure the ``splashback radius'' of the dark halos of galaxy clusters with an accuracy of one percent, which can be used to study the modified gravity theories \cite{Adhikari18}. Besides, combined with data from spectroscopic surveys, CSST can also use the Galaxy Infall Kinematics (GIK) method to test modified gravitational theories \cite{Zu13}.

\subsubsection{Primordial perturbations}

The primordial metric perturbations of the Universe contain information about the early Universe, and are of great significance for exploring the origin of the Universe and the physical laws at ultra-high energy scales. In the standard model, the large-scale primordial metric perturbations of the Universe originate from the inflation \cite{Riotto02,Tsujikawa03}. The power spectrum of the primordial perturbations given by the simplest single-field slow-roll inflation, referred to as the ``primordial power spectrum'', is very close to a power-law, which is also supported by the CMB observations. 

Besides, the primordial perturbations in the single-field slow-roll model are also very close to a random Gaussian field, and the localized non-Gaussian parameter $f_{\rm NL}^{\rm local}$ is of the order of the slow-roll parameter, which is at most the order of ${\mathcal O}(10^{-2})$. Current observations with $f_{\rm NL}^{\rm local}$ accuracy $\sim{\mathcal O}(1)$ are still far from detecting the non-Gaussianity of the single-field slow-roll model. Accurately measuring these properties of primordial perturbations is of great significance for detecting the origin of the Universe and testing different cosmological models.

At present, the most effective method to study the primordial perturbation is to look for ``non-trivial events'' beyond the single-field slow-roll model. These effects often leave detectable features in the power spectrum and bispectrum of the primordial perturbation. Except for a small amount of information overlap with the CMB lensing and the late integrated Sachs-Wolfe (ISW) effect, the data of the LSS almost independently contains information about the primordial perturbation of the Universe. 

Compared with the two-dimensional CMB data, the matter density field of the Universe is three-dimensional, and in principle contains more information, especially at linear scales. For instance, the primordial non-Gaussianity can affect the galaxy bias and make it scale-dependent at linear scales \cite{Dalal08}, which can be used to constrain $f_{\rm NL}^{\rm local}$. Therefore, at least at linear scales, it is reliable to use the LSS or galaxy clustering data to constrain the property of primordial perturbations. Using the galaxy clustering data in the CSST spectroscopic survey \cite{Gong19}, it is found that the $1\sigma$ error of $f_{\rm NL}^{\rm local}$ can reach $\sim 6$, which is comparable to the CMB result \cite{Planck20}.

The scalar spectral index $n_{\rm s}$ and the running of the spectral index $\alpha_{\rm s}={\rm d}n_{\rm s}/{\rm d\, ln}k$ of the primordial power spectrum also can be tightly constrained by the CSST 3$\times$2pt, peak statistics of weak gravitational lensing, and galaxy spec-$z$ clustering observations, with an accuracy higher than 3\% \cite{Gong19, Lin22, Miao23, Lin24}.

\subsubsection{Hubble parameter}

The Hubble constant $H_0$ characterizes the current expansion rate of the Universe, and the Hubble parameter $H(z)$ is used to describe the evolution of the expansion rate of the Universe at different redshifts. They directly affect the age, distance, and formation and evolution of the structure of the Universe. Therefore, the precise determination of $H_0$ and $H(z)$ is of great significance to cosmological research.

The direct measurement of the Hubble constant is mainly based on the distance ladder of the nearby Universe (e.g. Cepheid variables, Tip of the Red Giant Branch (TRGB) \cite{Rizzi07}, SNe Ia, RR Lyrae, Type II supernovae, etc.), the time delay of strong gravitational lensing, and other methods, and thus is limited to the observation of the low-$z$ Universe. On the other hand, by analyzing the CMB data, $H_0$ can be obtained by fitting the parameters of the cosmological model. However, there are significant differences in the $H_0$ results measured by different methods, especially the latest results based on the distances of Cepheid variables and SNe Ia differ by more than 5$\sigma$ from the results derived from the CMB data \cite{Riess22,Planck20a}, i.e. the so-called ``Hubble tension''.
If this is indeed caused by the evolution of the Universe, rather than some unknown systematics, it means that there are new physics to be discovered beyond the standard model of cosmology.

The differential age, also known as the cosmological chronometer, is one of the main methods for directly measuring Hubble parameter, and the results of the measurements are often referred to as observational Hubble parameter data (OHD) \cite{Zhang14}. However, due to the large measurement error, it is necessary to use large galaxy samples to improve the accuracy of the OHD reconstruction.

CSST will carry out precise measurements of $H_0$ and OHD based on Cepheid variables, TRGB, galaxy surface brightness fluctuations (SBF) \cite{Tony88}, low-redshift SNe~Ia and strong gravitational lensing \cite{Suyu17}, and the differential age of the Universe. It is expected to obtain important results in revealing the nature of the Hubble tension and accurately measure the evolution of the Universe.\\

In addition to the probes mentioned above, other cosmological probes, such as AP effect \cite{Xiao23}, galaxy-ellipticity correlation \cite{Xu23}, and cosmic optical background \cite{Cao22b}, also can be used in the CSST photometric and spectroscopic surveys to test cosmological models or theories, and constrain the cosmological parameters. 

CSST also has excellent synergy with other telescopes in the cosmological study. Compared to other Stage-IV survey telescopes, such as $Euclid$, RST and LSST, CSST has some significant advantages. For example, CSST-SC has high dynamic image quality ($R_{\rm EE80}<0.15''$), near-UV observation ($NUV$ band), and multi-band coverage (7 photometric and 3 spectroscopic bands), which can obtain high-resolution images, accurate photo-$z$ estimation, and adequate spectral information for cosmological and other studies. This provides a good foundation for cooperation with other Stage-IV telescopes, and has excellent complementarity in observational bands (e.g. near-UV and near-IR bands) and other aspects \cite{Cao18,Liu23}. In addition, CSST can also conduct joint observations with telescopes or detectors in other bands or methods, such as CMB experiments \cite{Wang23}, radio telescopes like Five-hundred-meter Aperture Spherical Telescope (FAST) \cite{Deng22} and MeerKAT \cite{Jiang23} for detecting 21-cm signal, and the future gravitational wave detectors \cite{Song24}. Considering all these observations, CSST is expected to make great contributions in the above cosmological fields.

\subsection{Galaxy and AGN}

Galaxies are the basic units that make up the Universe. Their properties and morphological structures are rich in diversity, and they also constitute part of the cosmic ecosystem. In addition, SMBHs at the center of galaxies co-evolve with their host galaxies through AGN feedback. Uncovering the key processes of galaxy evolution and interaction with SMBHs is one of the core scientific goals in astrophysics. In this field, by making use of the unique capabilities of CSST, e.g. large sky area, multi-band coverage, high spatial resolution, slitless spectral capability, multiple instruments, etc., and considering other space and ground observations, we can build large and complete photometric and spectroscopic samples of galaxies and AGNs covering the low-$z$ to high-$z$ Universe. We will comprehensively understand the various properties, physical processes, and cosmological evolution of galaxies, AGNs, and their central black holes, and try to solve a series of frontier issues on the formation and evolution of galaxies.

\subsubsection{AGN and massive black hole}

The SMBH accretion during the AGN phase can produce intense energy release and influence the galaxy evolution. They are also the key to galaxy evolution and material circulation. In recent years, with the breakthrough of observational technology and the development of multi-band survey projects, the exploration of the physical nature of AGNs and SMBHs has entered a new stage. 

Research on the AGN structure, accretion process and radiation mechanism will not only help reveal the intrinsic connection between black hole growth and galaxy co-evolution, but also provide key observational evidence for understanding the issues such as AGN feedback, matter transport and dark matter distribution. Meanwhile, the application of multi-band data will promote the classification and physical modeling of AGN light curves, accretion disk and special types (such as changing-look AGN, binary black hole system, etc.), and achieve a full range of analysis from statistics to microscopic processes. Using the high-angular resolution multi-color images, high-precision multi-color photometric and siltless spectroscopic data of the CSST main survey and supplemented by the data from MCI, IFS and other telescopes, it enable us to carry out research in the following four aspects.

\begin{figure}[H]
\centering
\includegraphics[scale=0.3]{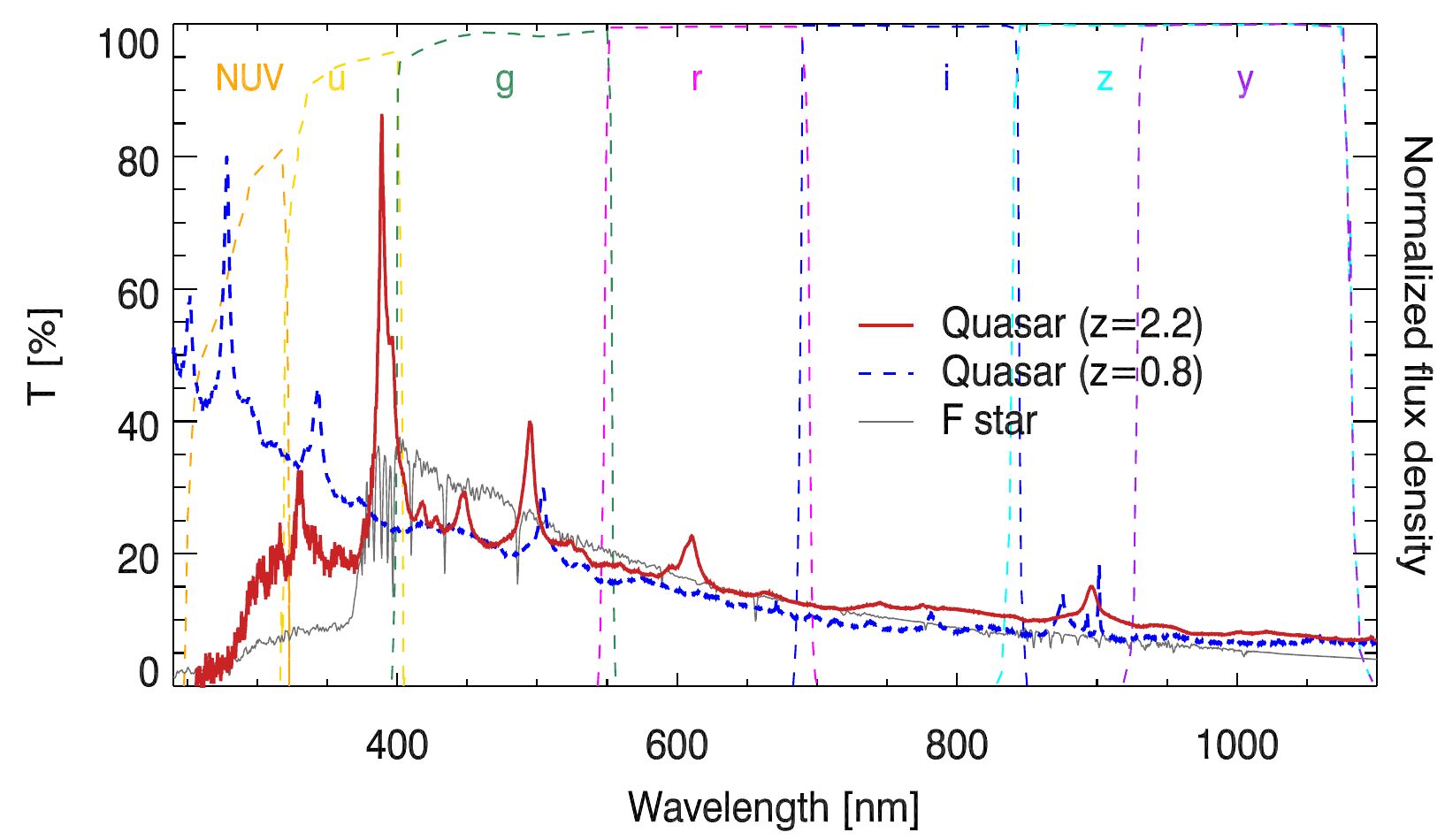}
\caption{The QSO spectra at $z=0.8$ (blue dashed curve) and $z=2.2$ (red solid curve), and F-type star spectrum (gray solid curve), compared to the CSST transmission curves of the seven photometric bands (i.e. $NUV$, $u$, $g$, $r$, $i$, $z$, and $y$). The $NUV$ and $u$ bands of the CSST are important for estimating photo-$z$ of quasars and distinguishing them  from stars.} 
\label{fig:QSO_spec}
\end{figure}

First, a large sample of AGNs and quasars (or quasi-stellar object, QSO) will be constructed, and their statistical properties and cosmological evolution can be studied in details. The CSST slitless spectroscopic survey will identify quasars brighter than 21 mag (per resolution element, $\sim23$ mag for a spectroscopic band), create a quasar identification catalog, and conduct physical and statistical analysis. On the other hand, the magnitude limit of the CSST multi-color imaging survey is about 26 mag, which is several magnitudes deeper than that of the slitless spectroscopic survey. This indicates that a considerable number of quasar candidates cannot be identified by their spectra. Therefore, for candidates with a high predicted probability (such as $p({\rm QSO})>0.9$) between 21 and 26 mag, their photo-$z$ can be estimated (see Figure~\ref{fig:QSO_spec}), and used in a series of subsequent studies.
In addition, by combining the CSST main survey data with the X-ray survey data, it is expected that a large sample of reliable AGNs will be identified. Based on the CSST 400 deg$^2$ deep-field survey data, we can obtain a AGN sample in the nearby Universe, especially a sample of low-mass AGNs.

Based on these large AGN samples, it is expected that the average relationship between black hole mass and halo mass can be established, and its redshift evolution can be obtained. By comparing with simulation data, it will also provide accurate constraints on the co-evolution model of SMBHs and host galaxies. With the greatly improved constraints on the galaxy stellar mass surface density, it can significantly enhance the measurement accuracy and provide effective constraints on the theories of the AGN triggering and quenching, as well as its evolution.

Second, the structure, accretion, radiation mechanism and light variation of AGNs and quasars will be probed. The advantages of CSST AGN survey are high spatial resolution and multi-band observations in large sky area, which are helpful for the resolution and identification of large samples of low-luminosity AGNs and eliminating the UV luminosity contamination of host galaxies. Besides, the CSST slitless spectroscopic survey can observe broad emission lines such as H${\beta}$, Mg~II and C~IV. Repeated observations can simultaneously provide the spectra of bright sources and the changes in emission lines. At the same time, the narrow emission lines measured in slitless spectra can provide the physical parameters of narrow-line clouds, making it possible to determine AGN classification. 
Using the CSST high spatial resolution data, combined with theoretical models and multi-band light variation observations, we can study the accretion disk structure, black hole accretion process and radiation mechanism of AGN. We will analyze the physics behind the special light variation behavior, and further explore the complex interaction between accretion physics and black hole spin and feedback effects.

\begin{figure}[H]
\centering
\includegraphics[scale=0.47]{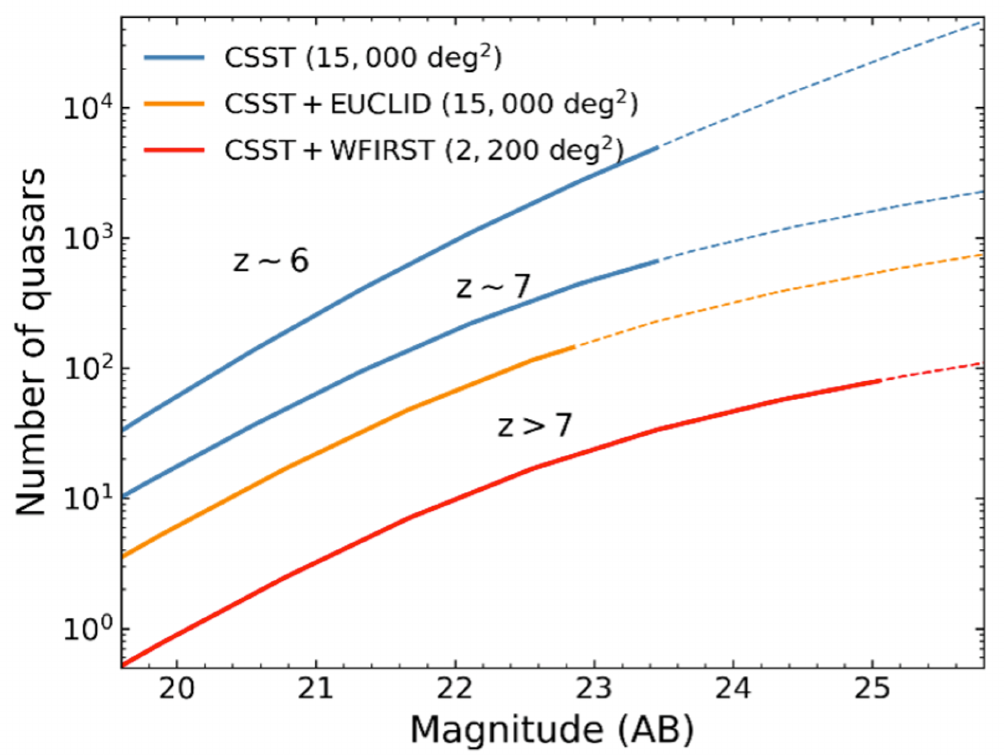}
\caption{The number of high-$z$ quasars that CSST can discover as a function of magnitude. In CSST main survey, more than 4,000 quasars at $z\simeq6$ and about 700 quasars $z\simeq7$ can be discovered. Combining the CSST optical survey data with $Euclid$ and RST or WFIRST infrared data, quasars at higher redshifts can be discovered. The solid curves show the depths that can be reached when searching for quasars (note that these depths are not equivalent to the CSST survey depths).} 
\label{fig:AGN_num}
\end{figure}

Third, with the help of multi-band data provided by CSST and other telescopes, special AGNs and quasars (such as changing-look AGNs) can be searched and confirmed. Their physical origins and evolution history will be explored, and new observational support for black hole growth, galaxy feedback mechanisms, and large-scale structure formation can be provided. Using the CSST multi-color photometric  and slitless spectroscopic data, as well as ground-based and space-based multi-band data (such as $Euclid$, RST etc.), CSST  will discover a large number of high-$z$ quasars ($\sim4000$ at $z\simeq6$ and $\sim700$ at $z\simeq7$) with the blue-end dropout technology (see Figure~\ref{fig:AGN_num}), based on the luminosity functions of quasars at $z\simeq6$ and $z>6.5$ \cite{Jiang16aa,Wang19aa}. The CSST wide-field survey will detect tens of thousands of lensed quasar candidates and further confirm strong lensed quasars through follow-up observations to explore a series of scientific issues involving foreground galaxies, quasar physical properties, and Hubble constant measurements. CSST also can detect thousands of new changing-look AGNs and a batch of narrow-line Seyfert 1 galaxies (NLS1s) and blazars, and will greatly advance the studies of these special AGNs.

Besides, CSST will effectively detect binary black hole systems, taking advantage of high spatial resolution and high-precision photometric observations. CSST is expected to detect millions of quasars and obtain about 100 million emission-line galaxies in the slitless spectroscopic survey. Theoretical calculations show that the fraction of supermassive binary black hole systems is about 5\%-10\% at low redshifts and increases at high redshifts \cite{Volonteri03}. Using high spatial resolution and large FoV of CSST, it is possible to detect binary AGNs from 100 kpc to 100 pc, and obtain the redshift evolution of their distribution, as well as search for tidal traces left by galaxy mergers. Using the slitless spectra, the broad line regions carried by binary AGNs or binary black holes at sub-pc scales can be analyzed in the time domain through reverberation mapping observations. It is expected to identify close binary AGNs with sub-pc spacing and measure their orbital parameters. The late process of galaxy mergers, especially the relationship between black hole mergers and the surface brightness distribution of the bulge clusters, can be obtained. Furthermore, the binary black hole sample will be constructed, and the relationship with pulsar timing array (PTA) observations can be established.

\subsubsection{Black hole-galaxy co-evolution}

SMBHs are found to be ubiquitous in the centers of massive galaxies. The correlation between SMBHs and host galaxies reveals that there is a tight connection between the two in terms of mass growth and evolution \cite{Kormendy13}. Based on this, a picture of the co-evolution of SMBHs and host galaxies was proposed \cite{Zhuang23}. It is the key to understanding the mass growth and co-evolution of black holes and galaxies to study the physical properties and evolutionary characteristics of AGN-galaxy systems, and it is also one of the important scientific goals of current and future multi-band observational projects.

Using the CSST multi-color photometric and slitless spectroscopic data from SC, supplemented by MCI and IFS, and data from other telescopes, we can study the morphological structure and star formation of AGN host galaxies, the relationship between AGN SMBHs and their host galaxies, the triggering of AGN, and the feedback of AGN.

At present, the difficulties in studying the host galaxies of AGNs mainly lie in two aspects. First, the radiation and morphology of stars in the host galaxies are difficult to observe; second, the strong emission lines produced by AGNs prevent the use of the calibration method based on inactive galaxies to estimate the star formation rate of host galaxies. Based on the CSST multi-band high-spatial resolution images, slitless spectra, IFS data, etc., we will effectively separate the galactic components and AGN components of host galaxies, efficiently observe and classify the morphology and environment of host galaxies, and measure the star formation rate and star formation history of host galaxies, revealing the co-evolution process of SMBHs and host galaxies.

High-$z$ quasars play an important role in studying the formation of black holes and the evolution of galaxies. Finding and studying the SMBHs in the early Universe can put the most powerful constraints on seed black holes \cite{Wang21}. A variety of methods (such as color, morphology, light variation, etc.) can be used to identify AGNs and quasars with redshifts ranging from 0 to 7. Their photo-$z$ and spec-$z$ can be measured, a highly complete large sample can be constructed, and the statistical properties (such as luminosity function, black hole mass function, and clustering) and cosmological evolution will be studied.

The AGN triggering mechanism is an important part of studying the co-evolution of galaxies and black holes. How kpc-scale matter loses a large amount of angular momentum and falls into the accretion disk (0.01pc scale) is the main problem in understanding the triggering of AGN . Major mergers between gas-rich galaxies are considered to be the main mechanism for triggering high-luminosity quasars \cite{Treister12,Fan16}. By using the survey data from CSST, large-sample studies can be carried out to statistically investigate the correlation between major mergers and the AGN triggering with different luminosities.

AGN feedback is another important field for studying the co-evolution of black holes and galaxies. The radiation, outflows, and jets generated by black hole accretion may affect the galactic bulge, or even the gas and stars of the entire galaxy, inhibiting or promoting star formation \cite{Fabian12}. In addition, the color, morphology, and quenching of star formation of galaxies may be closely related to the central black hole \cite{Bluck22}. The correlation between the physical properties of black holes and galaxies is an important means to study AGN feedback \cite{Heckman14}. However, the current research on high-$z$ quasars is severely limited by the number of quasars, so that we cannot conduct statistical analysis on large samples. Based on the large FoV and high spatial resolution of the CSST survey data, we expect to discover a large number of high-$z$ quasars and study the evolution of their central SMBHs and host galaxies as well as their relationship with the surrounding large-scale environment.

\subsubsection{Low-redshift galaxies}

The low-$z$ Universe is an ideal place to understand the components of galaxies and the transformation or interaction between them, and also provides a local benchmark for the study of the formation and evolution of medium- and high-$z$ galaxies. Galaxies are composed of stars of various ages, ISM, circumstellar medium, and massive black holes. These different components form larger structures in space, such as disks, spiral arms, bars, and rings. CSST-SC multi-band photometric imaging and slitless spectroscopic wide-field survey and IFS observations will provide large-sample continuous spectrum images and spectral data with high spatial resolution, which can be used to study the statistical properties of different galaxy components and structures, and understand the physical mechanisms of driving the formation and evolution of low-$z$ galaxies, by combining with high-precision numerical simulations. Besides, the diversity of low-$z$ galaxy types and galaxy environments also provides a key foundation for further comprehensive understanding of galaxy formation and evolution theory.

The CSST main survey is expected to measure the morphology and structure properties of young globular clusters of different galaxy types and environments within hundreds of Mpc. The high-quality imaging capability of the CSST will provide an unprecedented large sample of galaxy nuclear star clusters (NSCs). The high-precision photometric capability will obtain the NSC color index, radius, mass, stellar population, star formation history, morphology, and complete distribution of NSC physical parameters \cite{Ashok23}. In this way, the correlation with the parameters of AGN dust ring can be established, the physical connection between the NSCs, bulges and massive black holes in a galaxy can be explored, and the physical process from the galaxy to the AGN transition region can be revealed.

The detailed structure of galaxies contains clues related to the evolution history of galaxies and provides the internal environment of current evolution. The CSST wide-field survey with high spatial resolution will greatly expand the sample and carry out statistical analysis of the detailed structure of galaxies, which includes developing small-scale structure characterization methods focusing on disk structure, studying formation of deflection structure of central symmetric disks, exploring the influence of non-central symmetric structure on the distribution of gas and star formation, investigating detailed structure of galaxies in galaxy groups and clusters, and probing thick disks.

The outer regions of disk galaxies contain unique information about the physical processes that affect galactic evolution. Using the CSST high spatial resolution images, we can detect star clusters and single bright stars in the outer disk of galaxies. Considering the large sky coverage of CSST, the data at different radii of galaxies can be stacked, so that the areas with low star surface density can be detected. This can be helpful to answer two questions: the extension scale of galaxies and the physical mechanism that causes the growth of stellar disks. In addition, many galaxies have extended ultraviolet disks (XUV disks) at their outer edges \cite{Lemonias11}. CSST can be  used to conduct UV-optical observations of a large number of XUV galaxies that can resolve single stars and study their physical properties \cite{Thilker23}.

Early-type dwarf galaxies, especially dwarf elliptical galaxies (dE), are the most numerous type of galaxies in low-$z$ galaxy clusters and  groups. The high spatial resolution and multi-band coverage of the CSST wide-field survey will be able to provide a high-quality image library of low-$z$ dE galaxies. It is expected to achieve a complete survey of NSCs in dE galaxies, a systematic study of the structure and stellar composition of dE galaxies from the center to the outer regions, and the detection of high-resolution three-dimensional structure of low-$z$ galaxy clusters and groups.

Star formation in galaxies is one of the main driving forces of galaxy growth and evolution, and is closely related to the mass structure, nuclear activity and other physical characteristics of galaxies. CSST's multi-color imaging survey provides near-UV high-resolution deep-field images, which can effectively detect and analyze small-scale ($<1$ kpc) star formation and distribution in low-$z$ galaxies. CSST multi-color imaging survey is also expected to quantify the relationship between the star formation rate and molecular gas density on the sub-kpc scale,  by exploring the group behavior and statistical properties of star formation in low-$z$ galaxies. Besides, CSST can conduct analyses on galaxies with strong radiation in the UV spectrum, which will help explain their origins \cite{Yi11}.

\subsubsection{Medium and high-redshift galaxies}

The morphology of a galaxy is tightly related to the merger and accretion history, intrinsic dynamics, star formation, and the activity of SMBHs \cite{Binney87,Kormendy04}. The galaxy morphology classification system (i.e., the Hubble sequence) has been established for nearly a hundred years \cite{Hubble26}, but its cosmological evolution process is still unclear. Studying the morphology and structure properties of galaxies at different redshifts is helpful to understand the influence of the intrinsic physical processes of galaxies and the external environment. It can build a bridge between low-$z$ and high-$z$ galaxies, and provide an important observational foundation for theoretical models of galaxy formation and evolution.

Understanding these important issues related to galaxy formation and evolution requires high-resolution images of a large number of galaxies in different redshift ranges. Compared with existing sky surveys, CSST, which has a deep observational depth (average 26 mag), a large survey area (17,500 deg$^2$) and a high spatial resolution (sub-arcsecond level, similar to HST), will be an ideal device for studying the origin of the Hubble sequence and the formation and evolution of galaxy structures. The CSST photometric survey is expected to obtain billions of multi-band high-resolution images of nearby Universes and even medium- and high-$z$ galaxies, which will greatly promote the understanding of galaxy formation and evolution and bring unprecedented opportunities to the field of galaxy morphology research.

Based on CSST observation, we can carry out research on automatic classification of medium- and high-$z$ galaxy morphology, multi-component structure decomposition, and redshift evolution of morphological and structural parameters, to understand the physical processes related to the formation and evolution of medium- and high-$z$ galaxies from the perspective of galaxy morphology and structure. 

When the resolution and S/N of galaxy images are low (such as when the galaxy is close to the detection limit), there are large errors in the measurement of its structural parameters. Therefore, it is necessary to use simulated images to test how the measurement errors of galaxy structural parameters vary with resolution and S/N. The mock galaxy images can be obtained by using galaxy image analysis code (e.g. GALFIT) or real multi-band galaxy images (e.g. HST CANDELS) \cite{Peng10,Schreiber17}. By applying the  measurement methods, we can obtain the trend of measurement deviations of different structural parameters of galaxies as the image resolution and S/N change, which will help to understand and correct the statistical analysis results of galaxy image structural parameters.

The morphology of a galaxy characterizes its internal structure and its evolutionary stage. Galaxies of different morphological types may have different formation and evolution processes. The CSST image survey will significantly increase the number of samples of medium- and high-$z$ galaxies, and study their formation and evolution history of the morphology. It is expected that the statistical distribution characteristics of galaxy morphology within the highest redshift of 2 can be constructed, and the overall evolution process of the Hubble sequence can be reproduced \cite{Xu23aa,Zhou22,Fang23,Li22aa}. Besides, galaxy morphological parameters have rich physical meanings, and they are related to the star formation history of galaxies, intergalactic interactions, and small-scale structures. 

It is expected that CSST will provide a standard star catalog for multi-band morphological parameter measurements of galaxies, which can be used to study the statistical properties of galaxy structural parameters at different redshifts and different bands. CSST can also construct a large sample of unique compact galaxies, analyze their morphological parameters, and further understand the formation process of compact galaxies. It is also intended to find more loosely structured dwarf galaxies with nuclei, and even ultra-diffuse galaxies with nuclei, and study their statistical properties, which will enable us to further understand the fate of dwarf galaxies after mergers.

CSST will study the formation and evolution of substructures of medium and high-$z$ galaxies \cite{Delmestre07,Yu18}. The multi-component decomposition and statistical analysis of medium- and high-$z$ galaxy images will be explored, and the deviation of multi-component decomposition can be evaluated by using CSST simulated image data. It is expect to obtain a multi-component decomposition catalog of a large sample of galaxy images, and conduct statistical analysis on physical parameters, such as structural parameters of bulges and disks, bulge mass fraction, and stellar population in galaxies of different masses and star formation rates, to reveal their formation and evolution history as a function of redshift. In addition, CSST will focus on the identification and measurement of bars and spiral arms in galaxies within $z=1$. By combining HST and JWST data, CSST can  identify and measure the bar structure in the early stages of disk galaxies and reveal the role of substructures in galaxies in their mass growth and structural evolution. CSST will also study the physical properties and evolution of clumps within galaxies.

\subsubsection{Galaxy evolution and assembly history}

The properties and morphology of galaxies evolve over cosmic time and also depend on the environment they are located in. The evolution of galaxies is driven by two different factors: internal and external. Studies have shown that the stellar mass of a galaxy is the most important factor in its internal evolution, while the effect of external environment at different scales is another important factor. Therefore, to fully understand the evolution of galaxies, we must deeply understand the role of environments at different redshifts and scales. CSST's wide- and deep-field multi-band photometric and slitless spectroscopic surveys and IFS observations will provide detailed data for studying galaxy evolution and the effects of environments at different scales. 

Using CSST multi-band high-resolution photometric and slitless spectroscopic surveys, combined with other multi-band survey data, we are able to measure important parameters, such as stellar mass, star formation rate, dust extinction, stellar population age, and metallicity, for a large number of galaxies at the low and medium redshifts \cite{Wilkinson17}. CSST can also measure the overall properties of a large number of complete galaxy samples, combined with galaxy pairs, merging galaxy statistics, and different local environment measurements, to construct the mass/luminosity function at different redshifts, as well as the scaling relationship between the mass-structure parameters-stellar population properties. Combined with numerical simulations or different types of galaxy evolution models, statistical modeling of these observations will greatly deepen our understanding of the galaxy stellar assembly history.

During the cosmic noon at $z=2\sim3$, the physical properties of galaxies have changed significantly (such as star formation, morphological structure, stellar mass, etc.). It is crucial to study whether the evolution of galaxies is dominated by mass or environment at different redshifts for understanding the overall picture of galaxy formation and evolution. CSST provides a high-resolution survey from UV to near-IR bands, by combining the observational data of other surveys, it can also obtain relatively accurate photo-$z$ and other related physical properties for medium- and high-$z$ galaxies. Therefore, the large sample  provided by CSST can put observational constraints on the turning point of galaxy evolution from mass-dominated to environment-dominated, thereby solving the relationship between the three factors of redshift, mass and environment in galaxy evolution, and further testing and constraining galaxy formation and evolution models.

Besides, since CSST can perform high-resolution wide-field survey, it will effectively probe the morphology and structure of galaxies and their dependence on the environment. CSST can carry out measurements of environments at different scales (including groups, filaments, voids, etc.), and study the impact of the environment on various properties of galaxies (including mass function, star formation rate, morphology, halo assembly history, gas, etc.), and their redshift evolution \cite{Lin19,Wang23aa,Peng10aa,Gu21}. CSST can also use its high-resolution photometric and spectroscopic survey to explore the formation and evolution history of galaxy groups and clusters, the formation and assembly history of the member galaxies, and the environmental effects of different redshifts and different types of galaxy groups/clusters on the formation and evolution of their member galaxies.

By measuring tangential shear or excess surface mass density (ESD) around galaxies, CSST can detect the average matter density profile for studying the connection between dark matter halos and galaxy properties. Combining different galaxy observational properties (stellar mass, structure, star formation rate, etc.) and theoretical modeling methods, it will characterize in detail the important properties of dark matter halos such as mass, concentration, and fraction of satellite galaxies of different galaxy groups, providing the foundation for establishing a more complete galaxy-dark matter halo model. In addition, CSST will also discover a group of extreme emission-line galaxies (EELGs) and other special galaxies in the wide-field survey, and it is expected to greatly improve the understanding of the structure, morphology, gas composition and evolution of these galaxies.

\subsubsection{High-$z$ galaxies and reionization}

The study of cosmic reionization and first galaxies is one of the most important astrophysical topics, and theory and observations show that galaxy clustering is a crucial factor for driving the reionization process. CSST will discover a large number of high-$z$ protoclusters, and it will help us understand the reionization process and the early cosmic environment of galaxy evolution when collaborate with $Euclid$ and RST. The key scientific issues of high-$z$ galaxies and reionization include identifying the main contributors to reionization, understanding how ionizing photons are generated and escape from galaxies, and the process of ionizing the surrounding neutral hydrogen. It will help us gain a deeper understanding of the reionization by studying the dynamics of gas, the luminosity function of bright sources, the chemical evolution of galaxies, the structure evolution of protoclusters, etc.

By utilizing the 17,500 deg$^2$ wide-field, 400 deg$^2$ deep-field, 9 deg$^2$ UDF of the CSST-SC surveys, and the 0.1 deg$^2$ XDF of the CSST-MCI observation, considering the high spatial resolution and slitless spectral capabilities of the CSST, the largest high-$z$ galaxy sample can be established, including the Lyman break galaxy (LBG) photometric sample, the Lyman-alpha emitter (LAE) spectral sample, and the bright LBG spectral sample. These samples can be used to accurately explore various characteristics of early star-forming galaxies and their cosmological evolution, study the LSS and evolution, and provide a powerful tool for probing the reionization.

In addition, combined with the deep fields of HST (such as XDF, CANDELS, CLASH, Frontier fields, etc.), as well as the deep fields of JWST, $Euclid$, RST, etc., CSST can probe red disk galaxies and high-$z$ galaxies exhibiting red colors in the near-IR from UV to optical rest bands, including the extinction properties of dust-rich galaxies, the UV morphology of red galaxies, and the high-resolution multi-band properties of clumps identified by JWST. It is also of great significance for understanding the extinction properties of galaxies at different redshifts, the history of galaxy mass growth, the formation of bulges, and the activities of central SMBHs.

CSST can search protoclusters and study LSS at high-$z$. It is expected that the CSST main survey will make the number of protoclusters at $z>4$ with collapse mass $\gtrsim10^{15}\ M_{\odot}$ (similar to the Coma Cluster) at least one order of magnitude larger than the existing sample. Based on this sample, we can classify protoclusters according to the structure and properties of member galaxies, and establish the evolution history of galaxy clusters from the early Universe to nearby Universe,  with the help of numerical simulations. By studying the statistical properties of member galaxies and correlation with the whole protocluster, we will be able to probe the differences in the spatial distribution and clustering of dark matter and baryons in the early Universe, and study the baryon circulation process of galactic matter and energy feedback and intergalactic medium (IGM) enrichment in a dense metal-poor environment. Based on the CSST sample and combined with the existing new generation of spectroscopic surveys (e.g. DESI, PFS), we can establish the relationship between the IGM, circumgalactic medium (CGM) of protoclusters and the cosmic structure, and carry out comprehensive three-dimensional information exploration using IGM absorption lines \cite{Li21aa}.

\begin{figure}[H]
\centering
\includegraphics[scale=0.33]{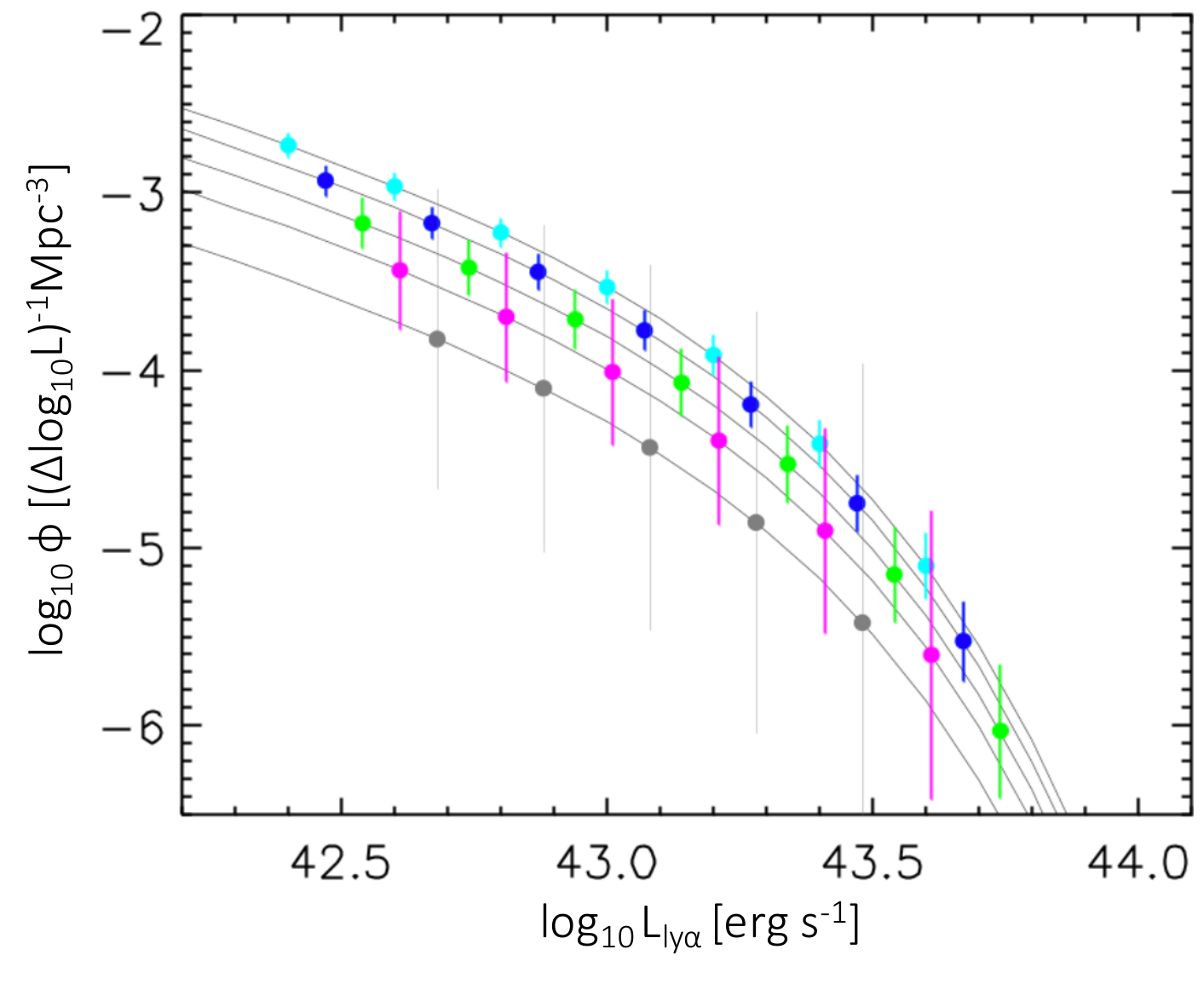}
\caption{Predicted Ly$\alpha$ luminosity functions from CSST high-$z$ LAE observation. The cyan, blue, green, purple, and gray data points denote the measured Ly$\alpha$ luminosity functions at $z=5.6$, 5.9, 6.2, 6.5, and 6.8, respectively. The theoretical Ly$\alpha$ luminosity functions are also shown in gray solid curves.} 
\label{fig:Lya_LF}
\end{figure}

The large sample of high-$z$ galaxies from CSST observation will greatly promote the study of cosmic reionization. CSST deep-field UV observations can directly detect ionizing radiation (e.g. Lyman continuum emission, LyC), which can increase the number of LyC emitting galaxies at $z=3\sim5$ by nearly one order of magnitude \cite{Yuan22}, which can give a strict upper limit on the average escape fraction of ionizing photons for estimating the contribution of galaxies to the reionization \cite{Naidu18, Wang25}. CSST will accurately measure the Ly$\alpha$ luminosity function at $z=5.5-7$ (see Figure~\ref{fig:Lya_LF}), and obtain a complete and large sample of LAE and LBG, which can improve the measurement accuracy of the fraction of neutral hydrogen in the IGM. The large sample of galaxies and AGNs at $z>5$ detected by CSST will be effectively combined with the data from next generation 21cm survey to explore the detailed ionization process of reionization.

\subsection{Milky Way and nearby galaxies}

Hundreds of billions of stars constitute a galaxy system with complex dynamic structures bounded by gravity, in which they participate deeply with gas and dust in various physical and chemical processes, such as nucleosynthesis, radiation transfer, hydromechanics, magnetic fields, molecular synthesis and dissociation. The overall stellar population in a galaxy reveal the process of star formation, evolution and death, and reflect the formation and evolution history of the galaxy as a whole. The complex motion trajectories of stars reveal the influence of dark matter on the galaxy scale. The ``fossil'' of chemical abundance inside a galaxy reflects the important process of nucleosynthesis in the early Universe. The ISM in a galaxy contains the distribution and evolution of chemical molecules in stars and is closely related to the formation of life. Part of the ISM diffuses to galaxy outskirts and intergalactic space and become warm gas, which is an important component of baryonic matter and a link in the ecological cycle of the Universe. Therefore, conducting detailed observational studies on star populations, gas and dust in a galaxy, and the galaxy as a whole has become a long-standing basic scientific issue in the field of astrophysics, and is of irreplaceable importance for understanding the formation and evolution of the Universe.

\subsubsection{Stellar population}

A stellar population is a group of evolutionarily related stars, whose physical characteristics are of great significance to the understanding star formation and galaxy composition, and is an important link between stars and galaxies. 
The stellar initial mass function of a stellar population is a basic distribution in astrophysics. It not only reflects the specific physical mechanism of star formation, but is also an important factor for understanding the formation and evolution of galaxies. Besides, it is believed to have a profound relationship with some rare physical phenomena (such as supernovae, gravitational wave events, etc.). 
On the other hand, the age, metallicity, and star formation history of stellar populations are also crucial to the study of galaxy formation and evolution. 

As a galaxy in the nearby Universe, the study of stellar populations in the Milky Way has made great progress in recent years. The release of the $Gaia$ astrometric catalog combined with ground-based stellar spectral survey data, such as the Large Sky Area Multi-Object Fiber Spectroscopic Telescope (LAMOST) \cite{Cui12} and SDSS \cite{Eisenstein05}, provides important observational data for the study of the Milky Way. The study of stellar populations in the Milky Way is mainly carried out in star clusters, field stars, nuclear bulges, etc.

Some studies indicate that the Milky Way may be special in the properties of star populations than other galaxies \cite{Lian23}, so the understanding of the formation and evolution history of galaxies cannot rely solely on the results of the Milky Way. The statistical results of star populations in different types of galaxies will be helpful to have a deeper understanding of the specific process of galaxy formation and evolution \cite{Dalcanton12}.

The CSST-SC will provide multi-color imaging and slitless spectroscopic data covering near-UV to near-IR. The CSST-MCI can observe three bands simultaneously, with a $7.5'\times7.5'$ FoV. These will be the two most important CSST scientific instruments for studying the stellar populations of the Milky Way and nearby galaxies.

Through CSST 10-year observation, it is expected that the largest and most complete stellar multi-color photometric catalog of 50-100 stellar-resolvable galaxies within 4 Mpc will be obtained. It includes Milky Way-like galaxies (e.g. M31), starburst galaxies (e.g. M82), active galaxies (e.g. M81), elliptical galaxies (e.g. Cen A), various dwarf galaxies (e.g. Large Magellanic Cloud (LMC), Small Magellanic Cloud
 (SMC), NGC300, and NGC55) and their surrounding single stars, satellite galaxies, star clusters, etc. For the member galaxies of key galaxy groups (e.g. M81 group, Cen A group, Sculpter group, Local Group, etc.) and intergalactic space, the photometric magnitude limit in $g$ and $r$ bands is at least 2 mag deeper than TRGB. It is expected that this data set will become one of the most important samples for understanding the formation and evolution of galaxies.

Based on this sample, great progress will be made in the properties of various stellar populations in nearby galaxies. Important observational results of the variation of the initial mass function in different interstellar environments will be obtained \cite{Li23aa}. The evolution of stellar populations in different galaxies and the characteristics and evolution of stellar halos in different types of galaxies can be accurately measured. The number and types of satellite galaxies, and the number, types and spatial distribution of star clusters around different types of galaxies will be probed. The fundamentals of satellite galaxies and star clusters of different types of galaxies will be established, providing unprecedented complete data for the origin, merger and accretion process of satellite galaxies and constraints on the formation and evolution of star clusters.

In addition, CSST can measure the overall proper motion of the member galaxies of the Local Group, and based on this, the dynamical structure of the Local Group can be given, which provides unique constraints on the dark matter distribution and dynamical evolution of the Local Group. CSST also will probe the outer halo of the Milky Way to understand its merger history, use different stellar populations as tracers to constrain the dark matter halo of the Milky Way, and confirm the boundaries of the stellar halo and thereby constrain the galaxy formation history. A larger sample of brown dwarfs within a few kpc from the Sun also can be collected for studying the properties and evolution of brown dwarfs.

\subsubsection{Extinction and dust}

The dust of various sizes, ranging from sub-nanometer to micron, is distributed in various environments from Solar System  to the early Universe, and plays an important role in astrophysics research \cite{Cardelli89}. Although interstellar dust accounts for only one thousandth of the baryonic mass of the Milky Way, it interacts with electromagnetic radiation in almost the entire frequency domain, thus comprehensively affecting all branches of astrophysics. The absorption and scattering of starlight by dust is collectively called interstellar extinction. The measurement of interstellar extinction and infrared re-radiation of the Milky Way and extragalactic galaxies is the key to restore the intrinsic brightness and SED of astronomical sources and determine the star formation rate on a galaxy scale.

To obtain the true luminosity and intrinsic color index of astronomical sources, as well as the photometric distance of stars, it is necessary to accurately eliminate the influence of extinction, especially extinction in the UV band. However, the lack of large-area UV survey data has made detailed research of UV extinction still challenging. In the CSST era, based on the CSST survey data from near-UV to near-IR, combined with other survey data, it is possible to carry out the systematic extinction study of the Milky Way and nearby galaxies.

Based on the CSST data, combined with existing spectroscopic, photometric, and astrometric data, we can determine the general extinction law of the Milky Way from near-UV to infrared bands. We can explore the differences in the extinction law under different interstellar environments due to the influence of ISM density, background radiation, etc., and obtain the spatial distribution of extinction. Finally, we will obtain the precise extinction law and spatial distribution of the Milky Way, providing a basis for CSST extinction correction. At the same time, CSST and other ground and space-based survey data can be used to obtain multi-band extinction curves and extinction maps of nearby galaxies, providing extinction corrections for the final star catalog products of CSST. It will reveal how the extinction law changes with the galactic environment, and achieve observational constraints on the construction of dust models for nearby galaxies.

Using the CSST multi-band photometric and slitless spectroscopic observations of the Milky Way and nearby galaxies in multiple sky regions and interstellar sight lines, combined with the infrared and millimeter data and interstellar element abundances from the Infrared Astronomical Satellite (IRAS), Spitzer and $Planck$, interstellar dust models can be constructed. The size distribution, chemical composition and other characteristics of dust can be obtained. The relationship between dust properties and the interstellar physical and chemical environment will be analyzed, and the relationship between dust extinction and other important interstellar observational features (such as interstellar diffuse belts, aromatic infrared radiation spectra, X-ray ``halos'', etc.) can be studied. The electronic absorption spectral characteristics of carbon dust in the CSST UV band, such as graphene, carbon nanotubes, nanodiamonds, fullerenes (such as C$_{60}$), carbon chains and non-crystalline carbon, will be examined \cite{Nie22,Li19,Li20}. In short, CSST will bring unprecedented opportunities for the study of extinction and dust physics.

\subsubsection{Galaxy structure}

The structure, formation and evolution of galaxies are the key to the formation of complex structures in the Universe, whose research depends on the observation of a large number of stars in galaxies. Although the Milky Way is currently the only disk galaxy that we can use a complete stellar population to study its star formation and structure growth process, the boundary of the Milky Way has not been observed so far, and the study of the high extinction region of the Milky Way needs further observations.

CSST can resolve stars in galaxies within a distance of Mpc from the Milky Way over a large sky area. It can also observe about 100  deg$^2$ of the bulge region of the Milky Way in a short time, which has both high extinction and dense star fields. In the next 10 years, CSST will be an important facility for the study of the bulge and halo of the Milky Way and nearby galaxies and probing the diversity of galaxies. It will provide a unique observational sample for verifying the cold dark matter model at small scales (from 3 kpc to 4 Mpc).

Through CSST observations, we can analyze the chemical properties and kinematic characteristics of the Galactic bar structure to distinguish halo stars from bulge stars. Combined with numerical simulations, the physical mechanism of the differences in the chemical structure of the bulge can be revealed and different bulge models will be verified. We also can study the co-evolution of the halo and bulge in the central region of the Milky Way through the characteristics of halo stars.

CSST will probe the structure and substructure of the stellar halo covering the border of the Milky Way. It will explore the Galactic gravitational potential through the motion of halo stars, accurately measure the dark matter distribution and the mass distribution of dwarf galaxies within 30 kpc of the Milky Way \cite{Wang22}. It is expected to find the boundary of the Galactic stellar halo, reveal the difference in the structure of the Galactic inner and outer halos, and constrain their evolutionary correlation. It will probe the accretion history of the Milky Way, and investigate its formation and evolution history.

The star clusters, giants and supergiants in the M31/M33 galaxies will be probed by CSST. It will measure their disk and halo structure, constrain the dark matter distribution of M31, and explore the impact of the ecological environment of the Local Group on the disk of M31/M33. CSST also will search for more dark dwarf galaxies/globular clusters and tidal streams, measure the overall proper motion of more than 50 galaxies in the Local Group. It is expected to discover a batch of dwarf galaxy candidates with low brightness and small size within 2 Mpc, greatly improve the completeness of the sample of nearby dwarf galaxies, and construct a more reliable luminosity function of satellite galaxies in the Local Group \cite{Qu23}. Using these data, we can construct a dynamical model of the Local Group, predict the interaction and merger of M31 and MW in the next four billion years, and infer the evolution and ultimate fate of the Local Group.

Additionally, CSST will detect tidal flows in nearby galaxy groups and establish the relationship between the evolution of galaxy structures and the environment. It is expected to reveal the shape distribution of multi-type galaxy halos and disks, deeply understand the relationship between galaxy structure and formation and its surrounding ecological environment, verify the cold dark matter model with a large galaxy sample, and understand the particularity of the Local Group of galaxies.

\subsubsection{Chemical evolution}

Since chemical abundance is the only parameter that does not change with position distribution and motion state, the chemical evolution of the Milky Way and nearby galaxies is not only crucial for explaining the early merger history of galaxies and tracing the co-evolution of different stellar population, but also provides key evidence for revealing the formation and evolution of different structures at the galaxy level and exploring the history of the abundance of cosmic material.

The chemical evolution of the Milky Way is mainly obtained by analyzing the medium and high resolution spectra of stars, so the detection distance is very limited (mainly within 5 kpc from the Galactic plane \cite{Majewski17}). However, the Galactic halo representing the early evolution of the Milky Way is more than 10 kpc away from the galactic plane \cite{Carollo10}, while the merger-dominated outer halo of the Milky Way is beyond 30 kpc. Hence, there is not enough chemical abundance data to trace galaxy early history at present.

The CSST survey, which has both deep observation and high spatial resolution, will change this situation. From its high-precision photometric and slitless spectroscopic data, we can extract information such as the chemical abundance, radial velocity and age of a large number of stars in the Galactic halo and nearby galaxies, search for extremely metal-poor stars in the Galactic outer halo and neighboring dwarf galaxies, and study the early evolution of galaxies. At the same time, using CSST high spatial resolution data, we can describe the chemical co-evolution of the Galactic halo, disk and bulge. Taking advantage of CSST survey depth,  we can compare the chemical evolution of the Milky Way with the surrounding satellite galaxies and M31, restore the dynamic process, and reveal the history of galaxy mergers, so as to deeply understand the properties of the first generation of stars and the history of chemical enrichment in the early Universe. 

Based on the large number of CSST star samples, we will obtain high-precision statistical ages of stellar populations. Using the CSST star coverage in the direction of the Galactic center, we can fill the gap in the study of the macroscopic stellar formation and evolution patterns in large-scale dense star fields. By analyzing the starquakes of high-luminosity red giants in the galactic center, we will construct the first high-precision and high-resolution asteroseismological samples in the galactic center. We expect to obtain the mass, age, and relationship between the vibration period and luminosity of such stars.

\subsubsection{Galaxy ecosystem}

The baryon cycle is one of the most critical and dynamic processes in the galaxy ecosystem. This cycle closely links galaxies, ISM, CGM, and IGM. The gas inflows, outflows, and feedbacks it presents form a complete dynamic system that drives galaxy evolution. The current theory suggests that the CGM is where the accretion inflow of LSS interacts with the outflow of galactic disks. Many important physical processes occur here, including accretion, feedback, cold and hot gas interaction and phase transition, which directly affect star formation and galaxy evolution. At large scales, the IGM constitutes the skeleton of the baryonic matter network. At small scales, the IGM profoundly affects the properties of the gas in the dark matter halo, and together with the CGM, it constitutes the background of galaxy evolution.

\begin{figure}[H]
\centering
\includegraphics[scale=0.32]{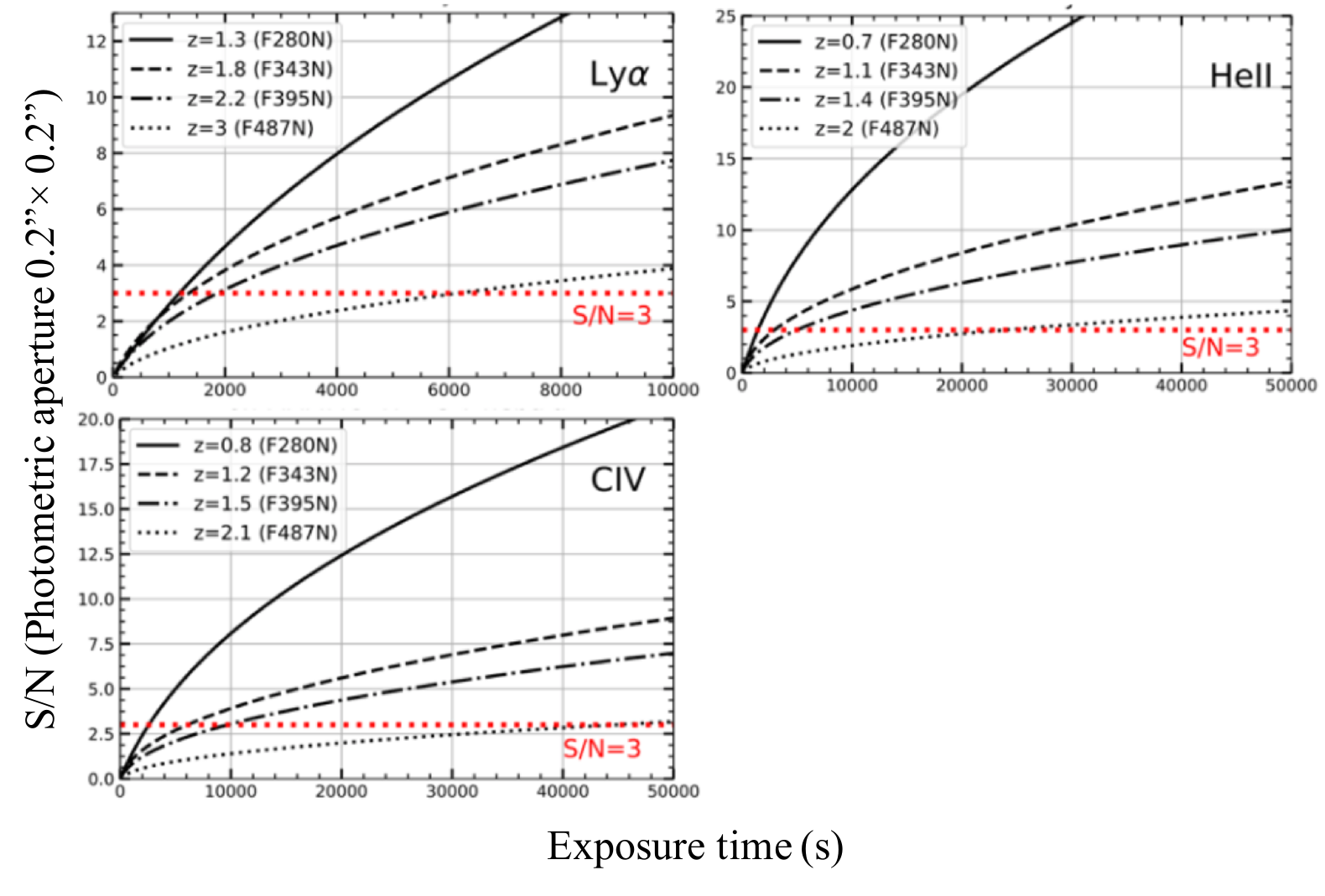}
\caption{The relationship between exposure time and S/N for CSST-MCI narrowband observations of diffuse radiation. The three panels correspond to the Ly$\alpha$, He~II, and C~IV emission lines. The red dashed lines indicate the effective detection with S/N=3.} 
\label{fig:MCI_SNR}
\end{figure}

Using the CSST main survey photometric and slitless spectroscopic samples, as well as observational data from the MCI, IFS, and TS, we will detect the CGM and IGM of nearby galaxies, and conduct narrow-band imaging and spectroscopic studies at different redshifts. Figure~\ref{fig:MCI_SNR} shows the relationship between the observation time and the expected S/N for narrow-band observations of Ly$\alpha$, He~II and C~IV emission lines at different redshifts by CSST-MCI. It can be seen that MCI can effectively detect weak diffuse radiation in a short exposure time. In addition, the slitless grating of CSST is also an important device for detecting faint diffuse line radiation. Through spectral stacking, the S/N can be significantly improved, and the ability of CSST to detect ISM/CGM/IGM radiation can be enhanced. Based on higher spectral resolution, more stable background control and better background subtraction, IFS is a more effective tool for observing faint diffuse gas radiation than narrowband and slitless gratings. It can not only image, but also display the dynamics of gas, allowing us to obtain deeper detection in the same time. The FoV of IFS allows us to perfectly explore from the fine structure of nearby galaxies to the dark matter halo of high-$z$ galaxies.

Since MCI has high spatial resolution and large FoV, it can perform a detailed study of the first link in the baryon cycle: the distribution of baryons and metal components in galaxies. By studying the history of star formation and comparing stellar population models \cite{Weisz14}, it is possible to infer the mass of all baryonic matter and metals produced by a galaxy throughout its entire star formation history \cite{Zheng20}. Compared with HST WFC/WFC3, the advantage of MCI is that its larger FoV can cover more of the galactic disk area of nearby galaxies in a short period of time, providing an excellent observational basis for the measurements of star formation history and radial distribution gradient of metals on the galactic disk. Using the MCI and IFS observations, combined with the existing HST/COS observational data, we can make detailed estimates of the baryonic composition and metallicity of the CGM/IGM of the Milky Way and nearby galaxies.

As a component of the CGM, the intracluster light (ICL) has witnessed the unique assembly history of the most massive dark matter halo in the Universe and serves as a good observational proxy for the mass and two-dimensional projection shape of the dark matter halo. The CSST main survey (and deep-field/UDF) can not only provide a large number of galaxy cluster samples for image stacking research, but also provide UV band images ($NUV$ and $u$ bands) that are sensitive to the properties of stellar populations. When combined with ground-based data, CSST can also help reduce the image confusion problem. Besides, the CSST UV observation can perform the first extended stellar disk search over large samples in space. Cooperating the CSST optical and UV observations with the highly sensitive observations of FAST can effectively resolve the origin of the cold gas in the CGM. The high spatial resolution and large FoV of CSST-SC, combined with the sensitivity of multiple narrow-band filters of CSST-MCI and the spectral and kinematic information of CSST-IFS, can promote the CGM research of low-$z$ AGN.

CSST will obtain a large number of quasar and galaxy samples, which can verify the structure of cosmic web, and statistically study the properties of IGM in the cosmic web by stacking spectra. By collaborating CSST optical observations with observations in other bands such as X-rays, we can systematically study special interstellar objects (such as planetary nebulae, X-ray binaries, etc.) and ionized gas emission regions, opening a new window for studying the physical properties and interactions of ICM. In addition, the CSST multi-band wide-field  survey is very suitable for estimating the spatial distribution and total extinction effect of intergalactic dust (IGD) through the statistical changes of the background galaxy color distribution. By utilizing the observational capabilities of the MCI and IFS, it is expected that the properties of emission line gas at the IGM/CGM interface of massive dark matter halos in the rapid formation stage ($z>2$) will be directly observed, and quantitative metallicity and kinematic measurements will be used to probe the IGM/CGM boundary region, which is currently poorly understood.

\subsection{Stars}

Stars are the most common objects and the main form of visible matter in the Universe. Stellar evolution and catastrophic explosion lead to the chemical evolution of galaxies and the enrichment of cosmic metal elements. Therefore, stellar physics is the cornerstone of modern astrophysics. The understanding of the formation, structure and evolution of stars is not only the foundation to study the Milky Way and the Universe, but also the starting point for exploring whether exoplanets are habitable.

CSST will obtain photometric data of billions of stars and hundreds of millions of stellar spectra during the ten-year survey. Its high spatial resolution ($0.074''$/pixel) and deep survey depth ($r\simeq26$ mag) will enable us to observe single stars in the Milky Way and dozens or hundreds of nearby galaxies, such as the Andromeda Galaxy and the Triangulum Galaxy. By utilizing the unique advantages of CSST in the near-IR and near-UV bands and combining with other survey projects (such as LSST, HST, $Gaia$, etc.), we can obtain multi-color photometry and proper motion of single stars in the Local Group. At the same time, slitless spectra can also provide effective temperature, metal abundance, surface gravity acceleration, low-precision radial velocity and other information for some stars ($r<23$ mag). CSST wide wavelength coverage ($255-1000$ nm) can enable observations to cover more star types and study the basic parameters of stars more comprehensively.

\subsubsection{Star formation}

Star formation is one of the most important astrophysical processes. Interstellar gas, especially molecular gas, is the basic material for the formation of stars. Therefore, it is crucial to study the evolution of interstellar gas for understanding star formation and galaxy evolution. Protoplanetary disks are produced during star formation, and planets are born in protoplanetary disks. The evolution of protoplanetary disks directly determines the properties of planetary systems.

Using CSST-TS, we can conduct THz spectrum mapping observations on different samples, including nearby molecular clouds, objects at different evolutionary stages, the central molecular belt of the Milky Way, and M31. We can obtain 492 GHz fine structure line of neutral carbon and rotational transitions of HDO in the same band. We will study gas circulation in the Milky Way and M31 by combining spectral data (such as CO, CII, HI, radio recombination lines, etc.) from other instruments, radiation transfer models, and astrochemical simulations.

Using CSST-TS broadband deep spectral surveys of different types of objects, such as star-forming regions, late-stage stars, and near-Earth comets (opportunistic), we can obtain high-sensitivity spectral data in the new window of the THz band. In combination with the molecular spectrum database and astrochemical models measured in the laboratory, we will carry out spectral line identification and analysis, determine the chemical composition and physical properties of astronomical sources, and search for new molecules.

Combined with mature accretion models, the accretion rate of young stars can be measured by using CSST UV data. The disk wind can be detected by observing the distribution of [OI]$\lambda$6300 using the CSST-IFS, and the generation mechanism of the disk wind will be  analyzed by comparison with the relevant model \cite{Ercolano10,Wang19}.

\subsubsection{Massive stars and late evolution}

Massive stars are one of the main sources of heavy elements, UV radiation fields and stellar feedback in the Universe, and they have a significant impact on the formation and evolution of galaxies. However, the evolution of massive stars is more complicated than that of low-mass stars, and many related physical processes are still under discussion. Improving the evolution model of massive stars is not only related to stellar physics, but also has important scientific significance for many fields of astrophysics, since it profoundly affects the chemical evolution of galaxies and star formation.

Based on the CSST main survey data, as well as the MCI and IFS equipment, the study of massive stars will be promoted from two aspects: pre-main sequence evolution, main sequence and late evolution. The extremely deep imaging depth, high spatial resolution and wide-band spectral coverage of the CSST will provide rich data for further development in this field.

For the pre-main sequence evolution of massive stars, young embedded clusters in the Galactic disk provide good observational samples. The angular diameter of these embedded clusters is comparable to the FoV of MCI. The high-resolution images of MCI can resolve each member star in these embedded clusters. Even in the case of severe extinction, MCI can detect massive young stars with an extinction of 20 mag within a range of 3 kpc. The wide spectral coverage of the IFS ($0.35-1.0$ $\mu$m) is conducive to the precise measurement of the spectral type, extinction and luminosity of massive young stars.
Using light variation data from multiple telescopes, such as CSST, Kepler, and the Transiting Exoplanet Survey Satellite (TESS) \cite{Ricker15}, combined with $Gaia$ and LAMOST data, we can study the internal structure and evolution of massive stars through asteroseismological methods, especially to advance our understanding of the laws of internal rotation and angular momentum transfer of stars.

Using CSST multi-color photometric data, combined with the United Kingdom Infrared Telescope (UKIRT), the Two Micron All-Sky Survey (2MASS) and other near-IR data and $Gaia$ data, red supergiants (RSG) in nearby galaxies can be identified, and the surface gravitational acceleration and luminosity of massive stars will be inferred through CSST time series data. A systematic study can be conducted on the late-stage massive star populations in nearby galaxies (within about 10 Mpc) with a metallic abundance difference of about 60 times, and the effects of stellar physical parameters (such as initial mass function, rotation, convection, binary stars, light variation, mass loss of stellar wind, etc.) and the galaxy environment on their properties can be explored.

Besides, we can compare images of the same sky region taken by CSST and HST at different times to search for ``disappearing'' stars (failed supernovae), providing direct observational constraints on massive stellar evolution models and supernova explosion theories \cite{Reynolds15}.

\subsubsection{Binary star evolution and special objects}

Current studies have shown that about half of the stars are located in binary star systems. In modern astrophysics, binary stars are involved in various fields, such as planets, stars, galaxies, cosmology, high-energy astrophysics and gravitational wave astronomy. Binary star evolution is the inevitable way to form SNe~Ia, and massive binary star evolution is the main way to produce compact binary stars, such as binary stellar black holes, binary neutron stars, black hole-neutron stars, black hole-white dwarfs, and neutron star-white dwarfs. In addition, binary star evolution also explains most of the mysteries about stars, such as the classic Algol paradox, spectra of symbiotic stars, and the formation of barium stars. Besides, binary stars are also believed to related to cosmic reionization \cite{Han20} and candidate sources of gravitational waves \cite{Brown16,Wu18,Li21}.

The study of the basic properties of binary stars is highly dependent on the establishment of binary star samples. Most of the current sky survey projects can only effectively detect a single type of binary star, so there is a natural inhomogeneity and a strong selection effect between samples. CSST has an inherent advantage in building a larger and more complete binary star sample. By utilizing the difference between the single/binary star energy spectra, CSST's high-precision multi-color photometric data and slitless spectroscopic data can efficiently detect unresolvable companion stars. 

CSST's wide wavelength range ($255-1000$ nm), especially the coverage of the UV band, can cover binaries with larger masses and mass ratios, and realize accurate measurement of the parameters of the primary and companion stars. The magnitude limit of about 26 mag enables the sample to cover at least LMC/SMC (100 kpc), and is expected to cover M31/M33 (1Mpc), and greatly enrich the binary star sample of $10 < a < 1000$ AU. Resolvable binaries can be obtained through CSST's high-resolution imaging and milliarcsecond-level astrometric information (proper motion and parallax). The design of multi-color imaging and slitless spectroscopic survey arrangement can ensure that the binary star sample has good intrinsic homogeneity. CSST can construct a binary star sample with the largest number, good consistency, and high-precision stellar physical parameters.

CSST will be a milestone in understanding the formation and evolution of special stars. Since binary stars are generally hotter and very sensitive to near-UV light, the CSST main survey can effectively identify special samples, the number of which may be several orders of magnitude higher than the currently known number. At the same time, the number of spectra obtained by CSST is also expected to be increased by more than one order of magnitude, which is very helpful for studying and solving the origin of special stars from a large sample statistical perspective. Besides, since the radiation of accreting white dwarfs is mainly concentrated in the UV to X-ray band, CSST has a unique advantage in searching accreting white dwarfs. CSST also is expected to detect more hypervelocity star samples, since the radial velocity accuracy estimated by CSST spectrum is about tens of kilometers per second \cite{Sun21}, which can meet the search requirements for hypervelocity stars..

\subsubsection{Asteroseismology and stellar activity}

At present, the internal physical laws of some widely existing active phenomena of stars (such as magnetic fields, material accretion and ejection) are still not clear. Asteroseismological observations can reconstruct the internal structure of stars, constrain microscopic physical processes, such as convection and material transport processes, and open a window to solve these problems. The habitability of exoplanets depends on the activity level of the host star, among which stellar flares and coronal mass ejections (CMEs) have a huge impact on the environment and composition of the planetary atmosphere. This is particularly important for the planetary system of M-type dwarfs. Since the habitable zone of M-type dwarfs is closer to the host star, the intense stellar activity is likely to be the ``central engine'' of the co-evolution of the host star and the planet.

CSST will conduct multi-color imaging and slitless spectroscopic 17,500 deg$^2$ wide-field surveys, covering billions of stars in the Milky Way. CSST's 400 deg$^2$ multi-color imaging deep-field observations and a small number of time domain observations will constitute time series data for many stars. In addition, CPI-C and IFS provide the possibility of observing stellar CME events. Based on these data products of CSST with deep magnitude limits, high observational accuracy and high spatial resolution, we can deeply study the magnetic field, accretion, CME and pulsation of stars, and discover new samples of active stars for detecting the diversity of various stellar activities. Using these research results, more comprehensive observational constraints can be obtained to improve the existing stellar structure and evolution models \cite{Zhang21}.

Stellar activity is manifested as anomalous absorption and emission of activity indicator spectral lines and changes in stellar luminosity, reflecting energy changes caused by magnetic fields and accretion activities. CSST near-UV and optical spectroscopic and photometric observation data contain important stellar activity information, which can include TiO absorption band, and H$\alpha$ and Mg~II h\&k spectral lines. Using these data, we can detect and characterize the magnetic field activities of stars in the photosphere and chromosphere, discover novel activity samples in extremely cool dwarfs and young stars, explore the diversity of stellar activities, and establish accurate statistical relationships between stellar activities and basic stellar parameters based on complete samples.

Besides, the CSST bulge region data are suitable for analyzing the relationship between the accretion process and magnetic field activity on young stars and related evolutionary characteristics, as well as capturing stellar white-light flare events \cite{Yang23}. IFS spectroscopic data can be used to analyze the H$\alpha$ spectral line profile to detect stellar CME events with large explosions. CPI-C can detect planetary systems with a contrast greater than $10^{-8}$, which is sufficient to ensure direct detection of stellar CME events for nearby active star samples. CSST bulge observations also will discover a large number of $\delta$ Scuti variables, which can be helpful to further improve stellar structure and evolution models.

\subsubsection{Compact objects}

Compact stars are the final objects formed in the late stage of stellar evolution, including white dwarfs, neutron stars and black holes. Due to their special physical properties, such as extremely high density and pressure, and extremely strong magnetic field, they can be regarded as natural extreme physics laboratories. Compact stars have always been an important topic in astronomical research, and are related to many fundamental physical and astrophysical processes.

In the CSST main survey, about 1-10 million high-confidence white dwarf candidates are expected to be discovered, and the distribution laws of their parameters, such as luminosity, mass and radius, will be constructed. A group of white dwarfs with special research significance will be searched and confirmed, such as white dwarfs with debris disks and planetary systems, and their formation and evolution mechanisms will be studied. Using the CSST-SC Galactic center direction and MCI time domain data, a group of pulsating white dwarfs with brightness of 19-25 mag can be discovered, thereby providing constraints on their internal structure and physical state in different cooling belts.

Taking advantage of the large FoV and deep magnitude limit of the CSST, a complete sample of accreting neutron stars can be established, especially a sample of medium-mass X-ray binaries, thereby constraining their evolution to pulsar binary systems and the radiation mechanism of accreting neutron star systems.
By combining the CSST and eROSITA survey data, and considering the optical and UV band data of Pan-STARRS, $Gaia$, GALEX, etc., we can search for the candidates of compact object through the phase space diagram of multi-band flux ratio, and analyze newly discovered black hole candidate sources. 

In addition, we can select hot subdwarfs, white dwarfs and other types of objects, construct the elliptical light curves using CSST bulge time-domain photometric data, and calculate the mass functions of binary stars. Furthermore, the mass of invisible stars will be calculated by estimating the mass of visible stars, and a batch of quiescent stellar-mass black hole candidates can be discovered.

\subsection{Exoplanets}

Exoplanets are a frontier and important field in astronomy. From the early 1990s to the early 21th century, the radial velocity method was mainly used to detect exoplanets. Since the 21st century, the transit method (such as Kepler, TESS, etc.) has become the main means of discovering exoplanets \cite{Lissauer19,Ricker15}. Combined with microlensing, direct imaging, astrometry and other methods, a number of medium- and long-period exoplanets have been discovered, improving the completeness of medium- and long-period exoplanet samples. At present, more than 5,000 exoplanets have been discovered, and a relatively rich sample of exoplanets has been accumulated. In addition, combined with the spectral detection of the atmosphere of exoplanets, the spectral observation of the atmospheric components and contents are carried out to help us understand the evolution of planetary atmospheres. Existing and future space telescopes will further expand the planetary samples, improve the completeness of detection, characterize the atmospheric characteristics of exoplanets, and help us understand the diversity of planetary systems and the universal laws of formation and evolution.

CSST has the advantages of high spatial resolution, large aperture and FoV, and its CPI-C also has direct imaging function. Using CSST to detect  exoplanet systems will help us to discover special exoplanet systems, expand exoplanet samples, and test the formation and evolution theory of planets.

\subsubsection{Direct imaging detection}

Direct imaging technology can truly ``see'' planets by directly observing planetary radiation, providing the possibility for studying characteristic signals of planetary atmospheres, and is a key technology for confirming signals of extrasolar life. So far, ground-based exoplanet imagers have discovered a total of about 100 exoplanets, most of which are more than 10 times the mass of Jupiter. Most of the planetary systems are very young ($1-200$ Myr) and have effective temperatures close to or exceeding 1000 K. These planets also have very strong radiation, which makes the imaging contrast requirements relatively low ($10^{-4}\sim10^{-6}$ for infrared band). At present, ground-based observations cannot detect mature, low-temperature ``cold'' exoplanet. 

The CPI-C module of CSST takes advantage of the space environment without atmospheric disturbance and the observational bands, combined with the off-axis structure of the CSST without central obstruction, to break through the existing ground-based high-contrast imaging capabilities. It can further enrich the number and types of exoplanets detected, and provide important observational evidence for the formation and evolution mechanism of planets, laying an important technical foundation for directly detecting ``Earth-like'' planets in the habitable zone.

CPI-C produces an ultra-high contrast imaging workspace of $10^{-8}$ through pupil modulation \cite{Ren07,Ren10,Dou10} and precise control of wave aberration \cite{Dou16}. It can obtain images of exoplanets in multiple bands from optical to near-IR, and then derive the atmospheric spectra and physical properties of these planets through spectral fitting. With an exposure time of more than 1 hour, it can achieve a sufficient S/N ($\ge5\sigma$) to observe a 25 mag planet around a 5 mag star.

The imaging contrast of the CPI-C is two orders of magnitude better than that of the ground, and it can realize the scientific observation of ``cold'' exoplanets around nearby solar-like stars (F, G, K types) for the first time. It is expected to conduct a high-contrast imaging survey of stars within 1,000 nearby stars, and search for planetary systems the size of Jupiter and Neptune within a range of less than 1 AU to 5 AU from the stars. The discovered candidates need to be confirmed by the method of proper motion analysis. If the candidate moves around a star, it will have similar annual parallax and proper motion as the star. In this way, background objects can be eliminated from the candidates.

After identifying a planet, the CSST CPI-C obtains the distribution of reflected light energy of the target planet through multi-band photometry. It can derive the cloud layer, abundance, radius, effective temperature, surface gravity, mass and other physical parameters of the target planet by fitting the reflected spectrum \cite{Dou14,Lacy19}. Using multiple observations of the CPI-C at different times, the orbit of the planet can also be fitted to determine the dynamic mass of the planet \cite{Bowler18}.
Besides, CPI-C can be used to perform direct imaging observations of protoplanetary disks, obtain scattered light images in the optical and near-IR bands, and obtain high-precision imaging of protoplanetary disks.

\subsubsection{Microlensing event detection}

The microlensing method is currently the best method for detecting ``cold'' exoplanets and rogue planets (or free-floating planets, FFPs) that are not bound by the gravity of stars \cite{Mao91,Gould92}. Since 2003, the microlensing method has discovered more than 200 exoplanets. These planets occupy relatively unique positions in parameter space, greatly improving our understanding of ``cold'' exoplanets and rogue planets  \cite{Suzuki16,Zhu21}.

Compared with ground-based microlensing surveys, the large-aperture space telescope CSST with a large FoV and high spatial resolution has a natural advantage in microlensing observations. The magnitude limit that CSST can reach is deeper, so it can detect more stars with smaller radii. More stars mean more microlensing events are expected, and stars with smaller radii imply that CSST can detect planets with smaller masses (including rogue planets).

\begin{figure}[H]
\centering
\includegraphics[scale=0.4]{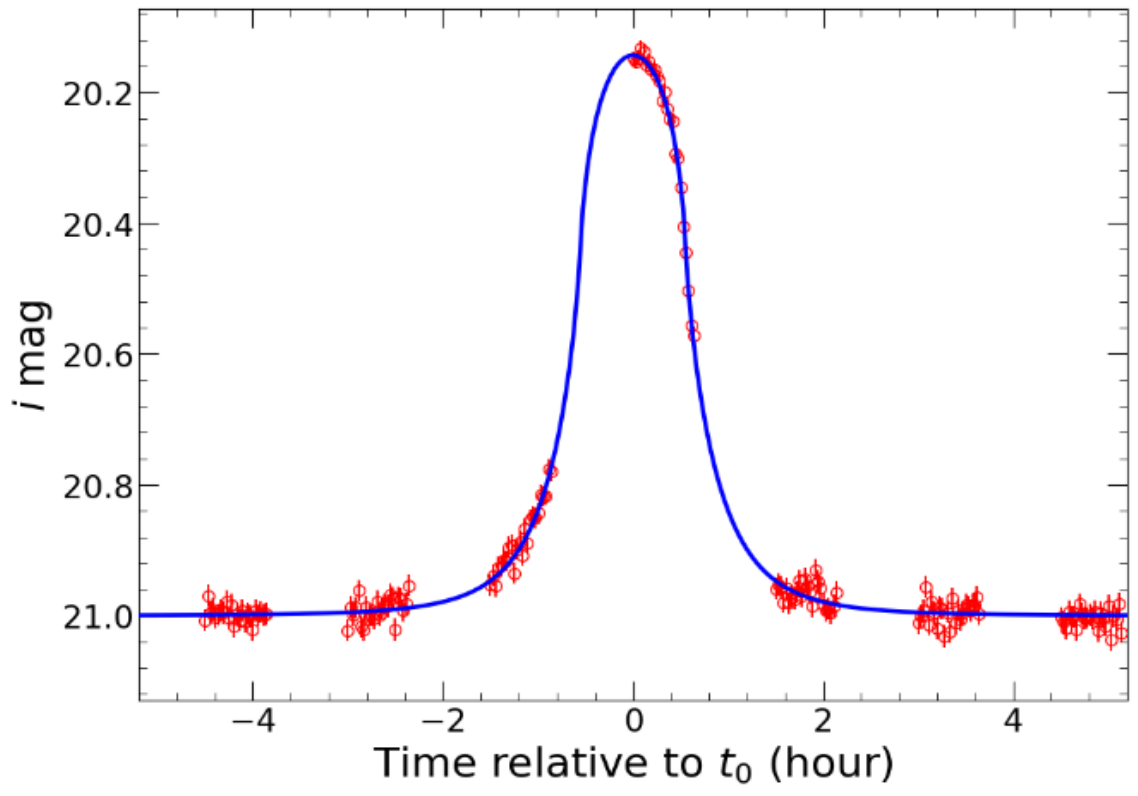}
\caption{Simulated CSST microlensing event with finite source effect caused by a rogue planet (FFP) of Mars mass.} 
\label{fig:FFP}
\end{figure}

Based on the known luminosity distribution function of bulge stars and the currently estimated rogue planet mass distribution function \cite{Gould22}, we expect that  a CSST total observation time (i.e. exposure plus readout time) of 15 days will be able to detect about 20 microlensing events caused by free-floating planets of Mars mass and smaller  (see Figure~\ref{fig:FFP}), and roughly two dozen more massive free-floating planets.

We also expect to detect about 50 signals from ``cold'' planets gravitationally bound to their host stars. These ``cold'' planets are about $1-30$ AU from their host stars and have masses as low as about Mars mass. This type of exoplanet is almost impossible to detect by other detection methods or ground-based microlensing surveys, and they can play an important role in testing the core accretion model of planet formation theory. It is expected that these extremely low-mass rogue planets and ``cold'' planets that CSST can detect will greatly expand the known planetary parameter space, and have very important scientific value for understanding the formation and evolution of planetary systems.

\subsubsection{Transiting exoplanet system detection}

As the main method for discovering exoplanets, the transit method has the characteristic of batch searching for exoplanets with high efficiency. At present, nearly 5,000 exoplanets have been confirmed by the transit method, providing a rich sample for statistical research on exoplanets and revealing various dynamic and physical characteristics of exoplanet systems. All known transiting planets are mainly in the Kepler field, as well as in the solar neighborhood \cite{Zhu21}. The basic principle of the transit method is to measure the tiny periodic photometric changes caused by the planet passing over the surface of the star through high-precision time-series photometric observations, thereby obtaining orbital characteristics, such as planet radius and orbital period, which helps us understand the properties of exoplanets.

The transit method for searching exoplanets requires long-term observation of a large number of stars in the same area of the sky. CSST has a large FoV and deep magnitude limit, which can monitor many stars simultaneously and search for transit signals. Meanwhile, CSST has a high angular resolution and can observe areas with dense star fields, such as the vicinity of the galactic bulge and star clusters. With a larger aperture than Kepler, CSST can detect transiting planets around farther and fainter bulge stars, which can reveal the planet formation mechanism in different environments throughout our galaxy. We find that with an exposure of 300 seconds, the CSST can measure stars with a photometric accuracy of 0.5\% for magnitudes below 20, and its detection depth is higher than that of current space telescopes which use the transit method to detect planets, such as the Kepler apace telescope and TESS \cite{Lissauer19,Ricker15}. With an observational time of 7 days, more than 100 transiting exoplanet candidates can be detected.

In summary, CSST has a large FoV, high resolution and high photometric accuracy, which can help search for planets in Galactic bulge, star clusters, star streams and other star-dense areas. These areas were previously difficult to search for planetary signals due to the limitations of detection methods. Therefore, CSST observations can allow us to have a deeper understanding of the characteristics of exoplanet systems in different environments in the Milky Way, and provide new clues for our understanding of the formation and evolution of planetary systems.

\subsubsection{Exoplanet atmosphere}

Transit spectroscopy, direct imaging and Doppler spectroscopy are the best methods for studying the atmospheres of exoplanets \cite{Madhusudhan19}. Although planets are too dark and too small, their size is not that different from their host stars. When a planet moves in front of a star, the star light is partially blocked, so it appears to be darker, and the degree of darkening depends on the wavelength. The ``transit method'' is continuously tracking stars, searching for the phenomenon of periodic dimming of stars and measuring the degree of dimming of planetary systems in multiple bands, which can be used to detect planets and obtain the atmospheric spectrum of planets. When a planet moves behind its host star, the radiation of the planet is completely blocked. By comparing the total radiation of the entire system at this time with that before the planet is blocked, the radiation of the planet itself or its reflection can be obtained. This is called the ``secondary eclipse method''.

The CSST-SC includes a slitless spectrometer with a resolution of about 200 and a wavelength coverage of $\sim$0.25 to 1 $\mu$m, which can be used to obtain the transmission or emission spectra of planetary atmospheres. The CSST-IFS  can provide a resolution of $\ge1000$, a wavelength coverage of $0.35-1.0$ $\mu$m transmission or reflection spectra, and can directly distinguish some wide-spaced planets. The CSST-MCI can provide three-color simultaneous high-precision light curves. These instruments have their own advantages and can be used for observation and research of exoplanet atmospheres.

Considering the CSST instruments and band coverage, cooperated with other instruments (e.g. JWST, HST, etc.), we expected to obtain transit spectra and multi-color phase curves of a total of $20-30$ hot and ultra-hot Jupiters, and perform detailed studies of their atmospheric properties. We will also conduct atmospheric studies of planets around K-type and M-type dwarfs, and use CSST to conduct follow-up optical spectroscopy or multi-color observations of exoplanets that have been observed by HST or JWST near-IR transmission spectroscopy. Besides, we will probe about 10 hydrogen ocean planets, super-Earths and warm planets, and use the IFS to perform direct spectroscopic detections of a wide range of young giant planets. Moreover, the CSST-SC spectrometer and MCI should enable effective detection of the atmospheres of most hot Jupiters, some super-Earths and wide-spaced young planets.

\subsection{Solar System objects}

In addition to the large objects such as the Sun, eight planets, dwarf planets and their satellites, the Solar System also has a large number of small objects, including asteroids and comets. Asteroids are divided into main-belt asteroids, Trojan asteroids, Kuiper belt objects, Centaurs and near-Earth asteroids according to their distribution locations. Comets can be divided into short-period comets, long-period comets, main-belt comets and interstellar comets according to their origins. At present, the number of small objects discovered in the Solar System exceeds 1.31 million, but there are still a large number of small objects that have not been discovered. Small objects are the remnants from the formation of the Solar System, and have well preserved the original information. Studying small objects is helpful for studying the distribution of water in the Solar System and the origin of water and life on Earth, and is helpful for studying the formation and evolution of the Solar System.

\subsubsection{Kuiper belt objects}

Kuiper Belt Objects (KBOs) are located beyond the orbit of Neptune and contain important clues of the formation and evolution of the Solar System. Research on these objects can provide the most important evidence for understanding the evolution of planetary orbits, the stability of the Solar System, and the space environment.

The number of KBOs discovered so far (about 2,000) is far less than the number of main-belt asteroids discovered (more than 700,000) \cite{Barucci08,Batygin19}. In addition, their orbit determination accuracy is low, and the data of physical and chemical properties on their surface are scarce, which greatly limits the understanding of the origin and motion properties of KBOs. The high magnitude limit and multi-band photometric capabilities of CSST provide an excellent opportunity to observe KBOs. In addition, CSST observations  with high orbital inclinations will help reveal the authenticity of the ninth planet.

Under the limitation of the CSST main survey, it is still expected to obtain orbital data of hundreds of KBOs, greatly increasing the number of KBO samples. CSST can obtain multi-color photometric data of hundreds of KBOs, reveal the essence of the correlation between KBO colors and their orbital characteristics, and understand the formation and orbital evolution of objects in the outer Solar System.

CSST can perform surveys in the high ecliptic latitude region (or a narrow band region) to search for small objects with orbital inclinations greater than 30 degrees. It is expected that dozens of high-inclination objects will be discovered during the CSST main survey, and a more detailed distribution and multi-color photometric data of such objects will be obtained to reveal the relationship between their color index and orbit.

\subsubsection{Active asteroids / main-belt comets}

Some asteroids will have comet-like activities, which are called active asteroids \cite{Hsieh06,Jewitt11,Hsieh12,Drahus15,Jewitt15}. Active asteroids can be divided into main-belt comets and broken asteroids according to the different driving mechanisms of their activity. At present, there are 42 active asteroids/main-belt comets discovered, of which only 15 are main-belt comets whose activity is driven by water ice sublimation. Studying active asteroids/main-belt comets is helpful for studying the distribution and activity mechanism of active asteroids/main-belt comets, the source of water on Earth, the thermal history and composition of the Solar System, and provides some constraints on the protosolar disk model.

Active asteroids/main-belt comets are difficult for ground-based equipment to identify their activity characteristics. CSST has the strong optical power and a large FoV, making it very suitable for performing search and observation on main-belt comets. CSST will study the physical properties and activity of active asteroids/main-belt comets, reveal the activity mechanisms, compare the activity differences between them, and analyze the causes and mechanisms of the differences. We will explore the orbital characteristics and spatial distribution of active asteroids/main-belt comets, develop a CSST-based active asteroid/main-belt comet search and data processing program, and discover new active asteroids/main-belt comets.

\subsubsection{Physical properties and activity of comets}

Comets are the remnants left over from the formation of the Solar System, which are rich in water ice and organic matter. Studying comets is helpful for exploring the origin of the Solar System and the source of water and life on Earth. The activity mechanism of comets is different at different heliocentric distances. Within 5 AU, its activity can be explained by the standard model, that is, the volatilization drive of water ice, but outside 5 AU, its activity is mainly caused by the volatilization of volatile gases.

Comets have a wide variety of morphological features during their orbits. CSST has a large FoV and high resolution, making it very suitable for studying the morphology and activity of comets. We also can use the CSST main survey to carry out observations of comets, and build a database of comet physical parameters by combining other survey data. Besides, The short-time scale ($5-30$ s) exposure full-band observation/monitoring mode of the CSST-MCI can measure the light variation of comets with high spatial resolution at the minute level, which can be used to study the rotation period and activity evolution of comets. The start and end positions of comet activity are one of the important parameters for comparing comet activity. Since the imaging depth of the XDF observation of CSST-MCI can reach 30 magnitude in optical band, CSST-MCI can greatly expand the heliocentric distance at which comets are active.

\subsubsection{Identification and data processing of asteroids}

At present, the number of asteroid samples with known shape, rotation and other parameters accounts for a very low fraction of the total number of asteroids. Such data volume is far from meeting the needs of theoretical research on the origin and evolution of asteroids. Compared with $Gaia$ and $Euclid$, CSST has a larger FoV, is equipped with more observation modules, and has stronger observation capabilities. Therefore, CSST can observe more and fainter small Solar System objects. Using the photometric and spectroscopic data of asteroids obtained by the CSST main survey, it is expected to make a significant  contribution in the research of asteroid physical parameters, spectral type classification, origin and evolution.

The CSST small object observational data are mainly obtained by utilizing the most advanced machine deep learning technology to realize the identification and measurement of the small object signals. CSST will provide multi-color photometric and spectroscopic data of a large number of faint asteroids. In order to realize the inversion analysis of asteroid physical parameters, we can combine CSST data with other space and ground observational data to establish a new asteroid photometric model and inversion method.

\subsubsection{Discovery and physical properties of natural satellites}

Most of the natural satellites of the large planets are irregular satellites, which are usually very dim ($>20$ mag) and difficult to observe using ground-based telescopes \cite{Porco05,Cooper08,Weaver06,Sheppard03,Sheppard10,Sheppard19}. The study of these faint satellites is of great significance to the origin and evolution of the Solar System, and also plays a key role in the navigation of deep space exploration. 

The large FoV, high resolution and  sensitivity of the CSST-SC make it possible to discover new natural satellites. By rationally utilizing and planning the observational mode of CSST, it is expected to increase the observational data of faint natural satellites, greatly promoting the reidentification, new discovery and dynamics research of faint satellites.

\subsection{Astrometry}

CSST, with its wide wavelength coverage and high angular resolution, is expected to provide a strong supplement to space- and ground-based optical observations. CSST is designed to observe celestial objects fainter than 18 mag, potentially reaching depths up to 26 mag. It is expected to extend $Gaia$'s tomographic mapping of the Milky Way to much fainter limits, providing milliarcsecond-level astrometric accuracy. By enabling precise measurements of stellar proper motions and parallaxes, CSST is expected to complement and extend $Gaia$'s results into deeper and darker regions of the sky, opening new possibilities for high-precision astrometric research \cite{Liao24,Fu23}. This extension will enrich the key data of the astronomical database, including parameters such as position, parallax and proper motion. 

Leveraging its high-precision astrometric measurements, the CSST is anticipated to extend the $Gaia$ celestial reference frame to substantially fainter magnitudes, thereby establishing the CSST celestial reference frame as the most densely populated and deepest in magnitude limit among existing celestial reference frames. Meanwhile, given CSST's observational cadence and astrometric precision, its data are expected to serve as a valuable complement to the $Gaia$ catalog for sources in the 18-21 magnitude range, thereby contributing significantly to the maintenance and refinement of their astrometric parameters. Considering its astrometric accuracy, CSST can provide key information, such as motion and distance, required for the discovery and mass measurement of binary star systems. In addition, CSST is also expected to provide effective observational data for orbital improvement and classification of small Solar System objects, detection of gravitational deflection effects of stars with large proper motion, and search for special objects, such as intermediate-mass black holes.

Therefore, CSST's milliarcsecond-level astrometric data will be used to extend from the establishment of celestial reference frames to astrometric binary star detection, Solar System asteroid dynamics, etc. With its excellent astrometric accuracy, CSST will make great contributions to astrometry, especially in the construction of reference frames and related astrometric scientific application research. In addition, CSST will provide high-precision astrometric data for other research fields in astronomy and enhance our understanding of the Universe.

\subsubsection{Construction of extremely deep celestial reference frame}

The celestial reference frame is an essential space reference in the fields of astronomy, geodesy, deep space exploration, navigation, and geodynamics research. The establishment and unification of high-precision space references have important scientific significance. At present, the International Celestial Reference Frame (ICRF) consists of more than 4,000 extragalactic radio sources. In 2022, the $Gaia$-CRF optical celestial reference frame was built, which consists of 1.6 million quasar sources. $Gaia$ successfully built a high-precision celestial reference frame in the optical band with a high density (about 42 stars/deg$^2$). The CSST survey depth exceeds 25 mag, and it is expected that the reference sources observed per square degree will reach the order of 1,000, which is expected to build the world's densest and deepest celestial reference frame.

To this end, it is essential to evaluate the completeness and reliability of existing quasar identification methods, and to develop reference source selection techniques, specifically adapted to the observational characteristics of CSST data. To enhance the overall stability of the CSST ultra-deep celestial reference frame, the quasar sample should be systematically reviewed, flagged, and rigorously classified based on their suitability for inclusion. Anomalous quasars must be accurately identified and excluded to ensure the integrity of the reference frame \cite{Wu22,Wu23,Wu24}. The astrometric performance, including the alignment between the CSST celestial reference frame and the $Gaia$-CRF, will be systematically investigated. Thanks to its broad wavelength coverage and high precision measurements, CSST is well-positioned to make significant contributions to the development of a multiband celestial reference frame \cite{Yao24}.

\subsubsection{Maintaining and updating the $Gaia$ catalog}

The $Gaia$ mission delivers star catalog with unprecedented astrometric precision $-$ approximately 1 milliarcsecond for objects in the 18 to 21 magnitude range \cite{Prusti16}. However, the accuracy of these catalogs, particularly in terms of stellar position, deteriorates over time due to the increasing uncertainty in astrometric parameters. Incorporating new observations is therefore essential to maintain, or even enhance, the initial level of precision \cite{Gai22}.

The CSST observations cover $Gaia$ targets of $18-21$ mag, and the combination of common target astrometric observational data between $Gaia$ and CSST can improve the accuracy of the proper motion and parallax solution results of the targets. We will optimally combine $Gaia$ astrometric epoch data with CSST astrometric epoch data, and solve the relevant astrometric parameters in the same set of equations.

\subsubsection{Astrometry discovery and mass measurement of binary star systems}

The mass distribution and eccentricity of binary star systems play a vital role in understanding the formation and evolution of stars. In particular, compact objects with companion stars, including black holes, neutron stars, white dwarfs, etc., are the most important objects in stellar evolution, which are laboratories with extreme physical conditions and play a pivotal role in astronomical research. By using astronomical measurement methods and detecting the position changes of a large number of objects, compact binaries can be efficiently discovered. Accurately tracking the sky position of stars can accurately measure the orbits of binary stars, thereby solving the mass and providing a large sample mass distribution.

We will first use CSST astrometric data and multiple indicators to construct a sample of astrometric binary candidates. Secondly, we will combine $Gaia$ and CSST astrometric position sequence data to confirm binary systems. Finally, we will integrate multi-dimensional observational data to reconstruct the three-dimensional motion of the visible companion star, and thus directly measure the mass. It is expected that a certain number of reliable astrometric binary candidates will be discovered, and a certain scale of binary mass samples will be constructed.

\subsubsection{Asteroid mass determination based on CSST}

Density is one of the most basic physical parameters of asteroids, and it is of great significance for revealing the composition and evolution of asteroids. Currently, there are relatively few asteroids with known density ($\gtrsim400$), which is mainly because there are relatively few asteroids with known mass. Accurate mass parameters are a necessary condition for accurate density measurement. Accurate asteroid mass can improve the modeling accuracy of the Solar System's gravitational field and thus improve the accuracy of the perturbation of the target. In addition, the accuracy of the current Mars ephemeris is mainly affected by the accuracy of the perturbation modeling of the main belt asteroids. Continuous improvement of asteroid mass will help improve the accuracy of the Mars ephemeris.

CSST observational data can be integrated with existing ground-based observational data and $Gaia$ asteroid observational data to improve the accuracy of asteroid mass measurement, and can improve the mass measurement of large-mass asteroids.

\subsubsection{Measuring the mass of stars with large proper motion based on gravitational deflection effect}

Mass is one of the most important physical parameters of stars and is also a decisive factor in stellar structure and evolution. The range of stellar mass can range from hundreds of times to several times the mass of the Sun. Most stars have a mass from 0.1 to 10 $M_{\odot}$, while large stars in the spiral arms of the Milky Way have a mass between 6 and 60 times the mass of the Sun. How to accurately measure the mass of stars is a fundamental problem in astronomy. For most isolated stars, the method of gravitational microlensing can accurately measure the mass of stars. The change in gravitational deflection caused by the motion of stars with large proper motion is an effective tool for directly determining their mass. The mass of stars with large proper motion can be calculated by accurately measuring the position change of background sources \cite{Sahu17,Kains17,Kluter20}.

Considering the survey depth of CSST, the density of objects in the FoV is very high. During the motion of stars with large proper motion, they will pass by a considerable number of background sources. So such opportunities will increase significantly, which will effectively enhance the probability of determining the mass of stars based on the change in gravitational deflection.

\subsubsection{Astrometric search for intermediate-mass black holes}

Current astronomical observations show that black holes are ubiquitous in the Universe. However, there is still a fundamental and long-lasting question about black holes that needs to be solved: Is the mass of black holes distributed continuously? More specifically, there is a huge gap between stellar mass black holes and SMBHs in the mass map of black holes. Do intermediate-mass black holes with masses between 100 and 100,000 $M_{\odot}$ exist? 

Globular clusters have dense environments for stars to merge and form intermediate-mass black holes. We can use CSST to search for intermediate-mass black holes in globular clusters, and try to give the clearest and most specific detection results.

\subsection{Transients and variable sources}

\subsubsection{Supernovae}

Supernovae are violent explosions caused by the core collapse of massive stars or the thermonuclear explosion of white dwarfs. The former include Type~II, Type~Ib and Type~Ic supernovae and various subtypes (i.e. core collapse supernovae, CCSNe), while the latter are mainly SNe~Ia. SNe~Ia are important cosmic distance probes, but the nature of their progenitors is still unclear \cite{Wang12}. In recent years, large-field surveys have led to the discovery of superluminous supernovae (SLSNe) with brightnesses of about 100 times that of ordinary CCSNe. The specific explosion physics and energy mechanism remain a mystery \cite{Gal-Yam12}. Exploring the progenitors and energy mechanisms of various supernovae and special transients is the core issue of current supernova research, and is also the key to understanding stellar evolution, supernova explosion mechanisms, and the formation and evolution of compact objects.

Although CSST main scientific goal is about cosmological and galaxy observation, the CSST-SC has the characteristics of high spatial resolution and large FoV, which is expected to make important progress in the detection of various types of transient sources in the Universe, including supernovae, during its ten-year scientific operation. The deep-field, UDF surveys based on the CSST-SC and the MCI-based XDF survey will be able to detect a large number of high-$z$ supernova samples (including SNe Ia and SLSNe) and several supernova gravitational lensing phenomena. These data are of great significance for further constraining the nature of dark energy and studying the formation and evolution history of stars in the early Universe.

\subsubsection{Gamma-ray bursts}

Gamma-ray bursts can generally be divided into two categories: long GRBs (duration time scale greater than 2 seconds) and short GRBs (duration time scale less than 2 seconds) \cite{Bargiacchi25}. The former originates from the core collapse of massive stars, while the latter is produced by the merger of binary compact stars. As the most violent stellar explosion in the Universe, the instantaneous radiation of GRBs is extremely bright. At the same time, gamma-rays have extremely strong penetrability in the ISM. These factors make GRBs detectable even at extremely long distances.

Under the guidance and coordination of high-energy satellites, CSST is expected to make deep observations of the optical counterparts, optical afterglows and host galaxies of high-$z$ GRBs, thus helping to determine their properties, such as redshift and metallicity. CSST-MCI observations can accurately characterize the evolution of the GRB afterglow energy spectrum, constrain the morphology and evolution of the electron energy spectrum, and reveal the universal law of relativistic shock wave acceleration of particles. At the same time, the detection depth of CSST can monitor the long-term evolution of GRB optical radiation from the early, middle and late stages, which plays a key role in revealing the possible evolution of the microscopic physical parameters of the shocked fluid (such as particle energy and magnetic field energy ratio). In addition, CSST can perform ToO observations (a small amount of total observational time) to follow up on the electromagnetic counterparts of gravitational waves, kilonovae of short GRBs, and high-$z$ GRBs.

\subsubsection{Fast radio bursts}

Fast radio bursts are the most energetic cosmic explosions in radio bands \cite{Lorimer07}, typically releasing an immense amount of energy in milliseconds, which can be the same as what the Sun produces in days or even months. To date, more than 860 cases have been officially reported, of which about 8\% have shown repetitive bursts \cite{Spitler16,Xu23} and a few are extremely active with hundreds to tens of thousands of bursts from a single FRB \cite{Li21bb,Zhang22}. In more than 40 cases, the host galaxies and redshifts are determined with arcsecond precision, confirming the cosmological origin of FRBs \cite{Chatterjee17}. However, their astrophysical origin remains a mystery and will be an important topic in frontier research in the next 5-10 years \cite{Zhang23}.

We can take advantage of CSST's large FoV, high sensitivity, and large sky coverage to focus on observational studies of the optical counterparts and host galaxies of FRBs. We expect to make breakthrough progress in key scientific issues, such as the origin and explosion mechanism of FRBs, and make important contributions for revealing the physical nature of such phenomena.

\subsubsection{High-energy neutrino events}

Neutrinos are one of the smallest units that make up the material world. In 2013, the IceCube Observatory detected high-energy neutrinos with energies above TeV from outside the Earth for the first time \cite{IceCube13}, marking the beginning of high-energy neutrino astronomy. These neutrinos may originate from high-energy cosmic ray particles, and the origin of cosmic rays is a major unresolved issue in the scientific community. Finding the origin of high-energy neutrinos and cosmic rays is an important topic in the intersection of astronomy and physics today \cite{Halzen17}.

Due to its large FoV and high sensitivity, CSST is very suitable for deep observations of the objects, such as blazars, supernovae, low-luminosity GRBs, TDEs, and starburst galaxies, to search for the origin of high-energy neutrinos.

\subsubsection{Tidal disruption events}

When a star accidentally moves to the vicinity of the SMBH at the center of a galaxy, and the distance is less than the tidal radius, it will be tidally disrupted. Some of its matter will be accreted by the black hole, producing bright electromagnetic flashes. The radiation peak is in the soft X-ray to UV band, and the duration is several months to several years. This phenomenon is called a black hole tidal disruption event. For a single galaxy, a TDE occurs only once every tens of thousands of years on average, and is an extremely rare transient source. TDE provides a powerful method to detect the majority of silent SMBHs in the Universe, including intermediate-mass black holes and even binary black holes. TDE also provides a new way to study the environment around quiet black holes \cite{WangT12,Jiang16}, thereby helping to understand nuclear activity. Due to the diversity of TDE observational manifestations and the relatively small and obviously biased observation samples at present, many key physics of TDEs are still not well understood, which seriously limits the scientific application of TDEs.

In the CSST era, various ground-based large-field surveys (e.g. the Wide Field Survey Telescope (WFST) \cite{Wang23bb} and LSST) and X-ray space satellites (e.g. Einstein Probe (EP) X-ray space telescope \cite{Yuan25}) have been routinely operated, and the TDE discovery rate will increase by orders of magnitude. CSST has its unique advantages as a space UV and optical telescope, and will discover TDEs in the deep-field survey, follow up important TDEs warned by other surveys, and study the host galaxies of TDEs.

\section{Summary}\label{sec:7}

CSST is China's flagship space optical astronomical project for the next decade. CSST includes five scientific modules: Multi-band Imaging and Slitless Spectroscopy Survey Camera, Multi-Channel Imager, Integral Field Spectrograph, Cool Planet Imaging Coronagraph, and THz Spectrometer. It will conduct in-depth research in many scientific fields and directions, such as cosmology, galaxies and active galactic nuclei, the Milky Way and nearby galaxies, stellar science, Solar System objects, exoplanets, astrometry, transients and variable sources. It is believed that with the joint participation and efforts of scientists in China and around the world, CSST will produce a large number of excellent results, and will play an important role in promoting the development of astronomy and understanding of the Universe.

\Acknowledgements{We acknowledge the support of National Key R\&D Program of China grant Nos. 2022YFF0503404 and 2020SKA0110402, the CAS Project for Young Scientists in Basic Research (No. YSBR-092), and NSFC-12473002. This work is also supported by science research grants from the China Manned Space Project with grant nos. CMS-CSST-2025-A02, CMS-CSST- 2021-B01, CMS-CSST-2021-A01, and CMS-CSST-2021-A03, CMS-CSST- 2021-A12, CMS-CSST-2021-B10.}

\InterestConflict{The authors declare that they have no conflict of interest.}



\end{multicols}

\begin{thebibliography}{99}

\bibitem {Cole05}  S. Cole, W. J. Percival, J. A. Peacock, et al., MNRAS, 362, 505 (2005).
\bibitem {Jones09} D. H. Jones, M. A. Read, W. Saunders, et al., MNRAS, 399, 683 (2009).
\bibitem {Eisenstein05} D. J. Eisenstein, I. Zehavi, D. W. Hogg, et al., ApJ, 633, 560, (2005).
\bibitem {Parkinson12} D. Parkinson, S. Riemer-SÞrensen, C. Blake, et al., Phys. Rev. D, 86, 103518 (2012).
\bibitem {Anderson12} L. Anderson, E. Aubourg, S. Bailey, et al., MNRAS, 427, 3435-3467 (2012).
\bibitem {Gil-Marin20} Héctor Gil-Marín, Julián E Bautista, Romain Paviot, et al., MNRAS, 498, 2492-2531 (2020).
\bibitem {DESI Collaboration16} DESI Collaboration: A. Aghamousa, J. Aguilar, S. Ahlen, et al., arXiv:1611.00036 (2016).
\bibitem {DESI Collaboration24} DESI Collaboration, A. G. Adame, J. Aguilar, S. Ahlen, et al., ApJ, 168,2 (2024).
\bibitem {Euclid24} Euclid Collaboration, Y. Mellier, Abdurro'uf, et al., arXiv:2405.13491 (2024).
\bibitem {Green12} J. Green, P. Schechter, C. Baltay, et al., arXiv:1208.4012 (2012).
\bibitem {LSST19} Z. Ivezic, S. M. Kahn, J. A. Tyson et al., ApJ, 873, 111 (2019).
\bibitem {Zhan11}  H. Zhan, Sci. Sin. Phys. Mech. Astron., 41, 1441 (2011).
\bibitem {Zhan21} H. Zhan, Chin. Sci. Bull., 66, 1290 (2021).
\bibitem {Cao18} Y. Cao, Y. Gong, X.-M. Meng, et al., MNRAS, 480, 2178-2190 (2018).
\bibitem {Gong19} Y. Gong, X. Liu, Y. Cao, et al., Astrophys. J., 883, 203 (2019).
\bibitem {Gong25} Y. Gong, H. Miao, X. Zhou, et al., Sci. China Phys. Mech. Astron., 68, 280402 (2025).
\bibitem {Zheng25} Z.-Y. Zheng, C. Xu, X. Liu, et al., arXiv: 2509.14691 (2025).
\bibitem {Cao22} Y. Cao, Y. Gong, Z.-Y. Zheng, and C. Xu, RAA, 22, 025019 (2022).
\bibitem {Li22} C. Li, Z. Zheng, X. Li, et al., RAA, 22, 095004 (2022).
\bibitem {Ratra88} B. Ratra \& P. J. E. Peebles, PhRvD, 37, 3406 (1988).
\bibitem {Caldwell02} R. R. Caldwell, Phys. Lett. B, 545, 23 (2022).
\bibitem {Feng05} B. Feng, X. Wang, X. Zhang, Phys. Lett. B, 607, 35 (2005).
\bibitem {Chevallier01} M. Chevallier \& D. Polarski, Int. J. Mod. Phys. D 10, 213 (2001).
\bibitem {Linder03} E. V. Linder, Phys. Rev. Lett. 90, 091301 (2003).
\bibitem {Brout22} D. Brout, D. Scolnic, B. Popovic, et al., Astrophys. J., 938, 110 (2022).
\bibitem {Rubin25} D. Rubin, G. Aldering, M. Betoule, et al., Astrophys. J., 986, 231 (2025).
\bibitem {Abbott25} T. M. C. Abbott, M. Acevedo, M. Aguena, et al., Astrophys. J. Lett., 973, L14 (2024).
\bibitem {Alam21} S. Alam, M. Aubert, S. Avila, et al., Phys. Rev. D, 103, 083533 (2021).
\bibitem {Abbott25b} T. M. C. Abbott, M. Acevedo, M. Adamow, et al., arXiv:2503.06712 (2025).
\bibitem {Burger24} P. A. Burger, L. Porth, S. Heydenreich, et al., A\&A, 683, A103 (2024).
\bibitem {Gomes25} R. C. H. Gomes, S. Sugiyama, B. Jain, et al., arXiv:2508.14018 (2025).
\bibitem {Aghanim18} N. Aghanim, Y. Akrami, M. Ashdown, et al., A\&A, 641, A6 (2020).
\bibitem {Calabrese25} E. Calabrese, J. C. Hill, H. T. Jense, arXiv:2503.14454 (2025).
\bibitem {Camphuis25} E. Camphuis, W. Quan, L. Balkenhol, et al., arXiv:2506.20707 (2025).
\bibitem {Karim25} M. Abdul-Karim, J. Aguilar, S. Ahlen, et al., arXiv:2503.14738 (2025).
\bibitem {Lodha25} K. Lodha, R. Calderon, W. L. Matthewson, et al., arXiv:2503.14743 (2025).
\bibitem {Gu25} G. Gu, X. Wang, Y. Wang, et al., arXiv:2504.06118 (2025).
\bibitem {Yao24a} J. Yao, H. Shan, R. Li, et al., MNRAS, 527, 5206-5218 (2024).
\bibitem {Zhou21} X. Zhou, Y. Gong, X.-M. Meng, et al., Astrophys. J., 909, 53 (2021).
\bibitem {Zhou22a} X. Zhou, Y. Gong, X.-M. Meng, et al., MNRAS, 512, 4593-4603 (2022).
\bibitem {Zhou22b} X. Zhou, Y. Gong, X.-M. Meng, et al., RAA, 22, 115017 (2022).
\bibitem {Lu24} J. Lu, Z. Luo, Z. Chen, et al., MNRAS, 527, 12140-12153 (2024).
\bibitem {Luo24a} Z. Luo, Z. Tang, Z. Chen, et al., MNRAS, 531, 3539-3550 (2024).
\bibitem {Luo24b} Z. Luo, Y. Li, J. Lu, et al., MNRAS, 535, 1844-1855 (2024).
\bibitem {Liu23} D. Z. Liu, X. M. Meng, X. Z. Er, et al., A\&A, 669, A128 (2023).
\bibitem {Han25} J. Han, M. Li, W. Jiang, et al., arXiv: 2503.21368 (2025).
\bibitem {Xiong24} Q. Xiong, Y. Gong, X. Zhou, accepted by ApJ, arXiv:2410.19388 (2024).
\bibitem {Miao23} H. Miao, Y. Gong, X. Chen, et al., MNRAS, 519, 1132 (2023).
\bibitem {Lin22} H. Lin, Y. Gong, X. Chen, et al., MNRAS, 515, 5743, (2022).
\bibitem {Lin24} H. Lin, F. Deng, Y. Gong, \& X. Chen,  MNRAS, 529, 1542 (2024).
\bibitem {Fan10} Z. Fan, H. Shan, J. Liu, ApJ, 719, 1408 (2010).
\bibitem {Liu15} X. Liu, C. Pan, R. Li, et al., MNRAS, 450, 2888 (2015).
\bibitem {Shan18} H. Shan, X. Liu, H. Hildebrandt, et al., MNRAS, 474, 1116 (2018).
\bibitem {Alcock79} C. Alcock, and B. Paczynski, Nature, 281, 358 (1979).
\bibitem {Kaiser87} N. Kaiser, MNRAS, 227, 1-21 (1987).
\bibitem {Zhou24} X. Zhou, Y. Gong, X. Zhang, et al., Astrophys. J., 977, 69 (2024).
\bibitem {Zhangy23} Y. Zhang, M. Chen, Z. Wen, and W. Fang, RAA, 23, 045011 (2023).
\bibitem {Song24a} Y. Song, Q. Xiong, Y. Gong, et al., MNRAS, 532, 1049-1058 (2024).
\bibitem {Song24b} Y. Song, Q. Xiong, Y. Gong, et al., MNRAS, 534, 128-134 (2024).
\bibitem {Song24c} Y. Song, Q. Xiong, Y. Gong, et al., Astrophys. J., 976, 244 (2024).
\bibitem {Song25} Y. Song, Y. Gong, Q. Xiong, et al., MNRAS, 538, 114-120 (2025).
\bibitem {Li23} S.-Y. Li, Y.-L. Li, T. Zhang, et al., Sci. China Phys. Mech. Astron., 66, 229511 (2023).
\bibitem {Wang24} M. Wang, Y. Gong, F. Deng, H. Miao, X. Chen, and H. Zhan, MNRAS, 530, 4288-4299 (2024).
\bibitem {Liu24} C. Liu, Y. Xu, X. Meng, et al., Sci. China Phys. Mech. Astron., 67, 119512 (2024).
\bibitem {Miao24} H. Miao, Y. Gong, X. Chen, Z. Huang, X.-D. Li, and H. Zhan, MNRAS, 531, 3991-4005 (2024).
\bibitem {Shi25} F. Shi, J. Tian, Z. Ding, et al., Sci. China Phys. Mech. Astron., 68, 249511 (2025).
\bibitem {SongR24} R. Song, K. C. Chan, H. Xu, \& W. Zheng, MNRAS, 530, 881-893 (2024).
\bibitem {Ding24} Z. Ding, Y. Yu, P. Zhang, MNRAS, 527, 3728-3740 (2024).
\bibitem {Aghamousa16} A. Aghamousa, J. Aguilar, S. Ahlen, et al., arXiv:1611.00036 (2016).
\bibitem {Cao24} X. Cao, R. Li, N. Li, et al., MNRAS, 533, 1960-1975 (2024).
\bibitem {Futamase01} T. Futamase, S. Yoshida, Progress of Theoretical Physics, 105, 887 (2001).
\bibitem {Treu10} T. Treu, ARA\&A, 48, 87 (2010).
\bibitem {Treu16} T. Treu, \& P. Marshall, A\&AR, 24, 11 (2016).
\bibitem {CaoX24} X. Cao, R. Li., N. Li, et al., MNRAS, 533, 1960 (2024).
\bibitem {Collett14} T. E. Collett, M. W. Auger, MNRAS, 443, 969 (2014).
\bibitem {Peccei77} R. Peccei, \& H. Quinn, PhRvL, 38, 1440 (1977).
\bibitem {Weinberg78} S. Weinberg, PhRvL, 40, 223 (1978).
\bibitem {Wilczek78} F. Wilczek, PhRvL, 40, 279 (1978).
\bibitem {Spergel00} D. N. Spergel, \& P. J. Steinhardt, Phys. Rev. Lett. 84, 3760 (2000).
\bibitem {Harvey19} D. Harvey, A. Robertson, R. Massey, \& I. G. McCarthy, MNRAS, 488, 1572–1579 (2019).
\bibitem {Adhikari22} S. Adhikari, A. Banerjee, K. K. Boddy, et al., arXiv:2207.10638 (2022).
\bibitem {Kamionkowski99} M. Kamionkowski, A. Kosowsky, Annu. Rev. Nucl. Part. Sci. 49,77 (1999).
\bibitem {Balantekin13} A. B. G. Balantekin \& M. Fuller, Prog. Part. Nucl. Phys. 71, 162 (2013).
\bibitem {Fukuda98} Y. Fukuda, T. Hayakawa, E. Ichihara, et al. Super-Kamiokande Collaboration, PhRvL, 81, 1562 (1998).
\bibitem {Ashie05} Y. Ashie, J. Hosaka, K. Ishihara, et al. Super-Kamiokande Collaboration, Phys. Rev. D 71, 112005 (2005).
\bibitem {Araki05} T. Araki, K. Eguchi, S. Enomoto, et al. KamLAND Collaboration, Phys. Rev. Lett. 94, 081801 (2005).
\bibitem {Abe08} S. Abe, T. Ebihara, S. Enomoto, et al. KamLAND Collaboration, Phys. Rev. Lett. 100, 221803 (2008).
\bibitem {Porredon22} A. Porredon, M. Crocce, J. Elvin-Poole, et al., Phys. Rev. D, 106, 103530 (2022).
\bibitem {Quintero25} C. Garcia-Quintero, H. E. Noriega, A. de Mattia, et al., arXiv:2504.18464 (2025).
\bibitem {Chebat25} D. Chebat, C. Yèche, E. Armengaud, et al., arXiv:2507.12401 (2025).
\bibitem {Cai16} Y.-F. Cai, S. Capozziello, M. De Laurentis, et al., Reports on Progress in Physics, 79, 106901 (2016).
\bibitem {Yan24} J.-H. Yan, Y. Gong, M. Wang, H. Miao, \& X. Chen, RAA, 24, 115013 (2024).
\bibitem {Chen22} A. Chen, Y. Gong, F. Wu, Y. Wang, \& X. Chen, RAA, 22, 055021 (2022).
\bibitem {Zhang07} P. Zhang, M. Liguori, R. Bean, \& S. Dodelson, PRL, 99, 141302 (2007).
\bibitem {Lombriser12} L. Lombriser, F. Schmidt, T. Baldauf, et al., PhRvD, 85, 102001 (2012).
\bibitem {Adhikari18} S. Adhikari, J. Sakstein, B. Jain, N. Dalal, B. Li, JCAP, 2018, 033 (2018).
\bibitem {Zu13} Y. Zu, D. H. Weinberg, MNRAS, 431, 3319 (2013).
\bibitem {Riotto02} A. Riotto, arXiv:hep-ph/0210162v1, (2002).
\bibitem {Tsujikawa03} S. Tsujikawa, arXiv:hep-ph/0304257, (2003).
\bibitem {Dalal08} N. Dalal, O. Dore, D. Huterer, A. Shirokov, Phys. Rev. D 77, 123514 (2008).
\bibitem {Planck20} Planck Collaboration:  Y. Akrami, F. Arroja, M. Ashdown, et al., A\&A,  641, A9 (2020).
\bibitem {Rizzi07} L. Rizzi, E. V. Held, I. Saviane, et al., MNRAS, 380, 1255 (2007).
\bibitem {Riess22} A. G. Riess, W. Yuan, L. M. Macri, et al., ApJL, 934, L7 (2022).
\bibitem {Planck20a} Planck Collaboration: N. Aghanim, Y. Akrami, M. Ashdown, et al., A\&A, 652, A6 (2020).
\bibitem {Zhang14} C. Zhang, H. Zhang, S. Yuan, T. J. Zhang, et al., RAA, 14, 1221 (2014).
\bibitem {Tony88} J. Tony, \& D. P. Schneider, AJ, 96, 807 (1988).
\bibitem {Suyu17} S. H. Suyu, V. Bonvin, F. Courbin, et al., MNRAS, 468, 2590 (2017).
\bibitem {Xiao23} L. Xiao, Z. Huang, Y. Zheng, X. Wang, X.-D. Li, MNRAS, 518, 6253-6261 (2023).
\bibitem {Xu23} K. Xu, Y.-P. Jing, G.-B. Zhao, and A. J. Cuesta, Nature Astronomy, 7, 1259-1264 (2023).
\bibitem {Cao22b} Y. Cao, Y. Gong, D. Liu, A. Cooray, C. Feng, and X. Chen, MNRAS, 511, 1830-1840 (2022).
\bibitem {Wang23} Z. Wang, J. Yao, X. Liu, D. Liu, Z. Fan, and B. Hu, MNRAS, 523, 3001-3017 (2023).
\bibitem {Deng22} F. Deng, Y. Gong, Y. Wang, S. Dong, Y. Cao, and X. Chen, MNRAS, 515, 5894-5904 (2022).
\bibitem {Jiang23} Y.-E. Jiang, Y. Gong, M. Zhang, et al., RAA, 23, 075003 (2023).
\bibitem {Song24} J.-Y. Song, L.-F. Wang, Y. Li, Z.-W. Zhao, J.-F. Zhang, W. Zhao, \& X. Zhang, Sci. China Phys. Mech. Astron., 67, 230411 (2024).
\bibitem {Jiang16aa} L. Jiang, I. D. McGreer, X. Fan, et al., ApJ, 833, 222 (2016).
\bibitem {Wang19aa} F. Wang, J. Yang, X. Fan, et al., ApJ, 884, 30 (2019).
\bibitem {Volonteri03} M. Volonteri,  F. Haardt, P. Madau, ApJ, 582, 559 (2003).
\bibitem {Kormendy13} J. Kormendy, \& L. C. Ho, ARA\&A, 51, 511 (2013).
\bibitem {Zhuang23} M.-Y. Zhuang, \& L. C. Ho, NatAs, 7, 1376 (2023).
\bibitem {Wang21} F. Wang, J. Yang, X. Fan, et al., ApJL, 907, 1 (2021).
\bibitem {Treister12} E. Treister, K. Schawinski, C.-M. Urry, et al., ApJL, 758, 39 (2012).
\bibitem {Fan16} L. Fan, Y. Han, G. Fang, et al., ApJL, 822, L32 (2016).
\bibitem {Fabian12} A. C. Fabian, ARA\&A, 50, 455 (2012).
\bibitem {Bluck22} A. F. L. Bluck, R. Maiolino, S. Brownson, et al., A\&A, 659, 160 (2022).
\bibitem {Heckman14} T. M. Heckman, \& P. N. Best, ARA\&A, 52, 589 (2014).
\bibitem {Ashok23} A. Ashok, A. Seth, P. Erwin,et al., ApJ, 958, 1 (2023).
\bibitem {Lemonias11} J. J. Lemonias, D. Schiminovich, D. Thilker, et al., ApJ, 733, 74 (2011).
\bibitem {Thilker23} D. A. Thilker, L. Bianchi, G. Meurer, et al., Formation and Evolution of Galaxy Disks, 396, 223 (2008).
\bibitem {Yi11} S. K. Yi, J. Lee, Y.-K. Sheen, et al., ApJS, 195, 22 (2011).
\bibitem {Binney87} J. Binney, \& S. Tremaine, Galactic Dynamics, (1987).
\bibitem {Kormendy04} J. Kormendy, \& R. C. Kennicutt, Jr., ARA\&A, 42, 603 (2004).
\bibitem {Hubble26} E. P. Hubble, ApJ, 64, 321 (1926).
\bibitem {Schreiber17} C. Schreiber, D. Elbaz, M. Pannella, et al., A\&A, 602, A96 (2017).
\bibitem {Peng10} C. Y. Peng, L. C. Ho, C. D. Impey, \& H.-W. Rix, AJ, 139, 2097 (2010).
\bibitem {Xu23aa} Q. Xu, S. Shen, R. S. de Souza, et al., MNRAS, 526, 6391 (2023).
\bibitem {Zhou22} C. Zhou, Y. Gu, G. Fang, \& Z. Lin, AJ, 163, 86 (2022).
\bibitem {Fang23} G. Fang, S. Ba, Y. Gu, et al., AJ, 165, 35 (2023).
\bibitem {Li22aa} R. Li, N. R. Napolitano, N. Roy, et al., ApJ, 929, 152 (2022).
\bibitem {Delmestre07} K. Menéndez-Delmestre, K. Sheth, E. Schinnerer, T. H. Jarrett, \& N. Z. Scoville, ApJ, 657, 790 (2007).
\bibitem {Yu18} S.-Y. Yu, L. C. Ho, A. J. Barth, \& Z.-Y. Li, ApJ, 862, 13 (2018).
\bibitem {Wilkinson17} D. M. Wilkinson, C. Maraston, D. Goddard, D. Thomas, \& T. Parikh, MNRAS, 472, 4297 (2017).
\bibitem {Lin19} X. Lin, G. Fang, Z.-Y. Cai, et al., ApJ, 875, 83 (2019).
\bibitem {Wang23aa} K. Wang, Y. Peng, Y. Chen, MNRAS, 523, 1268 (2023).
\bibitem {Peng10aa} Y. Peng, S. J. Lilly, K. Kovac, et al., ApJ, 721, 193 (2010).
\bibitem {Gu21} Y. Gu, G. Fang, Q. Yuan, S. Lu, \& S. Liu, ApJ, 921, 60 (2021).
\bibitem {Li21aa} Z. Li, B. Horowitz, \& Z. Cai, ApJ, 916, 20 (2021).
\bibitem {Yuan22} F. Yuan, Z. Zheng, P. T. Rahna, R. Lin, S. Zhu, Sci. China Phys. Mech. Astron, 52, 2 (2022).
\bibitem {Naidu18} R. P. Naidu, B. Forrest, P. A. Oesch, K.-V. H. Tran, \& B. P. Holden, MNRAS, 478, 791 (2018).
\bibitem {Wang25} X. Wang, H. I. Teplitz, B. M. Smith, et al., ApJ, 980, 74 (2025).
\bibitem {Cui12} X-Q. Cui, Y.-H. Zhao, Y.-Q. Chu, et al., RAA, 12, 1197-1242 (2012).
\bibitem {Lian23} J. Lian, M. Bergemann, A. Pillepich, G. Zasowski \& R. R. Lane, Nature Astronomy, 7, 951–958 (2023).
\bibitem {Dalcanton12} J. J. Dalcanton, B. F. Williams, D. Lang, et al., ApJS, 200, 18 (2012).
\bibitem {Li23aa} J. Li, C. Liu, Z. Y. Zhang, H. Tian, X. Fu, J. Li, \& Z. Q. Yan, Nature, 613, 460-462 (2023).
\bibitem {Cardelli89} J. A. Cardelli, G. C. Clayton, \& J. S. Mathis, ApJ, 345, 245 (1989).
\bibitem {Nie22} T. P. Nie, F.Y. Xiang, \& A. Li, MNRAS, 509, 4908 (2022).
\bibitem {Li19} Q. Li, A. Li, \& B. W. Jiang, MNRAS, 490, 3875 (2019).
\bibitem {Li20} Q. Li, A. Li, B. W. Jiang, \& T. Chen, MNRAS, 493, 3054 (2020).
\bibitem {Wang22} W. Wang, L. Zhu, Z. Li, et al., ApJ, 941, 2, 108 (2022).
\bibitem {Qu23} H. Qu, Z. Yuan, A. Doliva-Dolinsky, et al., MNRAS, 523, 1, 876 (2023).
\bibitem {Majewski17} S. R. Majewski, R. P. Schiavon, P. M. Frinchaboy, et al., AJ, 154, 94 (2017).
\bibitem {Carollo10} D. Carollo, T. C. Beers, M. Chiba, ApJ, 712, 692 (2010).
\bibitem {Weisz14} D. R. Weisz, A. E. Dolphin, E. D. Skillman, et al., ApJ, 789, 147 (2014).
\bibitem {Zheng20} Y. Zheng, A. Emerick, M. E. Putman, et al., ApJ, 905, 133 (2020).
\bibitem {Ercolano10} B. Ercolano, J. E. Owen, MNRAS, 406, 1553 (2010).
\bibitem {Wang19} L. Wang, X.-N. Bai, J. Goodman, ApJ, 874, 90 (2019).
\bibitem {Ricker15} G. R. Ricker, J. N. Winn, R. Vanderspek, et al., Journal of Astronomical Telescopes, Instruments, and Systems, 1, 014003 (2015).
\bibitem {Reynolds15} T. M. Reynolds, M. Fraser, G. Gilmore, MNRAS, 453, 2885-2900 (2015).
\bibitem {Han20} Z.-W. Han, H.-W. Ge, X.-F. Chen, et al., RAA, 20, 161 (2020).
\bibitem {Brown16} W. R. Brown, M. Kilic, S. J. Kenyon, et al., ApJ, 824, 46 (2016).
\bibitem {Wu18} Y. Wu, X. Chen, Z. Li, et al., A\&A, 618, A14 (2018).
\bibitem {Li21} J. Li, J. Wang, X. Lu, et al., ApJ, 919, 4 (2021).
\bibitem {Sun21} Y. Sun, D.-S. Deng, \& H.-B. Yuan, RAA, 21, 092 (2021).
\bibitem {Zhang21} L. Y. Zhang, G. Meng, L. Long, et al., ApJS, 253, 19 (2021).
\bibitem {Yang23} Z. Yang, L. Zhang, G. Meng, et al., A\&A, 669, A15 (2023).
\bibitem {Lissauer19} J. J. Lissauer \& J. Eisberg, New Astronomy Reviews, 04, 002 (2019).
\bibitem {Ren07} D. Q. Ren, \& Y. T. Zhu, The Publications of the Astronomical Society of the Pacific, 119, 1063 (2007).
\bibitem {Ren10} D. Ren, J. Dou, Y. Zhu, The Publications of the Astronomical Society of the Pacific, 122, 590 (2010).
\bibitem {Dou10} J. P. Dou, D.-Q. Ren, Y.-T. Zhu, Research in Astronomy and Astrophysics, 10, 189 (2010).
\bibitem {Dou16} J. P. Dou, \& D. Q. Ren, ApJ, 832, 84 (2016).
\bibitem {Dou14} J. P. Dou, D.-Q. Ren, X. Zhang, et al., Proceedings of the SPIE, 9147, 11 (2014).
\bibitem {Lacy19} B. Lacy, D. Shlivko, A. Burrows, ApJ, 157, 132 (2019).
\bibitem {Bowler18} B. Bowler, T. J. Dupuy, M. Endl, et al., ApJ, 155, 159 (2018).
\bibitem {Mao91} S. Mao, \& B. Paczynski, The Astrophysical Journal, 374, L37 (1991).
\bibitem {Gould92} A. Gould, \& A. Loeb, The Astrophysical Journal, 396, 104 (1992).
\bibitem {Suzuki16} D. Suzuki, D. P. Bennett, T. Sumi, et al., The Astrophysical Journal, 833, 145 (2016).
\bibitem {Zhu21} W. Zhu, \& S. Dong, Annual Review of Astronomy and Astrophysics, 59, 291 (2021).
\bibitem {Gould22} A. Gould, Y. K. Jung, K.-H. Hwang, et al., Journal of the Korean Astronomical Society, 55, 173 (2022).
\bibitem {Madhusudhan19} N. Madhusudhan, ARA\&A, 57, 617-663 (2019).
\bibitem {Barucci08} M. Barucci, H. Boehnhardt, D. Cruikshank, \& A. Morbidelli, The Solar System beyond Neptune, The University of Arizona Press (2008).
\bibitem {Batygin19} K. Batygin, F. Adams, M. Brown, et al., Physics Reports, 805, 1 (2019).
\bibitem {Hsieh06} H. H. Hsieh, \& D. Jewitt, Science, 312, 561 (2006).
\bibitem {Jewitt11} D. Jewitt, H. A. Weaver, M. Mutchler, et al., ApJ, 733, L4 (2011).
\bibitem {Hsieh12} H. H. Hsieh, B. Yang, \& N. Haghighipour, ApJ, 744,9 (2012).
\bibitem {Drahus15} M. Drahus, W. Waniak, S. Tendulkar, et al., ApJ, 802, L8 (2015).
\bibitem {Jewitt15} D. Jewitt, H. H. Hsieh, J. Agarwal, Asteroids IV, 221, 41 (2015).
\bibitem {Porco05} C. C. Porco, E. Baker, J. Barbara, et al., Science, 307, 1226 (2005).
\bibitem {Cooper08} N. J. Cooper, C. D. Murray, M. W. Evans, et al., Icarus, 195, 765 (2008).
\bibitem {Weaver06} H. A. Weaver, S. A. Stern, M. J. Mutchler, et al., Nature, 439, 943 (2006).
\bibitem {Sheppard03} S. S. Sheppard, \& D. C. Jewitt, Nature, 423, 261 (2003).
\bibitem {Sheppard10} S. S. Sheppard, \& C. A. Trujillo, Science, 329, 1304 (2010).
\bibitem {Sheppard19} S. S. Sheppard, C. A. Trujillo, D. J. Tholen, et al., ApJ, 157, 139 (2019).
\bibitem {Liao24} S. L. Liao, Z. X. Qi, \& J. C. Liu, Frontiers in Astronomy and Space Sciences, 11, 1396037 (2024).
\bibitem {Fu23} Z. S. Fu, Z. X. Qi, S. L. Liao, et al., Frontiers in Astronomy and Space Sciences, 10, 1146603 (2023).
\bibitem {Wu22} Q. Wu, S. L. Liao, X. Ji, et al., Frontiers in Astronomy and Space Sciences, 9, 822768 (2022).
\bibitem {Wu23} Q. Wu, S. Liao, Z. Qi, H. Luo, Z. Tang, \& Z. Cao, RAA, 23, 025006 (2023).
\bibitem {Wu24} Q. Wu, M. Scialpi, S. Liao, F. Mannucci, \& Z. Qi, A\&A, 692, A154 (2024).
\bibitem {Yao24} J. Yao, J. C. Liu, N. Liu, Z. Zhu, \& Z. W. Wang, RAA, 24, 085011 (2024).
\bibitem {Prusti16} T. Prusti, J. H. J. De Bruijne, A. G. Brown, et al., A\&A, 595, A1 (2016).
\bibitem {Gai22} M. Gai, A. Vecchiato, D. Busonero, Riva, et al., Frontiers in Astronomy and Space Sciences, 9, 1002876 (2022).
\bibitem {Sahu17} K. C. Sahu, J. Anderson, S. Casertano, et al., Science, 356, 1046-1050 (2017).
\bibitem {Kains17} N. Kains, A. Calamida, K. C. Sahu, S. Casertano, et al., ApJ, 843, 145 (2017).
\bibitem {Kluter20} J. Kluter, U. Bastian, \& J. Wambsganss, A\&A, 640, A83 (2020).
\bibitem {Wang12} B. Wang, Z. Han, New Astron. Rev., 56, 122 (2012).
\bibitem {Gal-Yam12} A. Gal-Yam, Science, 337, 927 (2012).
\bibitem {Bargiacchi25} G. Bargiacchi, M. G. Dainotti, \& S. Capozziello, New Astronomy Reviews, 100, 101712 (2025).
\bibitem {Lorimer07} D. R. Lorimer, M. Bailes, M. A. McLaughlin, et al., Science, 318, 777 (2007).
\bibitem {Spitler16} L. G. Spitler, P. Scholz, J. W. T. Hessels, et al., Nature, 531, 202 (2016).
\bibitem {Xu23} J. Xu, Y. Feng, D. Li, et al., Universe, 9, 330 (2023).
\bibitem {Li21bb} D. Li, P. Wang, W. W. Zhu, et al., Nature, 598, 267-271 (2021).
\bibitem {Zhang22} Y.-K. Zhang, P. Wang, Y. Feng, et al., RAA, 22, 124002 (2022).
\bibitem {Chatterjee17} S. Chatterjee, C. J. Law, R. S. Wharton, et al., Nature, 541, 58 (2017).
\bibitem {Zhang23} B. Zhang, Reviews of Modern Physics, 95, 035005 (2023).
\bibitem {IceCube13} IceCube Collaboration, Science, 342, 6161 (2013).
\bibitem {Halzen17} F. Halzen, Nature Physics, 13, 232 (2017).
\bibitem {WangT12} T.-G. Wang, H.-Y. Zhou, S. Komossa, et al., ApJ, 749, 115 (2012).
\bibitem {Jiang16} N. Jiang, L. Dou, T. Wang, et al, ApJ, 828, L14 (2016).
\bibitem {Wang23bb} T. Wang, G. Liu, Z. Cai, et al., SCPMA, 66, 109512, (2023).
\bibitem {Yuan25} W. Yuan, L. Dai, H. Feng, et al., SCPMA, 68, 239501, (2025).

\end{thebibliography}
\end{document}